\documentclass[fleqn,usenatbib]{mnras}

\usepackage{graphicx}	
\usepackage{amsmath}	
\usepackage{amssymb}	
\usepackage{multicol}        
\usepackage{bm}		
\usepackage{pdflscape}	
\usepackage{verbatim}
\usepackage{natbib}
\usepackage{newtxtext,newtxmath}
\usepackage{hyperref} 
\usepackage{xcolor}
\usepackage{float}
\usepackage[T1]{fontenc}
\usepackage{ae,aecompl}
\usepackage{soul} 

\defcitealias{Qin2019a}{Q19}

\title[The Origins of Scatter in the IRX Relation]{The Physical Origins of Scatter in the Dust Attenuation Scaling Relation of Star-Forming Galaxies: Star--Dust Geometry}

\author[Z. Lyu et al.]{
Zongfei Lyu,$^{1,2}$
X.~Z.~Zheng,$^{3}$\thanks{E-mail: xzzheng@sjtu.edu.cn}
Zhizheng Pan,$^{1,2}$
A. Katsianis,$^{4}$
Stijn~Wuyts,$^{5}$
Haiguang~Xu,$^{6}$
P. Cataldi,$^{7}$
\and 
Wenhao Liu,$^{1,2}$
Man~Qiao,$^{8}$
Dong~Dong Shi,$^{9}$
Yuheng~Zhang,$^{10,11}$
Shuang~Liu,$^{12}$
Run~Wen,$^{3}$
Chao~Yang$^{1,2}$
\\
$^{1}$Purple Mountain Observatory, Chinese Academy of Sciences, 10 Yuanhua Road, Nanjing 210023, China\\
$^{2}$School of Astronomy and Space Science, University of Science and Technology of China, Hefei 230026, China\\
$^{3}$State Key Laboratory of Dark Matter Physics, Tsung-Dao Lee Institute, Shanghai Jiao Tong University, Shanghai 201210, China \\
$^{4}$School of Physics and Astronomy, Sun Yat-sen University, Zhuhai Campus, 2 Daxue Road, Xiangzhou District, Zhuhai, China\\
$^{5}$Department of Physics, University of Bath, Claverton Down, Bath BA2 7AY, UK \\
$^{6}$School of Physics and Astronomy, Shanghai Key Laboratory for Particle Physics and Cosmology, Shanghai Jiao Tong University, Shanghai 200240, China \\
$^{7}$Universidad Nacional Aut\'onoma de M\'exico, Instituto de Astronom\'ia, A.P. 70-264, 04510, Ciudad de M\'exico, M\'exico\\
$^{8}$School of Physics and Electronic  Engineering, Jining University, Jining 273155, China \\
$^{9}$Center for Fundamental Physics, School of Mechanics and Optoelectronic Physics,
Anhui University of Science and Technology, Huainan 232001, China\\       
$^{10}$School of Astronomy and Space Science, Nanjing University, Nanjing 210093, China \\
$^{11}$Key Laboratory of Modern Astronomy and Astrophysics (Nanjing University), Ministry of Education, Nanjing 210093, China \\
$^{12}$College of Physics and Electronic Engineering, Huaibei Normal University, Huaibei 235000, China
}

\date{Accepted 2026 September 24. Received 2026 September 23; in original form 2026 June 8}
\pubyear{2026}

\begin{document}
\label{firstpage}
\pagerange{\pageref{firstpage}--\pageref{lastpage}}

\maketitle

\begin{abstract}
Characterising the physical drivers of dust attenuation in star-forming galaxies (SFGs) is essential for interpreting their spectral energy distributions and star formation histories. The infrared excess (IRX$\equiv$$L_{\mathrm{IR}}/L_{\mathrm{UV}}$) follows a universal scaling relation with metallicity, star formation rate, galaxy size, and inclination. However, the origin of the scatter around this relation remains poorly understood. We revisit this relation using $\sim$32,000 local SFGs from SDSS, GALEX, and WISE, and investigate why some galaxies deviate systematically from the best-fit relation. We find that the deviations are systematically linked to UV luminosity. UV-faint SFGs ($\log(L_{\mathrm{UV}}/\mathrm{L}_\odot)\le 9$) exhibit a median IRX excess of +0.23\,dex, while UV-bright SFGs ($\log(L_{\mathrm{UV}}/\mathrm{L}_\odot)\ge 10$) show a median deficit of $-$0.20\,dex relative to the relation defined by the dominant UV-intermediate population (86.38\,per\,cent of the sample). These offsets are not driven by metallicity, total infrared luminosity, or specific star formation rate. Instead, the deviations are closely linked to systematic differences in star--dust geometry. UV-faint SFGs are compact (median $R_{\mathrm{e}} = 2.90$\,kpc) and preferentially viewed edge-on (median $b/a = 0.41$), leading to high effective dust column densities along the line of sight. UV-bright SFGs are extended (median $R_{\mathrm{e}} = 5.99$\,kpc) and preferentially viewed face-on (median $b/a = 0.74$), allowing UV photons to escape efficiently. Galaxies with significant bulge components ($B/T > 0.4$) populate the envelope of the relation, exhibiting larger scatter. We conclude that the scatter in the IRX relation can be largely explained by variations in the three-dimensional star--dust geometry, with UV luminosity acting as an effective tracer of these geometric and physical differences.
\end{abstract}

\begin{keywords}
dust, extinction -- galaxies: evolution -- galaxies: ISM -- galaxies: star formation -- galaxies: statistics
\end{keywords}


\section{Introduction}\label{sec1}

The interstellar medium (ISM) of galaxies hosts complex interactions between gas, stars, and dust, which primarily determine the observable properties of these galaxies.  Despite comprising only $\sim 1$\,per\,cent of the ISM mass, interstellar dust plays a key role in the baryon and energy cycles. It provides a surface for molecule formation, acts as a coolant for star and planet formation, and facilitates the conversion of radiation pressure into mechanical energy of outflows. More importantly, dust essentially reshapes galactic spectral energy distributions (SEDs) by absorbing ultraviolet (UV)-optical light and re-emitting it in the infrared (IR) \citep{Galliano2021, Calura2025}. This process, known as dust attenuation, is one of the largest systematic uncertainties in the interpretation of galaxy SEDs. Accurate correction for dust attenuation is essential for deriving reliable estimates of key physical quantities such as star formation rates (SFRs), stellar masses ($M_\ast$), and star formation histories (SFHs) \citep{Kennicutt1998, Gadotti2010, Popescu2011, Iyer2020, Lower2022, Kouroumpatzakis2023, Markov2025}. Consequently, understanding the physical processes that govern dust attenuation, and how these processes scale with global and internal galactic properties, has become a central task in modern astrophysics.

Characterising dust attenuation in star-forming galaxies (SFGs) is profoundly influenced by the intrinsic UV slope of the stellar population, the spatial distribution of dust relative to stars (geometry),  and the optical properties of dust grains in the ISM \citep{Sachdeva2022, Lower2022, Hamed2023, Zhang2023}.  In the absence of resolved multi-wavelength imaging for distant galaxies, integrated empirical proxies are often used.  The infrared excess (IRX), defined as the ratio of total infrared luminosity ($L_{\mathrm{IR}}$) to observed UV luminosity ($L_{\mathrm{UV}}$), quantifies the fraction of bolometric UV light that is absorbed by dust and re-radiated in the IR, thereby serving as an effective integrated measure of dust obscuration \citep{Heckman1998, Meurer1999, Heinis2013, Bourne2017}. Alternative tracers of dust attenuation include the UV spectral slope $\beta_{\mathrm{UV}}$ (where $f_\lambda \propto \lambda^{\beta_{\mathrm{UV}}}$) and the Balmer decrement (e.g., H$\alpha$/H$\beta$). In the local Universe, IRX is observed to correlate with $\beta_{\mathrm{UV}}$, forming the well-known IRX--$\beta_{\mathrm{UV}}$ relation. However, this relation exhibits significant scatter and systematic variations with galactic parameters such as stellar mass, metallicity, and SFR \citep{Kong2004, Shivaei2020, Schulz2020, Popping2022, Bowler2024}.  Similarly, the Balmer decrement -- a standard nebular tracer -- exhibits systematic discrepancies from IRX depending on SFR surface density \citep[e.g.,][]{Qin2019b}. Recent results from the JADES and AURORA surveys with the James Webb Space Telescope (JWST) have further demonstrated a remarkable diversity in nebular attenuation curves at $z > 3$, suggesting that non-unity dust covering fractions and varying dust compositions must be accounted for \citep{Shapley2023, Maheson2024, Reddy2026}. In particular, studies of JADES galaxies at $2.7 < z < 7$ indicate that the relationship between nebular and stellar reddening is not universal, but evolves with galaxy properties \citep{Karthikeyan2026, Tsujita2026}. 

The physical origin of these deviations is rooted in the secondary parameter dependencies of attenuation. Foremost among these is gas-phase metallicity ($12+\log({\mathrm{O/H}})$).  Metallicity is a fundamental parameter governing the lifecycle of dust and thus the dust-to-gas ratio (DGR). A strong correlation therefore exists between the DGR and gas-phase metallicity, a relation that has now been traced from the local Universe out to cosmic noon and beyond \citep{DeVis2019, Casasola2020, Shapley2020, Popping2023, Konstantopoulou2024}, leading to a systematic metallicity-dependent attenuation, where low-metallicity systems appear UV-bright but IR-faint, resulting in an IRX deficit relative to more evolved systems \citep{Heckman1998, Reddy2018}.  Observations from surveys such as MOSDEF and recent JWST programs have robustly established that dust attenuation properties vary systematically with metallicity: lower-metallicity galaxies exhibit weaker 2175\,\AA\ bumps, steeper far-UV slopes, and generally lower attenuation for a given $\beta_{\mathrm{UV}}$ \citep{Shivaei2020, Shivaei2025}.  

While the global dust attenuation of SFGs depends on the dust content and chemical composition, the geometry of stars and dust has emerged as perhaps the most critical factor driving scatter in the IRX--$\beta_{\mathrm{UV}}$ relation and related attenuation diagnostics \citep{Salim2020, Lower2022, Mitsuhashi2024a, Reddy2026}. The size of a galaxy, often characterised by its half-light radius ($R_{\mathrm{e}}$), and its inclination (traced by the axis ratio $b/a$) are direct observational probes of this geometry. The collective evidence from recent observational and theoretical work makes a compelling case that dust attenuation is governed by a confluence of factors. Indeed, \citet[hereafter \citetalias{Qin2019a}]{Qin2019a} derived a universal IRX scaling relation (hereafter the IRX relation) for local SFGs, expressing IRX as a power-law function of $L_{\mathrm{IR}}$, $R_{\mathrm{e}}$, $b/a$, and gas-phase metallicity. This relation was successfully reproduced by an empirical two-component star--dust geometry model, in which both the diffuse ISM and the distribution of dense birth clouds are described by exponential discs \citep{Qin2024}. The parameters of this model, including the relative scale heights and optical depths of the two components, vary systematically with metallicity, providing a physical foundation for the empirical scaling relation. It is also worth mentioning recent numerical efforts to test \citetalias{Qin2019a}\,relation using cosmological simulations, which have helped to assess both its robustness and the limitations imposed by current galaxy formation and dust models (e.g. \citealt{Qiao2024} using EAGLE; Cataldi et al., submitted, using IllustrisTNG).

However, even with the inclusion of these parameters, a persistent residual scatter remains. Emerging evidence from the JWST and Atacama Large Millimeter/submillimeter Array (ALMA) observations has complicated the picture further.  At the highest redshifts ($z > 12$), the ALMA observations suggest low dust masses despite high star formation efficiency, pointing towards a rapid early enrichment of the ISM \citep{Mitsuhashi2026}. Observations of high-redshift ($z > 5$) galaxies have revealed an unexpected diversity in IRX values. Some early galaxies show dust emission significantly lower than predicted by local relations \citep{Fudamoto2020, Bowler2024}. Conversely, JWST has uncovered  populations of extremely dust-obscured galaxies that are invisible in deep UV surveys but dominate the IR energy budget \citep{Barrufet2023, Sun2506.06418}. 
These high-$z$ findings emphasise that the effective dust attenuation is governed by a confluence of factors, including metallicity-dependent grain growth and the evolving star--dust geometry \citep{Fujimoto2025, Lorenz2026}. In particular, the apparent decoupling of dust attenuation from inclination angle at z\,$\sim$\,1.3--2.6 suggests that dust is not simply distributed in a thin, planar disc but rather in a more isotropic, clumpy configuration, often described by the ``chocolate chip cookie'' model \citep{Lu2022, Lorenz2023, Zhang2023}. Such geometric complexity naturally explains the observed decoupling of attenuation from inclination and has profound implications for interpreting the UV colours and SEDs of high-$z$ galaxies \citep{Maheson2025}.

Nevertheless, a significant fraction of SFGs exhibit considerable scatter (${\sim} 0.2$\,dex) around the mean IRX relation derived by \citetalias{Qin2019a}. This scatter is not merely observational noise but likely encodes valuable physical information about deviations from the average star--dust geometry and stellar population properties. Several factors may contribute to this scatter. First, a substantial fraction of the observed $L_{\mathrm{IR}}$ in galaxies with relatively low sSFR may be powered by dust heated by evolved stellar populations rather than by young, UV-bright stars \citep{Bendo2012, Tailor2025}. Second, the structural parameter $R_{\mathrm{e}}$ is typically derived from optical $r$-band imaging, which primarily traces the older stellar population; the spatial distribution of young stars and dust, which are directly responsible for the UV-to-IR energy budget, may differ significantly from that of the old stars \citep{Bianchi2007, Smith2016}. Third, the presence of even a modest bulge component or non-disc morphology can alter the effective star--dust geometry in ways not fully captured by simple disc parameters \citep{Gadotti2010, Pastrav2013a}. Finally, sample selection effects and fitting methodologies can introduce artificial biases that inflate the apparent scatter \citep{SalimNarayanan2020, Qin2022}.

In this study, we revisit the universal dust attenuation scaling relation using a sample of $\sim$32,000 local SFGs drawn from SDSS, GALEX, and WISE. We systematically investigate the physical origins of the scatter and systematic offsets around the IRX relation, with particular focus on disentangling the roles of stellar population age (as traced by sSFR) and star--dust geometry (as traced by $R_{\mathrm{e}}$, $b/a$, and bulge-to-total ratio).  
The paper is organised as follows. In Section~\ref{sec2}, we present the galaxy sample and data. Section~\ref{sec3} presents our analysis of the dependence of IRX deviation on various galaxy properties, revealing a strong correlation with UV luminosity. In Section~\ref{sec4}, we provide a detailed discussion of the physical interpretation of our results, demonstrating why specific SFR is not the primary driver of scatter and highlighting the dominant role of star--dust geometry. We also address biases in sample selection and fitting methodology and explain why UV luminosity acts as an effective tracer of the systematic offsets. Finally, we summarise our main conclusions in Section~\ref{sec5}.

\begin{figure*}
	\centering
	\includegraphics[width=0.86\textwidth]{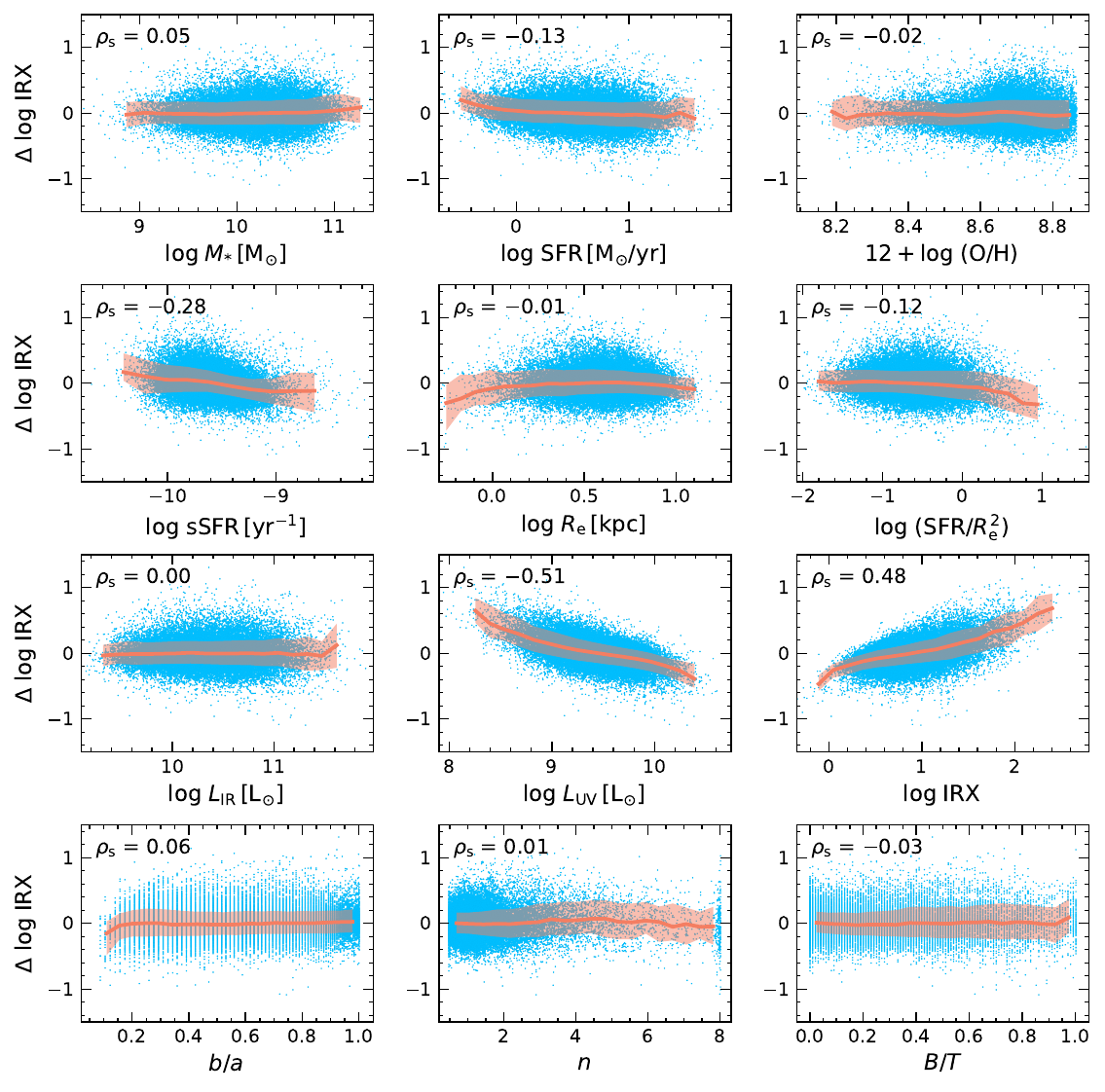}
	\caption{Relationships of IRX deviation with 12 physical and structural parameters for our sample of SFGs. In each panel, the solid curve indicates the running median while the shaded band encloses  the 16--84th percentiles. Spearman's rank correlation $\rho_{\mathrm s}$ is given to show the strength of the correlation in each panel.}
	\label{fig1}
\end{figure*}

\begin{figure*}
	\centering
	\includegraphics[width=0.79\textwidth]{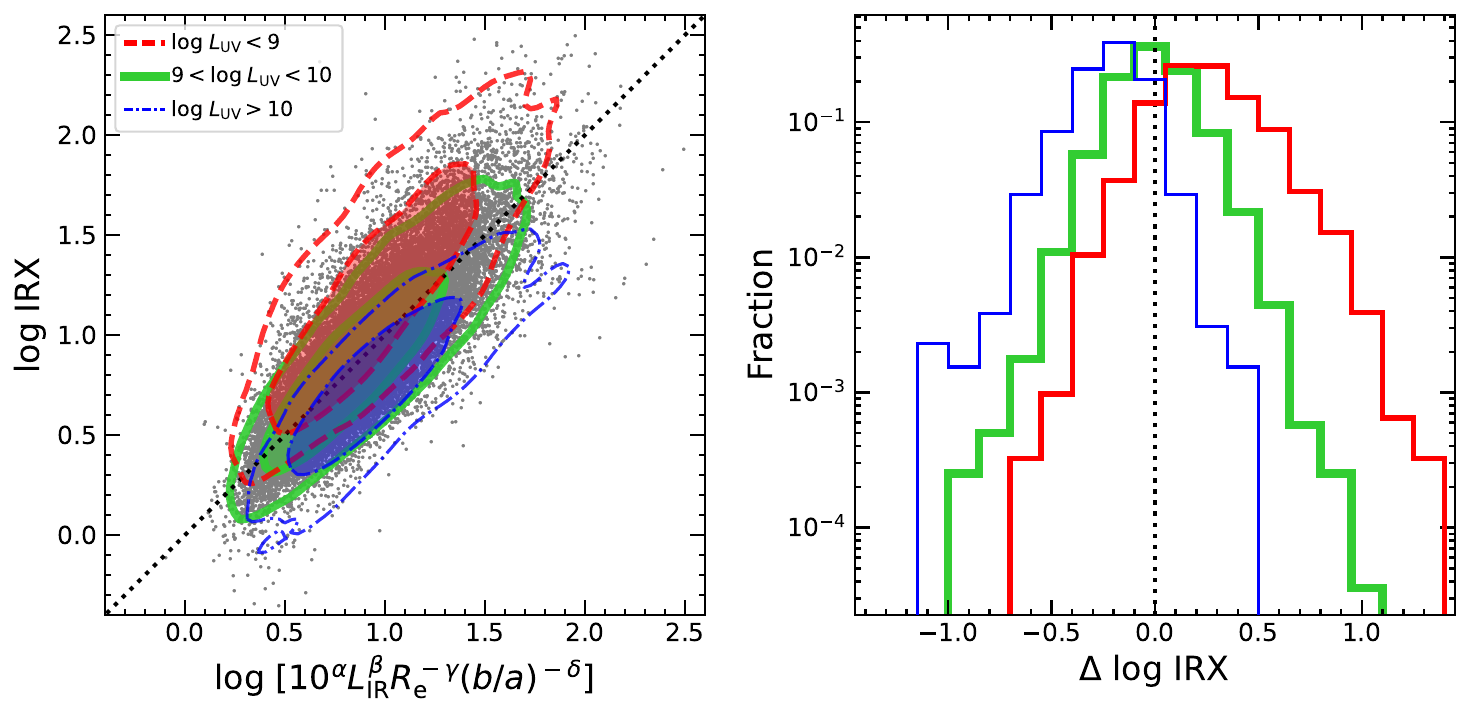}
	\caption{Left panel: comparison of the observed IRX with the inferred IRX in three $L_{\mathrm{UV}}$-selected  subsamples: UV-faint (red dash-dotted), UV-intermediate (green solid), and UV-bright (blue dashed). The two contours enclose 68 and 95 percentiles of the sample galaxies in each bin, respectively. The grey points represent local SFGs from \protect\citetalias{Qin2019a}. Right panel: the normalised distribution of IRX deviations in the three subsamples}
	\label{fig2}
\end{figure*}

\section{Galaxy Sample and Data} \label{sec2}

Our analysis is based on the SFG sample from \citetalias{Qin2019a}, consisting of 32\,354 local galaxies. This sample is selected using data from the Sloan Digital Sky Survey Data Release 10 \citep[SDSS DR10,][]{Ahn2014}, the Galaxy Evolution Explorer \citep[GALEX, ][]{Martin2005}, and the Wide-field Infrared Survey Explorer (WISE) All Sky Survey \citep{Wright2010}. We briefly summarise the key sample characteristics below.

The redshift range is restricted to $0.04<z<0.15$. The lower limit ensures SDSS fibers cover $>20$\% of total starlight in typical galaxies \citep{Kewley2005}, while also excluding galaxies with fiber coverage fraction $<0.2$ (fiber-to-total stellar mass ratio) -- both minimizing systematic differences between nuclear and global metallicity measurements. The upper limit reduces evolutionary effects. Emission lines (H$\alpha$, [N\,{\sc ii}]$\lambda$6584, H$\beta$, [O\,{\sc iii}]$\lambda$5007) meet signal-to-noise (S/N) thresholds of $>$20, 3, 3, and 2, respectively. These lines are used jointly in the BPT diagnostic \citep{Baldwin1981} to select SFGs and measure gas-phase metallicity traced by the Oxygen abundance 12+log(O/H). We require S/N $>2$ for WISE 22\,$\micron$ fluxes to estimate infrared (IR) luminosities. For GALEX NUV and FUV fluxes, detections are limited to S/N $>3$ to enable reliable ultraviolet luminosity calculation.

This study investigates the scatter in the IRX relation, and thus, the directly associated physical parameters include: $L_{\mathrm{IR}}$, $L_{\mathrm{UV}}$, 12+log(O/H), $R_{\mathrm{e}}$, and $b/a$. The $L_{\mathrm{IR}}$ (8--1000\,$\micron$) is derived from WISE 22\,$\micron$ fluxes using luminosity-dependent infrared templates from \citet{Chary2001}. The observed UV luminosity $L_{\mathrm{UV}}$ (1216--3000\AA) is calculated by integrating the best-fit galaxy SED template over the observed GALEX FUV, NUV, and SDSS $u$-band fluxes. Structural parameters derive from \citet{Simard2011}, who performed disc+bulge decomposition of SDSS $r$-band images. The semi-major axis half-light radius from the best-fitting $r$-band disc+bulge model is adopted as $R_{\mathrm{e}}$. The axis ratio, $b/a$, is taken from the disc component of the same model. We note that this axis ratio is calculated as $b/a = \cos(i)$, where $i$ is the disc inclination angle. Since the inclination angle $i$ in \citet{Simard2011} was restricted to integer degrees, the resulting $b/a$ values are discrete.

Additional parameters with potential secondary influence on the IRX relation include: $M_\ast$, SFR, specific star formation rate (sSFR$\equiv$SFR/$M_\ast$), S\'ersic index ($n$), and bulge-to-total ratio ($B/T$). $M_\ast$ comes from the MPA/JHU value-added catalog of SDSS DR10. SFR is estimated from IR and UV luminosities following \citet{Bell2005}. The quantity sSFR serves as a galaxy age proxy, where elevated values indicate a young stellar population or starburst activity. For galaxies with regular star formation histories, sSFR inversely correlates with mean stellar population age \citep{Chruslinska2024}. The parameters $n$ and $B/T$ are taken from \citet{Simard2011}, based on their pure S\'ersic and bulge+disc decompositions, respectively. We retain the 22\,$\micron$ single-band calibration for estimating $L_{\mathrm{IR}}$, following \citetalias{Qin2019a}.

\begin{figure*}
	\centering
	\includegraphics[width=0.79\textwidth]{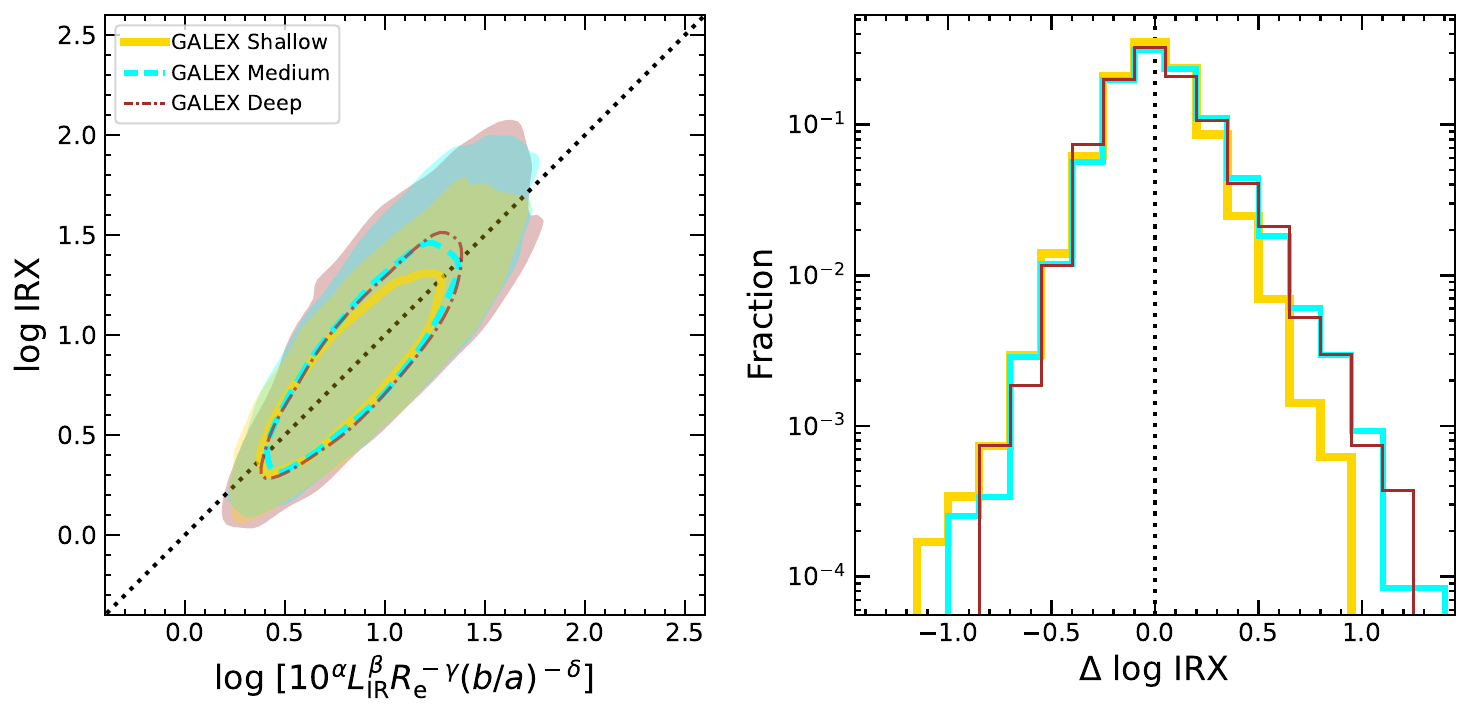}
	\caption{Left: distribution of sample galaxies in the IRX relation colour-coded by the depth of the GALEX UV imaging surveys: shallow (gold solid), medium (cyan dotted), and deep (red dash-dotted), containing 54.8, 36.8, and 8.4\,per\,cent of sample galaxies. The two contours enclose 68 and 95\,per\,cent of the sample in each bin, respectively. Right: corresponding distributions of IRX deviations ($\Delta \log \mathrm{IRX}$) for the same three survey-depth subsamples. Both panels show that the depth of the GALEX imaging survey does not introduce a systematic effect on the location or scatter of galaxies in the IRX relation.}
	\label{fig3}
\end{figure*}

\section{Results} \label{sec3}

\subsection{Dependence of IRX Deviation on Galaxy Properties}

The IRX deviation defined as $\Delta \log \mathrm{IRX}=\log \mathrm{IRX}_{\mathrm{observed}} - \log \mathrm{IRX}_\mathrm{predicted}$ is adopted to quantify the scatter of the IRX relation. Here, the reference relation from \citetalias{Qin2019a} is given by
\begin{equation} 
{\mathrm{IRX_{predicted}}} = 10^\alpha\, \left(\frac{L_{\mathrm{IR}}}{10^{10}\rm L_\odot}\right)^{\beta}\,\left(\frac{R_{\rm e}}{\rm kpc}\right)^{-\gamma}\,(b/a)^{-\delta},
\end{equation} 
where $\alpha$, $\beta$, $\gamma$ and $\delta$ are parametrized as linear functions of gas-phase metallicity. In order to investigate the physical origin of the scatter in the IRX relation, we examine possible correlations between $\Delta \log \mathrm{IRX}$ and a suite of observed and derived galaxy properties. The observed and derived parameters/quantities include 12+log(O/H), $L_{\mathrm{IR}}$, $L_{\mathrm{UV}}$, $R_{\mathrm{e}}$, $n$, $b/a$, $B/T$, $M_{\ast}$, SFR, sSFR, and SFR surface density (SFR/$R_{\mathrm{e}}^{2}$). Since the IRX relation itself is formulated using gas-phase metallicity, infrared luminosity, effective radius, and axis ratio, it is expected that $\Delta \log \mathrm{IRX}$ shows no significant dependence on these quantities, their combinations, or on $M_{\ast}$ (see \citetalias{Qin2019a} for details).

Figure~\ref{fig1} shows the relationships of IRX deviation with 12 physical parameters/quantities among local SFGs. We confirm that no correlation is found for the IRX deviation with $M_{\ast}$, 12+log(O/H), $R_{\mathrm{e}}$, $L_{\mathrm{IR}}$, $b/a$, $n$, and $B/T$, while weak tendencies are found with SFR, sSFR, and SFR/$R_{\mathrm{e}}^{2}$, $L_{\mathrm{UV}}$ and IRX. Particularly, the systematic trend with $L_{\mathrm{UV}}$ stands out as strongest. We notice that IRX, SFR, sSFR, and SFR/$R_{\mathrm{e}}^{2}$ are all combinations of two or three from $L_{\mathrm{IR}}$, $L_{\mathrm{UV}}$, $M_{\ast}$ and $R_{\mathrm{e}}$, and $L_{\mathrm{UV}}$ is the only one among the four parameter showing a correlation with IRX deviation. Therefore, these empirical tendencies suggest that the correlation of IRX deviations with sSFR and other parameters may originate indirectly from the underlying trend with $L_{\mathrm{UV}}$.

\subsection{Dependence of IRX Deviation on UV Luminosity}

To clarify the systematic trend of IRX deviation with $L_{\mathrm{UV}}$, we separate our sample SFGs into three bins based on UV luminosity: UV-faint with $\log (L_{\mathrm{UV}}/\mathrm{L}_{\odot}) \le 9$, UV-intermediate with $9 < \log (L_{\mathrm{UV}}/\mathrm{L}_{\odot}) < 10$, and UV-bright with $\log (L_{\mathrm{UV}}/\mathrm{L}_{\odot}) \geq 10$. These bins contain 3100, 27\,947, and 1307 galaxies, corresponding to 9.58, 86.38, and 4.04\,per\,cent of the total sample, respectively. The UV-intermediate SFGs clearly constitute the dominant population that defines the IRX relation. As shown in Figure~\ref{fig2}, the three UV-luminosity subsamples display clear, systematic offsets from the IRX relation. In the left panel, the scatter around the IRX relation is seen to depend systematically on UV luminosity. We find that UV-faint SFGs  exhibit a median excess in observed IRX of $+0.23$\,dex relative to the relation; the UV-intermediate SFGs follows the IRX relation closely, with scatter consistent with observational uncertainties; in contrast, UV-bright SFGs  show a median deficit of $-$0.20\,dex relative to the predicted IRX. The right panel of Figure~\ref{fig2} presents the distribution of $\Delta \log \mathrm{IRX}$ for each subsample. The median values are $0.23 \pm 0.23$, $-0.01 \pm 0.17$, and $-0.20 \pm 0.17$ for the low, intermediate, and high UV-luminosity bins, respectively. This three-bin scheme is adopted to highlight the outliers against the vast majority of normal SFGs. We point out that further subdividing the dominant UV-intermediate bin still yields the same systematic trend between $\Delta\log\mathrm{IRX}$ and UV luminosity.

Our sample incorporates GALEX UV photometry drawn from the \citet{Salim2016} catalogue, which combines data from surveys of varying depth: shallow (all-sky), medium, and deep. To evaluate whether these differing observational limits influence our results, we examine the distribution of $\Delta \log \mathrm{IRX}$ as a function of GALEX survey depth in the left panel of Figure~\ref{fig3}. The median values and interquartile ranges of $\Delta \log \mathrm{IRX}$ for the three survey depths, as quantified in the right panel of Figure~\ref{fig3}, are $0.01 \pm 0.19$, $0.01 \pm 0.22$, and $0.00 \pm 0.22$, respectively. These statistics demonstrate no significant variation across the different survey depths. We therefore conclude that the observed correlation between $\Delta \log \mathrm{IRX}$ and UV luminosity is robust and essentially independent of the depth of the GALEX observations.

\begin{figure*}
	\centering
	\includegraphics[width=0.79\textwidth]{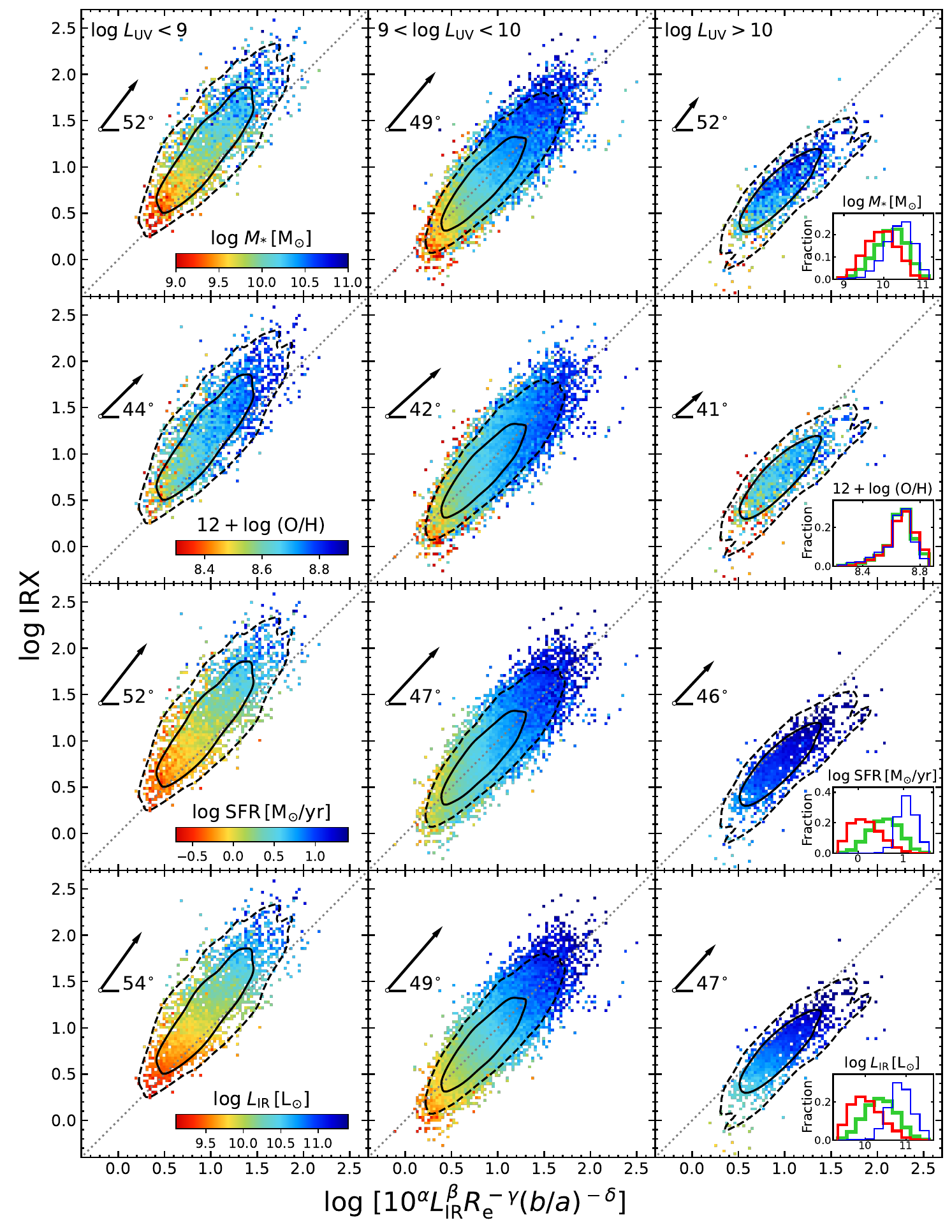}
	\caption{Comparison of the IRX relation for our sample of SFGs divided into three UV-luminosity bins: UV-faint with $\log L_{\mathrm{UV}} \le 9$ (left), UV-intermediate with $9 < \log L_{\mathrm{UV}} < 10$ (middle), and UV-bright with $\log L_{\mathrm{UV}} \ge 10$ (right), colour-coded by $M_\ast$, 12+log(O/H), SFR and $L_\mathrm{IR}$ from the top to the bottom. The solid and dashed contours enclose 68 and 95\,per\,cent of subsample galaxies in each bin. The arrow in each panel indicates the variation direction and magnitude (arrow length) of the given parameter across the IRX relation (see text for more details). The inset panels on the right present the distributions of the given parameters for the UV-faint (red), UV-intermediate (green-thick) and UV-bright (blue-thin) subsamples. The variations of the four parameters globally align with the IRX relation for all of the three UV-luminosity bins.}
	\label{fig4}
\end{figure*}

\begin{figure*}
	\centering
	\includegraphics[width=0.79\textwidth]{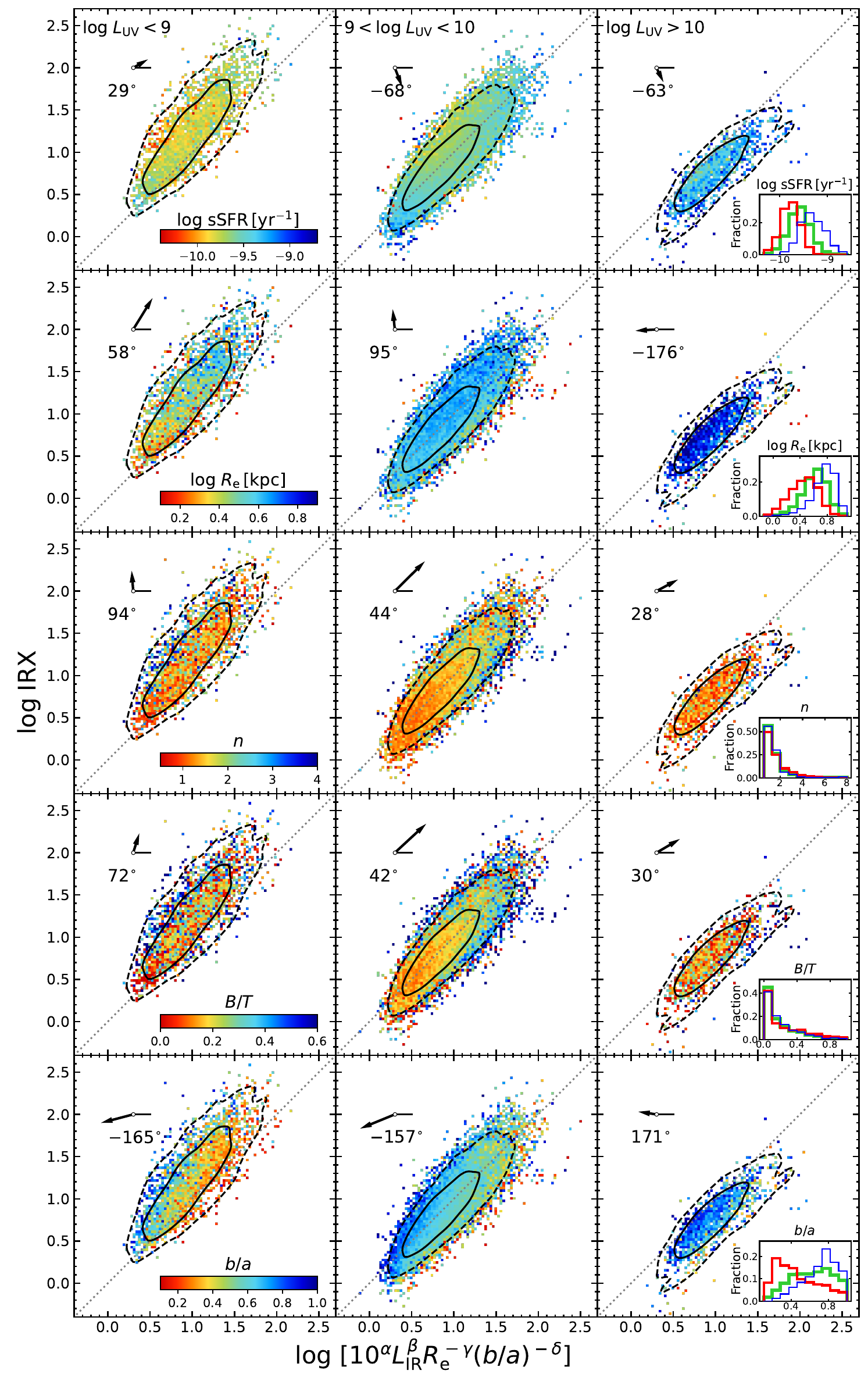}
	\caption{Comparison of the IRX relation for the UV-faint ($\log L_{\mathrm{UV}} \le 9$), UV-intermediate ($9 < \log L_{\mathrm{UV}} < 10$), and UV-bright ($\log L_{\mathrm{UV}} \ge 10$) subsamples from left to right, colour-coded by sSFR, $R_\mathrm{e}$, $n$, $B/T$ and $b/a$ from the top to the bottom. The contours, arrows and inner panels are set in the same way as in Figure~\ref{fig4}.}
	\label{fig5}
\end{figure*}

\begin{figure*}
	\centering
	\includegraphics[width=0.85\textwidth]{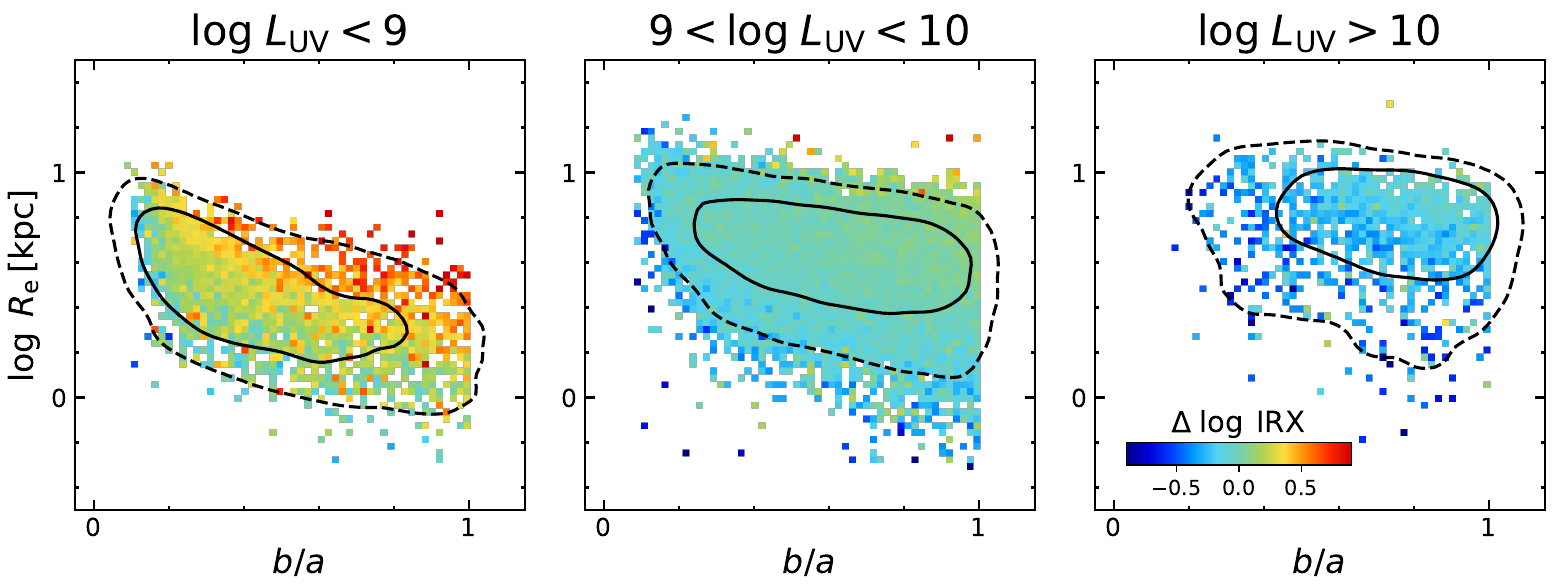}
	\caption{Relationship between effective radius and axis ratio for UV-faint (left), UV-intermediate (middle) and UV-bright SFGs from our sample. The solid and dashed contours enclose 68 and 95\,per\,cent of total number of SFGs in given subsamples.}
	\label{fig6}
\end{figure*}

\subsection{The Physical Properties of UV-Selected Galaxy Subsamples}\label{sec3.3}

Despite correcting for the systematic offsets identified within each UV-luminosity bin (Figure~\ref{fig2}), a significant dispersion in the IRX deviation still persists. We now investigate whether this residual scatter can be attributed to secondary physical parameters, and why UV luminosity itself is primarily coupled with the systematic offsets from the IRX relation. As previously noted, the UV-faint, UV-intermediate, and UV-bright subsamples constitute 9.58, 86.38, and 4.04\,per\,cent of the total sample, respectively. The systematic offsets of the UV-faint and UV-bright SFGs are the principal contributors to the overall scatter observed around the IRX relation.

To explore the distribution of physical properties within the IRX parameter space, we adopt a colour-coding scheme. The data are first binned into a two-dimensional grid spanning the IRX diagram. For each cell, we compute the median value of a given physical parameter; this median is then mapped to a colour according to the accompanying colour bar. The grid cell size is set to $0.03 \times 0.03$ in logarithmic units. Note that the number of sample galaxies in individual cells can vary by more than two orders of magnitude across the diagram. This colour coding approach enables us to trace systematic variations in physical parameters across the IRX relation and to identify potential drivers of the observed scatter.

Figures~\ref{fig4} and~\ref{fig5} present the distribution of various physical parameters as a function of deviation from the IRX relation, with the UV-faint, UV-intermediate, and UV-bright subsamples arranged from left to right. The inset panels on the right show the distribution of these parameters for each of the three UV-luminosity subsamples. To quantify the variation of a given parameter across the IRX diagram, we overplot an arrow connecting two characteristic loci. Specifically, we divide the data cells into a lower half and a higher half  in terms of the given parameter. For each half, we calculate its parameter-weighted centroid in the IRX diagram. The arrow is then drawn to connect the loci of the centroids.  The direction of the arrow indicates how the parameter trend aligns with the IRX relation; an orientation close to 45$\degr$ suggests that the parameter variation consistent with being directed along the underlying relation. The length of the arrow scales with the magnitude of this variation.

Figure~\ref{fig4} shows the distribution of stellar mass ($M_\ast$), gas-phase metallicity ($12+\log({\mathrm{O/H}})$), star formation rate (SFR), and total infrared luminosity ($L_{\mathrm{IR}}$) across the IRX relation for the three UV-luminosity subsamples. In nearly all cases, the arrows are oriented close to 45$\degr$, indicating that these parameters increase systematically along the IRX relation. This reflects the well-established scaling relations: more massive SFGs tend to exhibit higher star formation activity, higher metal enrichment, and correspondingly larger IRX values. However, while IRX correlates with $M_\ast$, $12+\log({\mathrm{O/H}})$, SFR and $L_{\mathrm{IR}}$ in accordance with these fundamental relations, the deviation from the IRX relation shows no such correlation. This behaviour is consistent across all three UV-luminosity subsamples, confirming that stellar mass, metallicity, SFR, and infrared luminosity are not the primary factors responsible for the deviations from the IRX relation.

In contrast, as shown in the inset panels of Figure~\ref{fig4}, the three UV-luminosity subsamples exhibit systematic differences: UV-faint SFGs appear to be statistically lower in SFR and $L_{\mathrm{IR}}$, while UV-bright SFGs show higher values relative to the UV-intermediate subsample. Stellar mass displays a similar but weaker trend. These trends naturally reflect the star-forming main sequence (MS): lower-mass galaxies inherently possess lower absolute SFRs, thereby generating less intrinsic UV radiation and heating less dust (yielding lower $L_{\mathrm{IR}}$), and vice versa for UV-bright systems. No significant differences are observed in gas-phase metallicity among the three subsamples, likely because the expected metallicity variations driven by stellar mass and SFR effectively cancel each other out, as dictated by the Fundamental Metallicity Relation.

Figure~\ref{fig5} presents the distribution of specific SFR and the structural parameters including $R_{\mathrm{e}}$, $n$, $B/T$ and $b/a$ across the IRX relation for the three UV-luminosity subsamples. In contrast to the parameters presented in Figure~\ref{fig4}, sSFR and these structural parameters do not exhibit a prominent scaling pattern along the IRX relation. The arrows, which display varied orientations and considerably shorter lengths, differ dramatically from those in Figure~\ref{fig4}, although a weak gradient can be seen with $n$ and $B/T$ only for the UV-intermediate subsample. Importantly, the distribution of these parameters in the IRX diagram reveals correlations with deviations from the IRX relation.It is evident from Figure~\ref{fig5} that the UV-faint and UV-bright SFGs exhibit statistically distinct distributions of sSFR, $R_{\mathrm{e}}$ and $b/a$ compared to the UV-intermediate SFGs (we discuss the observational biases driving this $b/a$ difference below), whereas the three subsamples show broadly similar distributions in $n$ and $B/T$. All three subsamples are predominantly composed of disc galaxies with negligible to modest bulge components, specifically characterised by $n < 2.5$ and $B/T < 0.3$. A S\'ersic index of $n$ < 2.5 is widely adopted as the threshold for disc-dominated systems, as lower values trace the extended, exponential-like light profiles typical of discs, whereas higher values indicate centrally concentrated spheroidal or bulge-dominated morphologies \citep{Shen2003, Graham2005}. Relative to the dominant UV-intermediate population, UV-faint SFGs tend to have lower sSFR, smaller $R_{\mathrm{e}}$ and lower $b/a$ (indicative of more edge-on orientations), while UV-bright SFGs exhibit higher values in these parameters.  Quantitatively, the median sSFR values are 0.168, 0.257, and 0.471\,Gyr$^{-1}$ for the UV-faint, UV-intermediate and UV-bright subsamples, respectively. The corresponding median $R_{\mathrm{e}}$ values increase from  2.90 to 4.24 to 5.99\,kpc, and the median $b/a$ values from 0.41 to 0.63 to 0.74. To ensure these size differences are not an artefact of the stellar mass--size relation, we compare the subsamples at fixed $M_*$ and SFR. Even after controlling for these variables, UV-faint galaxies remain systematically more compact than the UV-bright ones across the $M_*$--SFR plane, showing that the size difference is not driven by these scaling relations (see Appendix~\ref{AppendixB} and Figure~\ref{figB1} for detailed bin-by-bin comparisons). In terms of fractional distributions, the proportion of SFGs with $\log (R_\mathrm{e}/\mathrm{kpc})<0.6$ is 75\,per\,cent  among UV-faint, 44\,per\,cent among UV-intermediate, and 18\,per\,cent among UV-bright SFGs, whereas the fraction with $\log (R_\mathrm{e}/\mathrm{kpc})\geq 0.6$ increases from 25 to 56 to 82\,per\,cent across the same sequence. Similarly, the fraction of SFGs with $b/a<0.5$ declines from 64\,per\,cent (UV-faint) to 33\,per\,cent (UV-intermediate) to 15\,per\,cent (UV-bright), while those with $b/a\geq 0.5$ correspondingly rise from 36 to 67 to 85\,per\,cent. These systematic differences reflect a combination of intrinsic galaxy properties and selection effects. UV-faint SFGs are intrinsically less active and more compact, while UV-bright SFGs are intrinsically more active and more extended. The difference in b/a is instead driven by viewing angle: edge-on orientations maximise the line-of-sight dust optical depth and suppress the observed UV luminosity, whereas face-on orientations minimise dust attenuation, so that the former are preferentially observed as UV-faint and the latter as UV-bright.

An intriguing feature emerges in the distribution of $n$ and $B/T$. The envelope region of the IRX relation is predominantly occupied by SFGs with $B/T \gtrsim 0.4$, whereas the inner region exhibits a gradient from $B/T\sim 0$ to $B/T\sim 0.3$ along the IRX sequence, from its lower to upper end. This behaviour is most clearly seen in the UV-intermediate subsample, which constitutes the dominant population of our sample. Similar features are also present in the UV-faint and UV-bright subsamples, though they appear less pronounced owing to statistical fluctuations arising from the substantially smaller number of SFGs in these two subsets. The increase of $B/T$ along the IRX relation is consistent with the overall scaling relations governed by stellar mass. In contrast, the association of the envelope region -- characterised by large deviations -- with $B/T \gtrsim 0.4$ reveals that the presence of a substantial bulge component is linked to significant deviations from the IRX relation defined by disc-dominated SFGs.

The distribution of $n$ mirrors that of $B/T$, with the envelope region dominated by SFGs of a high S\'ersic index ($n \gtrsim 2.5$), while the inner region displays a gradient from $n\sim 1$ to $n\sim 2$. This correspondence is not surprising because a tight correlation exists between $B/T$ and $n$ for galaxies of regular disc+bulge morphologies.

The distribution of axis ratio $b/a$ in the IRX diagram also show some systematic trends across the three UV-luminosity subsamples. It has been pointed out that the majority (64\,per\,cent) of UV-faint SFGs have $b/a<0.5$, compared to the vast majority (85\,per\,cent) of UV-bright SFGs with $b/a>0.5$. In contrast, UV-intermediate SFGs span a broad range over $0.1<b/a<0.9$. Moreover, for both UV-faint and UV-intermediate SFGs, the distribution of $b/a$ in the IRX diagrams shows a decreasing gradient from the left to the right side of the IRX relation. This gradient is largely absent for UV-bright SFGs.

\begin{figure*}
	\centering
	\includegraphics[height=0.285\textwidth]{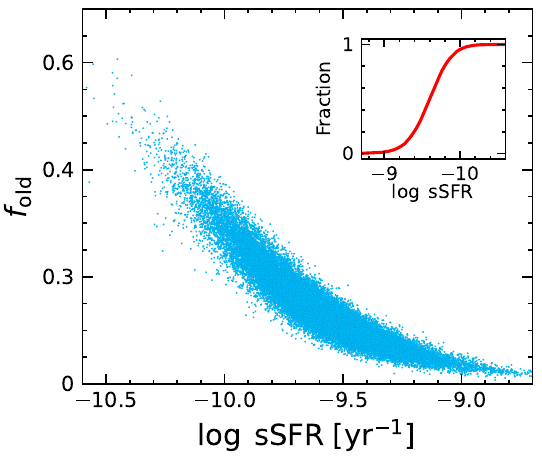}
	\includegraphics[height=0.285\textwidth]{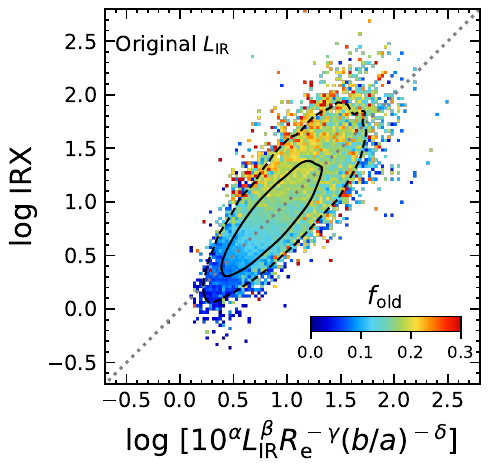}
	\includegraphics[height=0.285\textwidth]{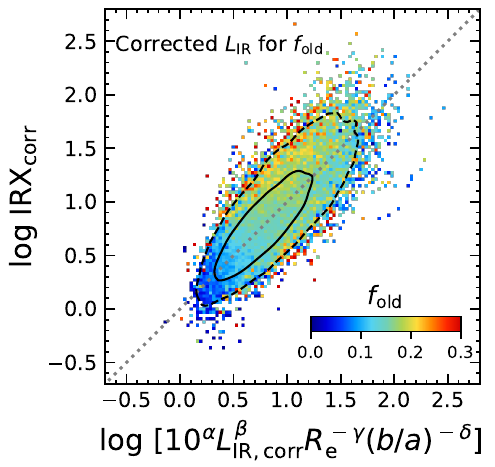}
	\caption{\textit{Left:} Fraction of infrared luminosity contributed by old stellar population $f_{\mathrm{old}}$ as a function of sSFR for our SFG sample. The inner panel shows the cumulative distribution of $f_{\mathrm{old}}$ from the high-sSFR end to low-sSFR end. \textit{Middle:} The IRX relation of our sample SFGs based on the total infrared luminosity. \textit{Right:} The modified IRX relation based on the corrected infrared luminosity for the contribution from the old stellar population. A regular $0.03 \times 0.03$\,dex grid is applied across the IRX diagram, and each cell is colour coded with the median $f_{\mathrm{old}}$ of sample SFGs within the cell. The solid and dashed contours enclose 68 and 95\,per\,cent of sample galaxies. With correction applied, these data points with relatively-high $f_{\mathrm{old}}$ move towards the low end of the IRX relation, and the dispersion around the IRX relation changes from 0.201\,dex to 0.199\,dex. In the modified IRX relation, the best-fitting normalization power index $\alpha$ becomes $\alpha=0.93\log(Z/\mathrm{Z}_\odot)+1.29$ relative to the original $\alpha=0.96\log(Z/\mathrm{Z}_\odot)+1.22$, while the other indices $\beta$, $\gamma$ and $\delta$ remain nearly unchanged.}
	\label{fig7}
\end{figure*}

We investigate whether observational or selection effects could account for the negative $b/a$ gradient across the IRX relation for UV-faint and UV-intermediate SFGs. To this end, we examine the relationships between axis ratio, effective radius and IRX deviation in Figure~\ref{fig6}. Few sample SFGs populate the bottom-left region of this diagram (characterised by low $R_\mathrm{e}$ and low $b/a$), whereas a notable excess appears in the top-left region (high $R_\mathrm{e}$ and low $b/a$). This excess of high-$R_\mathrm{e}$, low-$b/a$ SFGs becomes more pronounced with decreasing UV luminosity. Importantly, $R_\mathrm{e}$ generally increases with $M_*$ following the mass--size relation for SFGs. The scarcity of low-$R_\mathrm{e}$, low-$b/a$ systems likely reflects a combination of intrinsic structural properties and survey selection effects: compact, low-mass galaxies are intrinsically more spheroidal and thus lack very low axis ratios, while edge-on disc galaxies suffer enhanced dust attenuation along elongated sightlines, lowering their apparent surface brightness below the SDSS spectroscopic targeting limits. Conversely, the excess of high-$R_\mathrm{e}$, low-$b/a$ SFGs likely arises from an observational bias affecting massive SFGs, which appear fainter in the UV when oriented edge-on. We also note that the underdensity of extended, face-on, UV-faint galaxies in the upper-right corner may similarly stem from surface-brightness detection limits.

We caution that the presence of even a modest bulge component in a host disc can modify the observed axis ratio, particularly for edge-on orientations, causing the measured axis ratio to increase and the galaxy to deviate from the expectation for purely disc-dominated systems. However, this effect is unlikely to be responsible for the observed $b/a$ gradient, as SFGs with high $B/T$ are preferentially located in the outer regions of the IRX relation on both sides. Therefore, we attribute the negative gradient in $b/a$ to a combination of observational and sample selection effects: the IRX relation is primarily occupied by low-mass SFGs with intrinsically rounder morphologies and higher $b/a$ at its lower end, with a gradually increasing fraction of low-$b/a$ SFGs towards higher IRX.

Taken together, our findings suggest that structural deviations from a disc-dominated morphology contribute to the scatter around the IRX relation. We next examine how deviations in sSFR and $R_\mathrm{e}$ are associated with offsets from the IRX relation.

\subsection{IRX Offsets Correlated with Deviations in specific SFR and Galaxy Size}\label{sec:3.4}

Relative to UV-intermediate SFGs, UV-faint SFGs exhibit statistically smaller sizes and lower sSFRs, with a systematic offset towards higher IRX by $+$0.23\,dex. Conversely, UV-bright SFGs tend to be larger and host higher sSFRs, showing systematically lower IRX by $-$0.20\,dex. An offset in IRX may arise from a genuine difference in dust opacity, or from biases in using observed IRX as a tracer of dust attenuation -- for instance, due to unaccounted contributions from evolved stellar populations to $L_\mathrm{IR}$ or $L_\mathrm{UV}$. 

The observed $L_\mathrm{IR}$ is not powered exclusively by young stellar populations. In galaxies with significant evolved populations, optical and near-infrared photons from old stars can heat the dust, contributing non-negligibly to the total infrared budget even in the absence of strong UV emission \citep{Dale2005, Draine2007, Bendo2012, Katsianis2021}. Therefore, it is essential to disentangle the impact of both young and old stellar populations on the observed $L_\mathrm{UV}$ and $L_\mathrm{IR}$. 

\subsubsection{Dust Emission Powered by Evolved Stellar Populations}\label{sec:3.4.1}

To quantify the contribution of  evolved stellar populations to dust heating and, subsequently, to the total infrared luminosity, we employ the analytic star--dust geometry model developed by \citet{Qin2024}. By parametrizing the spatial distributions of stars and dust, this model has been shown to successfully reproduce the IRX relation observed among local SFGs. While retaining the best-fitting geometric parameters from that work, we introduce an explicit old stellar disc component to account for the differential attenuation experienced by different stellar populations.

We assume that a disc galaxy is composed of two distinct stellar populations: a young component (age $<100$\,Myr) with a constant SFH, and an old component with an exponentially declining SFH, an e-folding time of $\tau = 1$\,Gyr, and an age of $3$\,Gyr, corresponding to a mass-weighted age of $2.1$\,Gyr \citep{Salim2018, Nersesian2019, Nersesian2020, Law2021}. The old stellar population is spatially distributed as an exponential disc whose scale length and scale height are identical to those of the dust disc.  Note that the old stellar population is usually more extended than the young stellar population and the dust in nearby SFGs \citep{Xilouris1999, Yoachim2006, Popescu2011, DeGeyter2014, Casasola2017}. If the scale length and scale height of the old stellar population are twice of the those of the dust disc, then the radiation density heating dust could drop by 6.2\,per\,cent. This bias could be higher for  SFGs with lower sSFR.  Therefore, our assumption provides an upper limit for the contribution of the old stellar population in heating dust. 

Based on this configuration, the mass ratio between young and old stellar populations is constrained by the sSFR and the  lifetime of the young stellar population. Combining the intrinsic spectra from \citet{BC03} and the attenuation curve from \citet{Calzetti2000}, together with the energy balance principle, we derive the fraction ($f_{\mathrm{old}}$) of the total infrared luminosity powered by the old stellar population. Specifically, the infrared luminosity contributed by a given stellar population ($i \in \{\mathrm{young}, \mathrm{old}\}$) is calculated by integrating the difference between its intrinsic and attenuated spectral energy distributions following 
\begin{equation} 
L_{{\mathrm{IR}}, i} = \int L_{\lambda, i}^{\mathrm{int}} \left(1
	- 10^{-0.4 A_{\lambda, i}}\right) {\mathrm d}\lambda,
\end{equation} 
where $L_{\lambda, i}^{\mathrm{int}}$ is the intrinsic luminosity and $A_{\lambda, i}$ is the dust attenuation for that population. The fraction of dust emission powered by old stars is then simply given by 
\begin{equation} 
	f_{\mathrm{old}} = \frac{L_{\mathrm{IR, old}}}{L_{\mathrm{IR, young}} + L_{\mathrm{IR, old}}}.
\end{equation}

As shown in the left panel of Figure~\ref{fig7}, $f_{\mathrm{old}}$ follows a tight correlation with sSFR. It is obvious that a SFG with a lower sSFR contains a higher fraction of old stellar population and thus a higher  $f_{\mathrm{old}}$. The median value of $f_{\mathrm{old}}$ is  $0.13 \pm 0.08$, broadly consistent with the estimates of $f_{\mathrm{old}}$ for nearby SFGs in the literature \citep{Nersesian2019, Nersesian2020, Law2021, Paspaliaris2021, Calzetti2025}. The cumulative distribution of $f_{\mathrm{old}}$ along sSFR shows that  the vast majority  (95\,per\,cent) of our sample galaxies at $\log ({\mathrm{sSFR}}/{\mathrm{yr}^{-1}}) > -10$  have $f_{\mathrm{old}} \lesssim 0.3$, suggesting that the total infrared luminosity is dominantly powered by the young stellar population  for the majority of our sample galaxies.

The middle panel of Figure~\ref{fig7} shows $f_{\mathrm{old}}$ across the IRX relation.  We adopt a regular $0.03 \times 0.03$\,dex grid in the \(\log{\mathrm{IRX}}\) diagram, where each cell is colour-coded by the median $f_{\mathrm{old}}$ of galaxies within it, with redder colours corresponding to a higher contribution fraction.  It is clear that the sample SFGs with relatively high $f_{\mathrm{old}}$ (${\gtrsim} 0.3$) are located above the IRX relation. These SFGs  represent only  5\,per\,cent of the entire sample and tend to have relatively higher IRX at given $L_{\mathrm{IR}}$. After correcting for the contribution from old stellar populations,  the IRX relation for pure young stellar population is shown in the right panel of Figure~\ref{fig7}. The median value of $\log (L_{\mathrm{IR}}/\mathrm{L}_{\rm \odot})$ changes from $10.37 \pm 0.4$ to $10.29 \pm 0.4$ and the median of $\log {\mathrm{IRX}}$ changes from $0.93 \pm 0.35$ to $0.81 \pm 0.35$. The SFGs of high $f_{\mathrm{old}}$ are no longer preferentially concentrated above the IRX relation. The fraction of the SFGs in our sample with $f_{\mathrm{old}} > 0.3$ located above the IRX relation changes from 2.5\,per\,cent to 2.2\,per\,cent, while the fraction located below the relation changes from 1.9\,per\,cent to 2.1\,per\,cent.

We correct the total infrared luminosity for the contribution from old stellar population and update both $L_{\mathrm{IR}}$ and IRX. The IRX relation based on the corrected $L_{\mathrm{IR}}$ and IRX exhibits a scatter of 0.199\,dex, in comparison with the scatter of 0.201\,dex without correction applied, indicating that the dispersion around the IRX relation does not change significantly even if the infrared emission from dust powered by the old stellar population is removed. Taken together, these results suggest that  sSFR (or stellar age) is not associated with the driver of the scatter around the IRX relation.

\subsubsection{The Difference in Dust Opacity Associated with Star--Dust Geometry}

The systematic variation of IRX with UV luminosity is fundamentally linked to differences in effective dust opacity driven by galaxy size. For a fixed dust mass, assuming that the physical scale of the dust distribution broadly scales with the global optical size of the stellar disc (i.e. $R_{\rm dust} \propto R_{\rm e}$), a more compact galaxy with smaller optical $R_{\rm e}$ naturally corresponds to a higher average dust mass surface density, $\Sigma_{\rm dust} \propto M_{\rm dust}/R_{\rm e}^2$, and consequently a higher effective optical depth ($\tau \propto \Sigma_{\rm dust}$); conversely, a more extended region yields a lower dust surface density and a lower optical depth.  In our sample, UV-faint SFGs are predominantly compact: 75\,per\,cent have $\log(R_{\mathrm{e}}/\text{kpc}) < 0.6$, whereas UV-bright galaxies are significantly more extended, with 82\,per\,cent having $\log(R_{\mathrm{e}}/\text{kpc}) \geq 0.6$---a distribution that lies beyond the size range of UV-intermediate SFGs.  

The impact of size on opacity is amplified by inclination. UV-faint SFGs are not only compact but preferentially viewed at higher inclinations (median $b/a = 0.41$), maximizing the path length through the dusty ISM. Conversely, UV-bright SFGs are extended and predominantly face-on (median $b/a = 0.74$), allowing a larger fraction of UV light to escape unattenuated. This size--inclination synergy underscores the need for a three-dimensional description of dust geometry, as encapsulated in the two-component exponential disc model of \citet{Qin2024}.


 The  star--dust geometry model from \citet{Qin2024} includes two exponential discs as structural components for the diffuse and dense ISM, together with UV-emitting young stars.   For simplification, the evolved stellar populations are ignored as their contribution to dust heating is not significant in starburst galaxies.  The UV-emitting young stars are assumed to homogeneously mix with their birth clouds in the dense ISM, following the same exponential disc, while the diffuse ISM forms a more extended exponential disc.  

In this model,  five key parameters, $F_{\text{bc}}$, $C_{\text{bc}}$, $\tau_{\text{bc}}$, $\hat{R}$, and $\hat{H}$, are used to tune the star--dust geometry and decide the output model IRX.  Briefly, \(F_{\text{bc}}\) represents the fraction of total dust mass  in birth clouds. This parameter is a power-law function of gas-phase metallicity jointly determined by a constant \(F_{\text{bc},\odot}\) and a power-law index \(\eta\).   $\hat{R}$ is the scale length ratio of the UV-emitting stellar disc to the dust disc, describing their relative radial distribution: $\hat{R}<1$ indicates a more compact stellar disc compared to the dust disc, and vice versa. $\hat{H}$ is the scale height ratio of the stellar disc to the dust disc, describing their relative vertical distribution: $\hat{H}<1$ implies a thinner stellar disc than the dust disc. Since the radial scale ratio and vertical scale height ratio have highly similar effects on IRX, both parameters modulate attenuation efficiency by changing the average path length of UV photons through dust. We therefore follow the convention in \citet{Qin2024} and set \(\hat{R}\equiv\hat{H}\) to simplify model calculation, using a single parameter to describe the three-dimensional scale difference between the stellar and dust discs.  $\tau_{\text{bc},\odot}$ characterises the UV optical depth of individual birth clouds and quantifies the degree to obscure UV radiation from the embedded young stars.  $C_{\text{bc}}$ is the escape fraction of UV photons from birth clouds, determined by $C_{\text{bc},\odot}$ and a power-law index $\nu$ encoding the metallicity dependence, which regulates the fraction of UV radiation from young stars that escapes directly without being obscured by birth clouds.  The effective optical depth of birth clouds is determined by \(\tau_{\text{bc}}\) and \(C_{\text{bc}}\), and not affected by \(F_{\text{bc}}\). This model does not count mutual obscuration between birth clouds,  and all birth clouds have identical effective optical depth, indicating that the integrated IRX over all birth clouds equals that of a single birth cloud. The model IRX of the dense dust is independent of the global dust surface density of a galaxy. For the diffuse dust, the effective optical depth is governed by \(F_{\text{bc}}\), \(\hat{R}\), and \(\hat{H}\), corresponding to the content of diffuse dust and the relative spatial distribution of the diffuse dust disc with respect to the stellar disc. We refer readers to \citet{Qin2024} for more details about the model. 

\begin{figure}
	\centering
	\includegraphics[width=0.9\columnwidth]{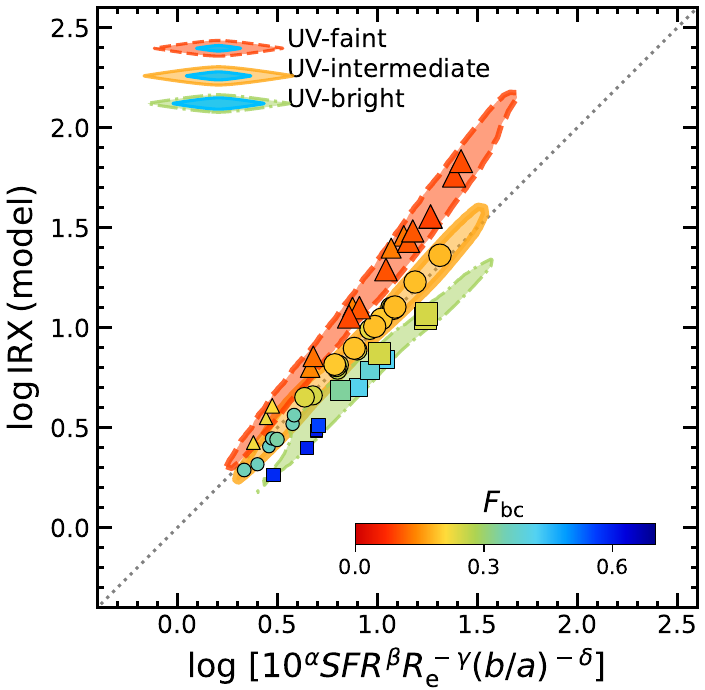}
	\caption{Comparison of predicted IRX from three sets of structural parameter for the star--dust geometry model best fitting the UV-faint (triangles), UV-intermediate (circles), and UV-bright (squares) SFGs, respectively. The symbols represent individual bins in each of the three subsample divided in parameter space. They are colour coded by $F_\mathrm{bc}$, and their size is scaled by gas-phase metallicity. The colour contours indicate the 68\,per\,cent distributions of the three subsamples, with colours corresponding to their median $F_\mathrm{bc}$ values. In the top-left corner disk symbols demonstrate the relative scales of the stellar (blue) and colour dust discs for the three subsamples. Their size is proportional to the median effective radius of each subsample.}
	\label{fig_geo}
\end{figure}

We suspect that the IRX offsets between the three UV-selected subsamples is mainly attributed to the systematic  differences in the three-dimensional star--dust geometry. To test this, we examine whether the IRX offsets between the three UV-selected subsamples (shown in Figure~\ref{fig2}) can be reproduced using the model with different sets of parameters.  Following \citet{Qin2024}, we set \(\hat{R}\equiv\hat{H}\), as their effects on IRX are highly similar. Note that $F_{\text{bc}}$ and $\hat{R}$ (and $\hat{H}$)  govern the geometry of the diffuse ISM component relative to the dense ISM component, while the model parameters $C_{\text{bc}}$ and $\tau_{\text{bc}}$ determine the dust attenuation of dense birth clouds. 

In practice, we construct test samples data by binning each of the three UV-selected subsamples in physical parameter space: three bins in $12+\log \rm{(O/H)}$ over [8.22,  8.87] with a bin width of 0.22, two bins of $[-0.5, 0.55]$ and $(0.55, 1.7]$ in $\log \rm{SFR}$, two bins of $[-0.2, 0.6]$ and $(0.6, 1.1]$ $\log R_{\mathrm{e}}$, and two bins of $[-0.9, -0.3]$ and $(-0.3, 1]$ in $\log (b/a)$. We take the median value of each parameter within each bin and exclude those with fewer than 3 objects, resulting in 17, 24, and 11 test samples for the three UV-selected  subsamples, respectively. These test samples represent the overall parameter distributions of the full sample.

Taking the best-fitting model parameters from \citet{Qin2024} as the reference, we tune each parameter individually to assess how the model IRX changes compared with the reference expectation in terms of the IRX relation (derived from global galaxy properties).  Our experiments reveal that parameters $F_{\text{bc}}$, $\hat{R}$ and $\hat{H}$ jointly control the overall degree of the IRX offset (see also figure~1 of \citealt{Qin2024}). They modulate the global dust attenuation by regulating the dust fraction in the diffuse ISM, exerting a similar effect across the full  range of IRX and manifesting as a systematic vertical shift in the offset distribution. In contrast, the scale ratio $\hat{R}$ mainly governs the IRX-dependence of the offset, with its influence becoming most pronounced at the high-IRX end.  Note that the predicted IRX from the IRX relation reflects the effective dust column density. Even for a fixed dust content, variations in the star--dust geometry can still change the predicted IRX because of the well-mixed distribution of stars and dust within a galaxy, other than a simple foreground screen geometry.

In doing so, we are able to successfully reproduce the IRX offsets of the UV-selected subsamples.  Figure~\ref{fig_geo} presents the model results matching the IRX offsets between the three UV-selected subsamples. We adopt the best-fit parameters for typical disc galaxies from \citet{Qin2024}, derived from a large sample of local SFGs, as our reference values. Under these reference parameters, model-predicted IRX values closely match those from the IRX relation, producing a typical model offset of \(0.01 \pm 0.03\)\,dex. This demonstrates that the reference model reliably reproduces the mean geometric behaviour of typical SFGs, which corresponds to the UV-intermediate SFGs that make up 86.7\,per\,cent of our sample. At Solar metallicity, the geometric parameters that best fit the UV-intermediate SFGs are $(F_{\mathrm{bc}}, \eta, \hat{R}) = (0.15, -1.69, 0.46)$. In contrast, the parameter set $(F_{\mathrm{bc}}, \eta, \hat{R}) = (0.11, -1.2, 0.35)$ generates an IRX offset of $\Delta\log\mathrm{IRX}_{\text{model}}=+0.23 \pm 0.10$\,dex for the UV-faint SFGs, while reproducing a negative offset of $\Delta\log\mathrm{IRX}_{\text{model}}=-0.20 \pm 0.03$\,dex for the UV-bright SFGs requires $(F_{\mathrm{bc}}, \eta, \hat{R}) = (0.26, -1.9, 0.65)$.

It is evident from Figure~\ref{fig_geo} that the IRX values derived from the reference star--dust geometry for UV-intermediate SFGs closely follow the universal IRX scaling relation. Statistically, the positive IRX offset (+0.23\,dex) for UV-faint SFGs arises from two geometric effects: a lower $F_{\text{bc}}$ leads to a larger fraction of dust in the diffuse ISM, increasing the effective column density, while a smaller disc of UV-emitting stars  relative to the dust disc increases the optical path across the diffuse ISM. Together, these two effects increase the effective optical depth, resulting in a higher IRX. Conversely, the negative IRX offset ($-$0.20\,dex) for UV-bright SFGs is attributed to the reversal of the two geometric effects:  a  higher $F_{\text{bc}}$ and a larger stellar-to-dust scale ratio  $\hat{R}$. The former results in a higher fraction of dust in dense birth clouds and thus a lower dust column density in the diffuse ISM. The latter leads to a smaller effective optical depth for UV-optical photons.  Therefore, the systematic changes in the star--dust geometry naturally account for the systematic IRX offsets between the three UV-selected subsamples and the observed IRX-dependent trend of the offsets.

\section{Discussion}\label{sec4}

The universal dust attenuation scaling relation, as formulated by \citetalias{Qin2019a} and further explored in this work, represents a significant step towards a unified description of dust obscuration across a wide range of galaxy populations and redshifts.  By linking the infrared excess (IRX) to a combination of global parameters -- IR luminosity, effective radius, axis ratio, and gas-phase metallicity -- the relation captures the bulk behaviour of dust reprocessing across a diverse galaxy population. However, as demonstrated in Sections~\ref{sec3} and~\ref{sec4}, a non-negligible intrinsic scatter persists, and systematic offsets are observed in galaxies with extreme UV luminosities. In this section, we dissect the physical origins of this scatter and these offsets, focusing on the role of star--dust geometry, the limitations of specific SFR as a driver, and the methodological biases that can influence the inferred scaling relations. We conclude by synthesizing these findings to explain why UV luminosity acts as an effective tracer of the systematic deviations from the IRX relation.

\subsection{IRX as A Probe of Dust Column Density}

The infrared excess is, at its core, a bolometric proxy for the fraction of starlight absorbed and re-radiated by dust. In the simple geometric limit of a foreground screen, IRX is directly related to the effective optical depth $\tau_{\mathrm{UV}}$ and the dust column density. However, the geometry of the interstellar medium (ISM) in real galaxies is complex. The two-component model, comprising a diffuse ISM and dense birth clouds, has become a standard framework for interpreting dust attenuation \citep{Charlot2000, Wild2011}. In this context, IRX is sensitive not just to the total dust mass, but to the covering fraction and spatial distribution of the dust relative to the heating sources.

Our analysis of the local SFG sample confirms that IRX scales tightly with the parameters identified by \citetalias{Qin2019a}, namely $L_{\mathrm{IR}}$, $R_{\mathrm{e}}$, and $b/a$, modulated by metallicity. This scaling reflects the underlying physics of dust column density. For a fixed stellar mass and SFR, a galaxy with a smaller $R_{\mathrm{e}}$ will have a higher stellar and gas surface density. If the dust-to-gas ratio is roughly constant at fixed metallicity \citep{Popping2023}, this translates to a higher dust mass surface density and, consequently, a higher column density of dust along typical sightlines. This geometric compression naturally increases the probability that UV photons will interact with dust grains, boosting IRX.

Furthermore, the dependence on axis ratio $b/a$ underscores the importance of the viewing angle in determining the effective column density. For an oblate, disc-like dust distribution, an edge-on orientation ($b/a \ll 1$) maximises the path length through the dusty ISM, increasing the effective attenuation for a given intrinsic dust mass \citep{Chevallard2013, Greener2020}. Conversely, a face-on orientation ($b/a \sim 1$) provides a minimal path length, allowing a larger fraction of UV light to escape. The fact that the IRX relation incorporates $b/a$ with a metallicity-dependent exponent \citepalias{Qin2019a} indicates that the relative ``flattening'' or ``clumpiness'' of the dust distribution may evolve with galaxy properties. Recent work with JWST has further emphasised this point: at high redshift, the dust distribution appears to be more isotropic or ``puffy'',  leading to a decoupling of attenuation from inclination \citep{Lorenz2023, Zhang2023, Lorenz2024}. In our local sample, the strong dependence on $b/a$ for UV-intermediate and UV-faint SFGs suggests that the classical disc-like geometry remains a valid description, but its impact on the effective column density is a dominant factor in the spread of observed IRX values.

The conversion from IRX to a physical dust column is complicated by the contribution of evolved stellar populations to the infrared luminosity. As shown in Figure~\ref{fig7}, the fraction of $L_{\mathrm{IR}}$ powered by old stars ($f_{\mathrm{old}}$) can range from $\sim$5\,per\,cent to above 30\,per\,cent,  depending on the SFH and geometry. In galaxies with low sSFR, the UV luminosity is intrinsically faint, and the infrared emission may be dominated by dust heated by the ambient radiation field of older stars \citep{Bendo2012, Tailor2025}. This effect artificially inflates the observed IRX relative to the effective attenuation of the young stellar population. Correcting for this $f_{\mathrm{old}}$ contribution is therefore essential for interpreting IRX as a probe of the dust column density relevant to ongoing star formation. Our correction, based on the \citet{Qin2024} geometry model, marginally reduces the systematic scatter in the residuals, confirming that only a negligible fraction of the scatter is attributable to this age-dependent heating bias. After this correction, the remaining scatter can be more confidently assigned to genuine variations in the star--dust geometry and the effective optical depth.

\subsection{Evaluating Specific SFR as a Driver of the Scatter}

It is a common assertion in the literature that variations in recent star formation activity, often traced by specific SFR (sSFR), are a primary source of scatter in attenuation relations \citep[e.g.,][]{Wild2011, Reddy2015, Salim2020}. The argument holds that galaxies with higher sSFR have a larger fraction of young, UV-bright stars embedded in their birth clouds, leading to a different balance between UV obscuration and IR emission compared to more quiescent systems. To systematically evaluate these pathways, we analyse both the direct and indirect impacts of sSFR. While we demonstrate that sSFR is not the primary driver of the systematic offsets observed in the IRX scaling relation, we also explore the secondary roles that star-forming activity can play in modulating both dust properties and galaxy structural evolution.

Within each of the three $L_{\mathrm{UV}}$ bins (corresponding to the columns in Figure~\ref{fig5}), we measure the Spearman rank correlation coefficient ($\rho_{\mathrm{s}}$) between $\Delta\log\mathrm{IRX}$ and sSFR. While the global sample exhibits a correlation between $\Delta\log\mathrm{IRX}$ and sSFR of $\rho_{\mathrm{s}} = -0.28$ (Figure~\ref{fig1}), controlling for $L_{\mathrm{UV}}$ via binning significantly suppresses this correlation to $\rho_{\mathrm{s}} = -0.08$, $-0.20$, and $-0.23$ for the UV-faint, UV-intermediate, and UV-bright regimes, respectively. This represents a dramatic suppression (by a factor of $\sim 3.5$) in the UV-faint regime where the residual correlation becomes negligible ($\rho_{\mathrm{s}} = -0.08$). The remaining weak correlation within the UV-intermediate and UV-bright bins ($\rho_{\mathrm{s}} = -0.20$ and $-0.23$) is not an intrinsic sSFR-driven effect, but is rather a projection of the residual $\sim 1$\,dex range in $L_{\mathrm{UV}}$ that each bin still spans, across which the dominant $L_{\mathrm{UV}}$ trend still partially operates. Indeed, subdividing the dominant UV-intermediate bin into narrower 0.5\,dex sub-bins $[9.0,9.5)$ and $[9.5,10.0)$ yields even weaker correlation coefficients of $\rho_{\mathrm{s}} = -0.12$ and $-0.14$, respectively, confirming that this residual correlation is merely a projection of the finite bin size. Visually, this is reflected in the top row of Figure~\ref{fig5}, where the color-coded sSFR field is relatively uniform within each panel, showing no prominent gradients across the IRX sequence.

Beyond this local intra-bin scatter, our analysis in Section~\ref{sec3} demonstrates that while UV-faint SFGs exhibit systematically lower sSFR and UV-bright SFGs exhibit higher sSFR (median values of 0.168, 0.257, and 0.471\,Gyr$^{-1}$ for the UV-faint, intermediate, and bright subsamples, respectively), the correction for $f_{\mathrm{old}}$ -- which directly accounts for the age-dependent heating of dust -- only marginally reduces the global scatter (from 0.201 to 0.199\,dex). More importantly, the systematic shifts of the UV-faint and UV-bright populations in the IRX relation (Figure~\ref{fig2}) cannot be eliminated by merely accounting for sSFR.

Although our empirical results demonstrate that sSFR-driven dust heating is not the dominant cause of the offsets, star-forming activity can still influence dust properties and attenuation through physical pathways not associated with heating. Specifically, a higher sSFR corresponds to a more intense UV radiation field, which can physically destroy or process small carbonaceous grains and polycyclic aromatic hydrocarbons (PAHs), thereby shifting the dust grain size distribution and altering the intrinsic dust opacity \citep{Allain1996, Jones2013}. This is consistent with observational evidence showing that the strength of the 2175\,\AA\ extinction bump and the attenuation curve slope correlate with sSFR \citep{Battisti2022, Zhou2023}. While these radiation-driven dust processing channels likely modulate the residual scatter within each UV-luminosity bin, they are secondary to the global offsets and cannot be easily disentangled from other structural effects with integrated photometry alone.

In addition, sSFR acts as an evolutionary tracer structurally coupled with galaxy size through morphological compaction. A decline in sSFR is typically accompanied by gas depletion and compaction into a more compact, bulge-dominated configuration \citep{Barro2013, Tacchella2019}, explaining the empirical coupling between star-forming activity and galaxy compactness. Beyond the coupling between sSFR and global galaxy structure, sSFR also relates to structure on sub-galactic (down to cloud) scales, in the sense that a higher sSFR implies that a larger portion of the stellar emission emerges from star-forming regions embedded in birthclouds with associated dust obscuration \citep{Wild2011, Qin2019b}. This multi-scale, local-to-global connection is supported by spatially resolved studies \citep{Greener2020, Lee2025}, which show that on kpc scales, dust attenuation is regulated by local star formation rate surface density ($\Sigma_{\mathrm{SFR}}$) and local dust column density. Indeed, \citet{Maheson2024} found that at cosmic noon, dust attenuation correlates more strongly with local surface densities of SFR and stellar mass than with global sSFR. Similarly, \citet{Zhang2023} and \citet{Lorenz2024} demonstrated that at high redshift, dust attenuation shows no dependence on inclination, pointing to a geometry where the global orientation is less important than the local, clumpy distribution of dust. On global scales, our integrated observables ($L_{\mathrm{UV}}$, $R_{\mathrm{e}}$, $b/a$) act as the integrated projection of these multi-scale ISM conditions. Consequently, these evolutionary pathways indicate that while sSFR is tied to various aspects of galaxy evolution and dust processing, the global systematic offsets in the IRX relation are not directly caused by sSFR-related activity, pointing instead towards structural and geometric factors as the dominant contributors to the scatter, as has also been argued for the local Universe in the context of the IRX--$\beta_{\mathrm{UV}}$ relation \citep{Salim2019}.

\subsection{Scatter Associated with Star--Dust Geometry}\label{sec4.3}

The preceding discussion points to star--dust geometry as the dominant factor underlying both the intrinsic scatter around the IRX relation and the systematic offsets observed in the UV-faint and UV-bright outlier populations. Geometry, in this context, encompasses two main aspects: (1) the relative spatial distribution of stars and dust (compact vs. extended, clumpy vs. smooth) and (2) the viewing angle (inclination).

Our results provide quantitative observational anchoring for this paradigm. UV-faint SFGs, which exhibit a median IRX excess of $+0.23$\,dex, are characterised by compact stellar distributions (median $R_{\mathrm{e}}=2.90$\,kpc) and a preference for edge-on orientations (median $b/a=0.41$). In such a configuration, UV-emitting young stars are concentrated within a small volume, likely embedded in a dense, compact dust reservoir. When viewed edge-on, the line of sight traverses a maximal path length through this dusty medium, leading to high effective optical depth and efficient conversion of UV photons into IR emission. This scenario is the local analogue of the ``optically thick'' or ``buried'' star formation seen in high-redshift submillimetre galaxies and compact star-forming systems \citep{Gomez-Guijarro2023, Mitsuhashi2024a}.

Conversely, UV-bright SFGs (median IRX deficit of $-0.20$\,dex) are characterised by extended stellar distributions (median $R_{\mathrm{e}}=5.99$\,kpc) and predominantly face-on orientations (median $b/a=0.74$). Here, young stars are spread over a large area with lower average dust surface density, and the face-on viewing angle provides the shortest escape path for UV photons. A larger fraction of the intrinsic UV luminosity therefore escapes unattenuated, leading to a low observed IRX. This behaviour is consistent with extended, low-surface-brightness discs \citep{Junais2024}. 

The distinct opacity regimes of compact and extended SFGs have important implications for fitting the IRX relation. We confirm that compact, UV-faint SFGs represent a distinct physical state of star--dust geometry rather than mere statistical outliers, a conclusion robust against sample density biases (see Section~\ref{sec4.4}). In summary, the systematic IRX offsets are not driven by metallicity or total SFR, but by the physical density of the dust screen regulated by galaxy size. The high-opacity environments of compact galaxies trap UV photons efficiently, while the diffuse nature of extended, face-on systems facilitates their escape. This geometric paradigm is essential for accurate attenuation corrections across cosmic time.

The importance of geometry over simple dust mass is further highlighted by the envelope of the IRX relation. As shown in Figure~\ref{fig5}, galaxies with a significant bulge component ($B/T > 0.4$) or high S\'ersic index ($n > 2.5$) preferentially populate the high-deviation envelope of the relation. A bulge represents a fundamentally different star--dust geometry compared to a thin disc. The stellar population in a bulge is typically older and more spheroidally distributed, while the dust may still reside in a disc or be distributed in a more spherical halo \citep{Sachdeva2022, Lu2023}. The superposition of a spheroidal UV source (even if weak) with a disc-like or patchy dust distribution creates a complex radiative transfer problem that deviates significantly from the assumptions of the scaling relation, which was calibrated primarily on disc-dominated systems.

For compact, edge-on systems, the effective covering fraction of dust is enhanced. This scenario is supported by recent work highlighting non-unity dust covering fractions as a primary driver of scatter in attenuation curves \citep{Reddy2026}. Geometric compression implies that young stellar populations in compact systems are embedded within a denser dust screen, leading to more efficient UV photon trapping and elevated IRX. This is consistent with findings from the ALPINE-CRISTAL-JWST survey, which indicate that at $z \sim 4$--$6$, nebular and stellar attenuation are heavily influenced by the compactness of the dust core relative to the stellar distribution \citep{Mitsuhashi2024a, Tsujita2026}. Similarly, galaxies with significant bulge components or high S\'ersic indices preferentially occupy the high-deviation envelope. In such systems, the central concentration of gas and dust creates a compact, high-opacity core. Even modest ongoing star formation in these regions results in elevated IRX, as UV photons must traverse the dense central ISM. This behaviour parallels the compact dust cores observed in high-redshift submillimetre galaxies \citep{Gomez-Guijarro2023} and is echoed in simulations of clumpy galaxy assembly at $z \sim 7$ \citep{Mawatari2026}, where spatially resolved dust clumps dominate the integrated attenuation signal.

Recent theoretical work further supports the primacy of geometry. Radiative transfer simulations demonstrate that variations in the relative scale heights of stars and dust, as well as the clumpiness of the ISM, can generate over an order of magnitude of scatter in the IRX--$\beta_{\mathrm{UV}}$ plane, even at fixed dust mass and metallicity \citep{Narayanan2018, Trayford2020, Ramnichal2026, Zhang2026}. The ``chocolate chip cookie'' model \citep{Lu2022}, in which dense, optically thick clumps (birth clouds) are embedded in a diffuse, low-attenuation matrix, naturally produces a wide range of attenuation curves and IRX values depending on the covering fraction of the clumps and the viewing angle. Our observational finding that the extremes of the IRX relation are populated by galaxies with extreme structural parameters ($R_{\mathrm{e}}$, $b/a$, and $B/T$) provides direct empirical validation for these geometry models.

\subsection{Biases in Sample Selection and Fitting Methodology}\label{sec4.4}

Any empirical scaling relation is susceptible to biases introduced by the sample selection function and the statistical methodology used to derive it. In the context of the IRX relation, these biases can artificially inflate the perceived scatter and shift the best-fit parameters. Our analysis identifies and quantifies several such biases.

\textit{Sample Selection and Observational Limits:} The construction of a multi-wavelength SFG sample, as used here and in \citetalias{Qin2019a}, imposes implicit limits on the range of observable attenuation. UV-selected samples are inherently biased against heavily obscured, edge-on systems where the UV flux may fall below the detection threshold. Our examination of GALEX survey depth in Figure~\ref{fig3} is reassuring: the distribution of IRX deviations does not shift systematically between shallow, medium, and deep surveys, indicating that our sample is not severely biased by UV flux limits. However, the scarcity of galaxies in the bottom-left region of the $R_{\mathrm{e}}$--$b/a$ diagram (Figure~\ref{fig6}) -- low-mass, edge-on systems -- suggests that such galaxies may be systematically missed due to their low apparent UV surface brightness. This selection effect could lead to an underestimate of the true scatter in the IRX relation at the low-mass, high-attenuation end.

\textit{Fitting Methodology and Density Weighting:} A subtle but important bias arises from the non-uniform distribution of galaxies in parameter space. Standard maximum likelihood estimation (MLE) methods are dominated by the high-density regime of the sample (i.e., the UV-intermediate population), which can artificially tilt the best-fit relation and mask the systematic offsets of outlier populations. To mitigate this, we implement a density-weighted MLE scheme. We tested several weighting strategies to suppress the dominance of the dense parameter space while avoiding over-amplification of sparse outliers (see Appendix~\ref{AppendixA} for detailed methodology and statistical tests). We find that an inverse square-root density weighting scheme effectively flattens the systematic trend in the residuals (reducing the residual tilt parameter from 0.09 to 0.02) without significantly impacting the overall bias and scatter. These results demonstrate that the density-weighted fitting successfully mitigates the tilt caused by the heavily unbalanced sample distribution, preventing the scaling relation from being exclusively anchored by the dominant population. However, the systematic offsets observed for the UV-faint and UV-bright populations are not artefacts induced by this statistical bias, but rather represent genuine physical features.

\textit{The Contribution of Old Stellar Populations:} As discussed in Section~\ref{sec:3.4}, the use of total $L_{\mathrm{IR}}$ as a proxy for the attenuation of young stars introduces a bias that depends on sSFR. Galaxies with low sSFR have a larger fractional contribution of old stars to dust heating ($f_{\mathrm{old}}$). By not correcting for $f_{\mathrm{old}}$, one would systematically overestimate the dust attenuation in low-sSFR galaxies, artificially creating a correlation between IRX deviation and sSFR. Our correction, based on the \citet{Qin2024} geometry  model, reduces this bias, as evidenced by the marginally decreased scatter of the IRX relation (from 0.201\,dex to 0.199\,dex). Future work should incorporate more sophisticated SED fitting or utilise mid-IR indicators to isolate the dust heated by young stars \citep[e.g.,][]{Calzetti2010, Shivaei2024}.

\textit{Metallicity Aperture Effects:} Finally, we note a potential aperture bias in metallicity measurements. Our sample selection requires a fiber covering fraction $>20$\,per\,cent, meaning that for galaxies near this threshold, the spectroscopic aperture predominantly probes the inner regions. Consequently, the fiber-based metallicities may not fully represent the global gas-phase metallicity, whereas $L_{\mathrm{IR}}$ and $L_{\mathrm{UV}}$ are globally integrated quantities. This spatial mismatch could introduce additional scatter into the derived IRX relation. However, given the generally flat metallicity gradients in typical star-forming galaxies and the limited availability of global metallicity measurements, we expect this aperture-induced discrepancy to be minor and unlikely to systematically bias our results. This effect will be rigorously examined in future work using IFU data.

\subsection{Why UV Luminosity Traces the Systematic Offsets from the IRX Relation?}\label{sec4.5}

A key result of this work is the discovery that deviations from the IRX relation ($\Delta \log \mathrm{IRX}$) correlate strongly with UV luminosity, with UV-faint SFGs exhibiting higher-than-predicted IRX and UV-bright SFGs showing lower-than-predicted IRX (Figure~\ref{fig2}). It can be understood as a natural consequence of the interplay between star--dust geometry and the non-linear nature of dust attenuation.

The UV luminosity of a galaxy is not a fundamental, independent parameter. It is an observed quantity that is itself a product of the intrinsic UV luminosity of the young stellar population ($L_{\mathrm{UV, int}}$) and the dust attenuation at UV wavelengths ($A_{\mathrm{UV}}$): $L_{\mathrm{UV}} = L_{\mathrm{UV, int}} \times 10^{-0.4 A_{\mathrm{UV}}}$. The IRX relation, by definition, relates the observed $L_{\mathrm{IR}}$ and $L_{\mathrm{UV}}$. Galaxies with extreme star--dust geometries will have extreme values of $A_{\mathrm{UV}}$ for their given global properties ($M_\ast$, SFR, metallicity), which will manifest as extreme values of the observed $L_{\mathrm{UV}}$.

The UV-faint outlier SFGs are mostly compact ($R_{\mathrm{e}} \sim 2.9$\,kpc) and edge-on ($b/a \sim 0.41$). This geometry results in a high effective dust column density and, consequently, a large $A_{\mathrm{UV}}$. The large attenuation drastically suppresses the observed $L_{\mathrm{UV}}$, pushing the galaxy into the UV-faint regime. The energy absorbed at UV wavelengths is re-radiated in the IR, contributing to a high observed $L_{\mathrm{IR}}$. The combined effect -- low $L_{\mathrm{UV}}$ and normal-to-high $L_{\mathrm{IR}}$ -- yields an observed IRX that is significantly higher than that predicted by a relation calibrated on average geometries (which would assume a smaller $A_{\mathrm{UV}}$ for the same $L_{\mathrm{IR}}$, $R_{\mathrm{e}}$, and $b/a$ combination). The low sSFR of these systems further exacerbates the effect: the intrinsic $L_{\mathrm{UV, int}}$ is already low due to a lack of massive young stars, and the severe attenuation pushes the observed $L_{\mathrm{UV}}$ to even fainter levels.

Conversely, the UV-bright outliers are extended ($R_{\mathrm{e}} \sim 6.0$\,kpc) and face-on ($b/a \sim 0.74$). Their geometry results in a low $A_{\mathrm{UV}}$, allowing a large fraction of the intrinsic UV light to escape. The observed $L_{\mathrm{UV}}$ is therefore high. The fraction of UV light absorbed is small, leading to a relatively low observed $L_{\mathrm{IR}}$ for the given SFR. The combination of high observed $L_{\mathrm{UV}}$ and relatively low $L_{\mathrm{IR}}$ yields an observed IRX that is significantly lower than predicted. Their high sSFR boosts the intrinsic $L_{\mathrm{UV, int}}$, further enhancing their observed UV brightness.

In essence, $L_{\mathrm{UV}}$ acts as an integrated, albeit indirect, probe of the geometric configuration of a galaxy. Because the geometry dictates the escape fraction of UV photons, the observed $L_{\mathrm{UV}}$ encapsulates information about the three-dimensional arrangement of stars and dust that is not fully captured by the simple combination of 2-D projected parameters $R_{\mathrm{e}}$ and $b/a$. The $R_{\mathrm{e}}$ and $b/a$ used in the IRX relation are derived from optical $r$-band imaging, which primarily traces the older, less obscured stellar population. The UV light, on the other hand, originates from the youngest stars and is most sensitive to the dust in their immediate vicinity. The deviations we observe are therefore telling us that the optical size and shape are imperfect proxies for the actual spatial distribution and scale ratios of dust and UV-emitting stars \citep[e.g.,][]{Bianchi2007, Casasola2017}. This is particularly true for galaxies with significant bulges or complex, clumpy ISM structures, where the $r$-band morphology is decoupled from the sites of active star formation \citep{Gadotti2010, Pastrav2013a}. Moreover, $L_{\mathrm{UV}}$ serves as a dual probe of both geometry and intrinsic dust physics. While the geometrical configuration determines the escape paths for UV photons, the intrinsic amount of UV emission (and hence the radiation field strength) also directly influences the grain size distribution and dust composition. In UV-bright galaxies, the intense radiation field may destroy small carbonaceous grains, altering the dust-to-gas ratio and the shape of the attenuation curve \citep{Battisti2022}. In UV-faint galaxies, the lower radiation field combined with high effective dust column densities allows dust to remain relatively unshielded or less processed in a more diffuse state. This dual sensitivity explains why $L_{\mathrm{UV}}$ is exceptionally effective at organizing the scatter in the IRX relation, as it encapsulates both the physical state of the attenuating dust and the three-dimensional distribution of the heating sources.

The local benchmark established here provides a crucial reference for interpreting JWST results. Studies of JADES galaxies at $z > 3$ have already begun to show that the diversity of nebular attenuation curves is  the physical scale of star formation \citep{Maheson2025}. As we push towards the first billion years of cosmic history, understanding how galaxy size regulates dust opacity will be vital for reconciling UV attenuation with IR emission \citep{Markov2025, Wijesekera2026}. The fact that this scatter can be organised by the simple, observed parameter $L_{\mathrm{UV}}$ is a powerful and practical result, suggesting that it acts as an efficient observational proxy for underlying variations in star--dust geometry. Future refinements of the universal attenuation relation may therefore benefit from incorporating more direct tracers of star--dust geometry.

The systematic differences in sSFR, galaxy size, and birth-cloud dust fraction $F_{\mathrm{bc}}$ between UV-faint and UV-bright SFGs are not independent but trace a coherent evolutionary sequence along the star-forming main sequence. The dependence of the size offset across the $M_\ast$--SFR plane (see Appendix~\ref{AppendixB}) provides direct empirical support for this sequence: evolutionary processes such as gas depletion and compaction shape the star--dust geometry, while the geometry itself strongly modulates attenuation and hence the IRX offset. Galaxies with high sSFR (UV-bright) are actively growing their discs outside-in, acquiring angular momentum and gas, which naturally leads to larger effective radii (median $R_{\mathrm{e}}=5.99$\,kpc). Conversely, low-sSFR galaxies (UV-faint) have largely exhausted their gas, causing the remaining gas and dust to concentrate in a compact, dense central region (median $R_{\mathrm{e}}=2.90$\,kpc). This size--sSFR correlation is well established: galaxies on the upper MS are extended, while those below the MS become more compact as outer discs quench first \citep{Wuyts2011, Brennan2017, Wangec2018, Jain2024}. Moreover, a decline in sSFR is often accompanied by bulge growth via a compaction phase---an inflow that builds a central bulge, reduces the effective radius, and leaves a compact configuration with lower sSFR but still significant dust mass \citep{Barro2013}. UV-faint SFGs thus represent the post-compaction, bulge-dominated phase, while UV-bright galaxies are still in the extended, disc-dominated phase.

\section{Summary and Conclusions} \label{sec5}

In this work, we have systematically investigated the physical origins of the scatter in the universal dust attenuation scaling relation using a sample of $\sim$32,000 local star-forming galaxies (SFGs) drawn from SDSS, GALEX, and WISE. By examining deviations from the IRX relation established by \citet{Qin2019a} and their correlation with a comprehensive suite of galaxy properties, we draw the following conclusions:

\begin{enumerate}

\item \textbf{UV luminosity systematically traces deviations from the IRX relation.} Our sample separates cleanly into three UV-luminosity regimes. The dominant UV-intermediate population ($9 < \log(L_{\mathrm{UV}}/{\mathrm{L}}_\odot) < 10$; 86.38\,per\,cent of the sample) defines the IRX relation with minimal systematic offset. In contrast, UV-faint SFGs ($\log(L_{\mathrm{UV}}/{\mathrm{L}}_\odot) \le 9$; 9.58\,per\,cent) exhibit a median IRX excess of $+0.23$\,dex, while UV-bright SFGs ($\log(L_{\mathrm{UV}}/{\mathrm{L}}_\odot) \ge 10$; 4.04\,per\,cent) show a median deficit of $-0.20$\,dex. These offsets are robust against GALEX survey depth variations and cannot be attributed to observational biases, indicating that UV luminosity traces fundamental differences in dust attenuation efficiency.

\item \textbf{Star--dust geometry plays a dominant role in accounting for the systematic offsets.} Although UV-faint SFGs have lower sSFR and UV-bright SFGs have higher sSFR, correcting for dust heating by old stellar populations reduces the global scatter by only 0.002\,dex (from 0.201 to 0.199\,dex) and does not eliminate the systematic offsets. Instead, the outlier populations exhibit distinct geometric and projection properties: UV-faint SFGs are intrinsically compact (median $R_{\rm e} = 2.90\,\text{ kpc}$) and are preferentially viewed edge-on (median $b/a = 0.41$), resulting in high effective dust column densities along the line of sight. Conversely, UV-bright SFGs are intrinsically extended (median $R_{\rm e} = 5.99\,\text{ kpc}$) and are preferentially viewed face-on (median $b/a = 0.74$), which minimises the line-of-sight dust attenuation and facilitates the escape of UV photons. Specific SFR acts as a tracer of evolutionary state, but the physical mechanism most strongly associated with the attenuation variations is the three-dimensional configuration of stars and dust.

\item \textbf{The two-component star--dust geometry model quantitatively reproduces the IRX offsets.} Using the analytic model of \citet{Qin2024}, we find that UV-faint SFGs require a lower fraction of dust in birth clouds ($F_{\mathrm{bc}}=0.11$) and a smaller stellar-to-dust scale ratio ($\hat{R}=0.35$), increasing the effective optical depth of the diffuse ISM. UV-bright SFGs require the opposite: higher $F_{\mathrm{bc}}=0.26$ and larger $\hat{R}=0.65$, reducing attenuation. These parameter variations naturally explain the observed IRX offsets of $+0.23$\,dex and $-0.20$\,dex, respectively, indicating that variations in star--dust geometry, rather than dust mass or metallicity, provide the dominant mechanism to reproduce the systematic deviations within this framework.

\item \textbf{Bulge components contribute to extreme scatter.} Galaxies with significant bulge contributions ($B/T > 0.4$ or S\'ersic index $n > 2.5$) preferentially populate the envelope region of the IRX relation, exhibiting larger deviations from the mean relation. This suggests that spheroidal components alter the star--dust geometry relative to pure disc systems, creating attenuation characteristics not fully captured by the disc-dominated parameters ($R_{\mathrm{e}}$, $b/a$, metallicity, SFR) used in the IRX relation.

\item \textbf{UV luminosity serves as an effective tracer of geometric configuration.} The observed correlation between $Delta\log{\mathrm{IRX}}$ and $L_{\mathrm{UV}}$ arises because UV luminosity integrates the combined effects of intrinsic stellar population brightness and dust attenuation. Extreme geometries produce extreme $A_{\mathrm{UV}}$ values, pushing galaxies into the UV-faint or UV-bright regimes. Since the IRX relation uses optical $r$-band structural parameters that trace older stars, the UV light -- originating from young stars and most sensitive to local dust -- provides complementary information about the geometry of active star formation regions.
\end{enumerate}

In conclusion, the universal dust attenuation scaling relation robustly describes the average behaviour of disc-dominated SFGs, but its scatter encodes rich physical information about deviations from this typical configuration. The systematic offsets are most naturally accounted for by variations in star--dust geometry -- specifically, the compactness of the stellar distribution, the inclination angle, and the presence of bulge components -- rather than by metallicity or total star formation rate. UV luminosity, through its sensitivity to both stellar population age and geometric attenuation, naturally traces these systematic deviations. By quantifying the physical origins of scatter and accounting for observational and methodological biases, this work provides a foundation for extending the universal attenuation relation to higher redshifts and more diverse galaxy populations, where resolved multi-wavelength observations with JWST and ALMA will enable direct constraints on the three-dimensional distribution of stars and dust.

\section*{Acknowledgements}

This work is supported by the National Key Research and Development Program of China (2023YFA1608100), the National Science Foundation of China (12233005, 12173088), the China Manned Space Program with grants nos. CMS-CSST-2025-A08 and CMS-CSST-2025-A20, in part by Office of Science and Technology, Shanghai Municipal Government (grant Nos. 24DX1400100, ZJ2023-ZD-001). A.K. acknowledges support by the National SKA Program of China (2025SKA0150104), the 100 talent program of the Sun Yat-sen University, and the Guangdong Basic and the Applied Basic Research Foundation with No. 2025A1515012670. DDS is supported by the start-up funding of the Anhui University of Science and Technology (2024yjrc104), the National Science Foundation of China (12303015) and the National Science Foundation of Jiangsu Province (BK20231106). S.W. acknowledges support from the Royal Society International Exchanges scheme (IES\textbackslash R2\textbackslash 242195).
M.Q. acknowledges support from the Natural Science Foundation of Shandong Province, China (Grant No. ZR2026QC2352Z).

\section*{Data Availability}

The data underlying this article will be shared on reasonable request to the corresponding author.


\bibliographystyle{mnras}
\bibliography{references}

@ARTICLE{qin2024,
       author = {{Qin}, Jianbo and {Zheng}, Xian Zhong and {Wuyts}, Stijn and {Lyu}, Zongfei and {Qiao}, Man and {Huang}, Jia-Sheng and {Liu}, Feng Shan and {Katsianis}, Antonios and {Gonzalez}, Valentino and {Bian}, Fuyan and {Xu}, Haiguang and {Pan}, Zhizheng and {Liu}, Wenhao and {Tan}, Qing-Hua and {An}, Fang Xia and {Shi}, Dong Dong and {Zhang}, Yuheng and {Wen}, Run and {Liu}, Shuang and {Yang}, Chao},
        title = "{Understanding the universal dust attenuation scaling relation of star-forming galaxies}",
      journal = {\mnras},
         year = 2024,
        month = feb,
       volume = {528},
       number = {1},
        pages = {658-675},
          doi = {10.1093/mnras/stad3999},
archivePrefix = {arXiv},
       eprint = {2312.16700},
 primaryClass = {astro-ph.GA},
       adsurl = {https://ui.adsabs.harvard.edu/abs/2024MNRAS.528..658Q}
}

@ARTICLE{Qin2019a,
       author = {{Qin}, Jianbo and {Zheng}, Xian Zhong and {Wuyts}, Stijn and {Pan}, Zhizheng and {Ren}, Jian},
        title = "{A universal relation of dust obscuration across cosmic time}",
      journal = {\mnras},
         year = 2019,
        month = jun,
       volume = {485},
       number = {4},
        pages = {5733-5751},
          doi = {10.1093/mnras/stz763},
archivePrefix = {arXiv},
       eprint = {1903.05121},
 primaryClass = {astro-ph.GA},
       adsurl = {https://ui.adsabs.harvard.edu/abs/2019MNRAS.485.5733Q}
}

@ARTICLE{Qin2019b,
       author = {{Qin}, Jianbo and {Zheng}, Xian Zhong and {Wuyts}, Stijn and {Pan}, Zhizheng and {Ren}, Jian},
        title = "{Understanding the Discrepancy between IRX and Balmer Decrement in Tracing Galaxy Dust Attenuation}",
      journal = {\apj},
         year = 2019,
        month = nov,
       volume = {886},
       number = {1},
          eid = {28},
        pages = {28},
          doi = {10.3847/1538-4357/ab4a04},
archivePrefix = {arXiv},
       eprint = {1909.13505},
 primaryClass = {astro-ph.GA},
       adsurl = {https://ui.adsabs.harvard.edu/abs/2019ApJ...886...28Q}
}

@ARTICLE{Hamed2023,
       author = {{Hamed}, M. and {Pistis}, F. and {Figueira}, M. and {Ma{\l}ek}, K. and {Nanni}, A. and {Buat}, V. and {Pollo}, A. and {Vergani}, D. and {Bolzonella}, M. and {Junais} and {Krywult}, J. and {Takeuchi}, T. and {Riccio}, G. and {Moutard}, T.},
        title = "{Decoding the IRX-{\ensuremath{\beta}} dust attenuation relation in star-forming galaxies at intermediate redshift}",
      journal = {\aap},
         year = 2023,
        month = nov,
       volume = {679},
          eid = {A26},
        pages = {A26},
          doi = {10.1051/0004-6361/202346976},
archivePrefix = {arXiv},
       eprint = {2309.01819},
 primaryClass = {astro-ph.GA},
       adsurl = {https://ui.adsabs.harvard.edu/abs/2023A&A...679A..26H}
}

@ARTICLE{Draine2007,
       author = {{Draine}, B.~T. and {Li}, Aigen},
        title = "{Infrared Emission from Interstellar Dust. IV. The Silicate-Graphite-PAH Model in the Post-Spitzer Era}",
      journal = {\apj},
         year = 2007,
        month = mar,
       volume = {657},
       number = {2},
        pages = {810-837},
          doi = {10.1086/511055},
archivePrefix = {arXiv},
       eprint = {astro-ph/0608003},
 primaryClass = {astro-ph},
       adsurl = {https://ui.adsabs.harvard.edu/abs/2007ApJ...657..810D}
}

@ARTICLE{Ahn2014,
       author = {{Ahn}, Christopher P. and {Alexandroff}, Rachael and {Allende Prieto}, Carlos and {Anders}, Friedrich and {Anderson}, Scott F. and {Anderton}, Timothy and {Andrews}, Brett H. and {Aubourg}, {\'E}ric and {Bailey}, Stephen and {Bastien}, Fabienne A. and {Bautista}, Julian E. and {Beers}, Timothy C. and {Beifiori}, Alessandra and {Bender}, Chad F. and {Berlind}, Andreas A. and {Beutler}, Florian and {Bhardwaj}, Vaishali and {Bird}, Jonathan C. and {Bizyaev}, Dmitry and {Blake}, Cullen H. and {Blanton}, Michael R. and {Blomqvist}, Michael and {Bochanski}, John J. and {Bolton}, Adam S. and {Borde}, Arnaud and {Bovy}, Jo and {Shelden Bradley}, Alaina and {Brandt}, W.~N. and {Brauer}, Doroth{\'e}e and {Brinkmann}, J. and {Brownstein}, Joel R. and {Busca}, Nicol{\'a}s G. and {Carithers}, William and {Carlberg}, Joleen K. and {Carnero}, Aurelio R. and {Carr}, Michael A. and {Chiappini}, Cristina and {Chojnowski}, S. Drew and {Chuang}, Chia-Hsun and {Comparat}, Johan and {Crepp}, Justin R. and {Cristiani}, Stefano and {Croft}, Rupert A.~C. and {Cuesta}, Antonio J. and {Cunha}, Katia and {da Costa}, Luiz N. and {Dawson}, Kyle S. and {De Lee}, Nathan and {Dean}, Janice D.~R. and {Delubac}, Timoth{\'e}e and {Deshpande}, Rohit and {Dhital}, Saurav and {Ealet}, Anne and {Ebelke}, Garrett L. and {Edmondson}, Edward M. and {Eisenstein}, Daniel J. and {Epstein}, Courtney R. and {Escoffier}, Stephanie and {Esposito}, Massimiliano and {Evans}, Michael L. and {Fabbian}, D. and {Fan}, Xiaohui and {Favole}, Ginevra and {Femen{\'\i}a Castell{\'a}}, Bruno and {Fern{\'a}ndez Alvar}, Emma and {Feuillet}, Diane and {Filiz Ak}, Nurten and {Finley}, Hayley and {Fleming}, Scott W. and {Font-Ribera}, Andreu and {Frinchaboy}, Peter M. and {Galbraith-Frew}, J.~G. and {Garc{\'\i}a-Hern{\'a}ndez}, D.~A. and {Garc{\'\i}a P{\'e}rez}, Ana E. and {Ge}, Jian and {G{\'e}nova-Santos}, R. and {Gillespie}, Bruce A. and {Girardi}, L{\'e}o and {Gonz{\'a}lez Hern{\'a}ndez}, Jonay I. and {Gott}, J. Richard, III and {Gunn}, James E. and {Guo}, Hong and {Halverson}, Samuel and {Harding}, Paul and {Harris}, David W. and {Hasselquist}, Sten and {Hawley}, Suzanne L. and {Hayden}, Michael and {Hearty}, Frederick R. and {Herrero Dav{\'o}}, Artemio and {Ho}, Shirley and {Hogg}, David W. and {Holtzman}, Jon A. and {Honscheid}, Klaus and {Huehnerhoff}, Joseph and {Ivans}, Inese I. and {Jackson}, Kelly M. and {Jiang}, Peng and {Johnson}, Jennifer A. and {Kinemuchi}, K. and {Kirkby}, David and {Klaene}, Mark A. and {Kneib}, Jean-Paul and {Koesterke}, Lars and {Lan}, Ting-Wen and {Lang}, Dustin and {Le Goff}, Jean-Marc and {Leauthaud}, Alexie and {Lee}, Khee-Gan and {Lee}, Young Sun and {Long}, Daniel C. and {Loomis}, Craig P. and {Lucatello}, Sara and {Lupton}, Robert H. and {Ma}, Bo and {Mack}, Claude E., III and {Mahadevan}, Suvrath and {Maia}, Marcio A.~G. and {Majewski}, Steven R. and {Malanushenko}, Elena and {Malanushenko}, Viktor and {Manchado}, A. and {Manera}, Marc and {Maraston}, Claudia and {Margala}, Daniel and {Martell}, Sarah L. and {Masters}, Karen L. and {McBride}, Cameron K. and {McGreer}, Ian D. and {McMahon}, Richard G. and {M{\'e}nard}, Brice and {M{\'e}sz{\'a}ros}, Sz. and {Miralda-Escud{\'e}}, Jordi and {Miyatake}, Hironao and {Montero-Dorta}, Antonio D. and {Montesano}, Francesco and {More}, Surhud and {Morrison}, Heather L. and {Muna}, Demitri and {Munn}, Jeffrey A. and {Myers}, Adam D. and {Nguyen}, Duy Cuong and {Nichol}, Robert C. and {Nidever}, David L. and {Noterdaeme}, Pasquier and {Nuza}, Sebasti{\'a}n E. and {O'Connell}, Julia E. and {O'Connell}, Robert W. and {O'Connell}, Ross and {Olmstead}, Matthew D. and {Oravetz}, Daniel J. and {Owen}, Russell and {Padmanabhan}, Nikhil and {Palanque-Delabrouille}, Nathalie and {Pan}, Kaike and {Parejko}, John K. and {Parihar}, Prachi and {P{\^a}ris}, Isabelle and {Pepper}, Joshua and {Percival}, Will J. and {P{\'e}rez-R{\`a}fols}, Ignasi and {Dotto Perottoni}, H{\'e}lio and {Petitjean}, Patrick and {Pieri}, Matthew M. and {Pinsonneault}, M.~H. and {Prada}, Francisco and {Price-Whelan}, Adrian M. and {Raddick}, M. Jordan and {Rahman}, Mubdi and {Rebolo}, Rafael and {Reid}, Beth A. and {Richards}, Jonathan C. and {Riffel}, Rog{\'e}rio and {Robin}, Annie C. and {Rocha-Pinto}, H.~J. and {Rockosi}, Constance M. and {Roe}, Natalie A. and {Ross}, Ashley J. and {Ross}, Nicholas P. and {Rossi}, Graziano and {Roy}, Arpita and {Rubi{\~n}o-Martin}, J.~A. and {Sabiu}, Cristiano G. and {S{\'a}nchez}, Ariel G. and {Santiago}, Bas{\'\i}lio and {Sayres}, Conor and {Schiavon}, Ricardo P. and {Schlegel}, David J. and {Schlesinger}, Katharine J. and {Schmidt}, Sarah J. and {Schneider}, Donald P. and {Schultheis}, Mathias and {Sellgren}, Kris and {Seo}, Hee-Jong and {Shen}, Yue and {Shetrone}, Matthew and {Shu}, Yiping and {Simmons}, Audrey E. and {Skrutskie}, M.~F. and {Slosar}, An{\v{z}}e and {Smith}, Verne V. and {Snedden}, Stephanie A. and {Sobeck}, Jennifer S. and {Sobreira}, Flavia and {Stassun}, Keivan G. and {Steinmetz}, Matthias and {Strauss}, Michael A. and {Streblyanska}, Alina and {Suzuki}, Nao and {Swanson}, Molly E.~C. and {Terrien}, Ryan C. and {Thakar}, Aniruddha R. and {Thomas}, Daniel and {Thompson}, Benjamin A. and {Tinker}, Jeremy L. and {Tojeiro}, Rita and {Troup}, Nicholas W. and {Vandenberg}, Jan and {Vargas Maga{\~n}a}, Mariana and {Viel}, Matteo and {Vogt}, Nicole P. and {Wake}, David A. and {Weaver}, Benjamin A. and {Weinberg}, David H. and {Weiner}, Benjamin J. and {White}, Martin and {White}, Simon D.~M. and {Wilson}, John C. and {Wisniewski}, John P. and {Wood-Vasey}, W.~M. and {Y{\`e}che}, Christophe and {York}, Donald G. and {Zamora}, O. and {Zasowski}, Gail and {Zehavi}, Idit and {Zhao}, Gong-Bo and {Zheng}, Zheng and {Zhu}, Guangtun},
        title = "{The Tenth Data Release of the Sloan Digital Sky Survey: First Spectroscopic Data from the SDSS-III Apache Point Observatory Galactic Evolution Experiment}",
      journal = {\apjs},
         year = 2014,
        month = apr,
       volume = {211},
       number = {2},
          eid = {17},
        pages = {17},
          doi = {10.1088/0067-0049/211/2/17},
archivePrefix = {arXiv},
       eprint = {1307.7735},
 primaryClass = {astro-ph.IM},
       adsurl = {https://ui.adsabs.harvard.edu/abs/2014ApJS..211...17A}
}

@ARTICLE{Martin2005,
       author = {{Martin}, D. Christopher and {Fanson}, James and {Schiminovich}, David and {Morrissey}, Patrick and {Friedman}, Peter G. and {Barlow}, Tom A. and {Conrow}, Tim and {Grange}, Robert and {Jelinsky}, Patrick N. and {Milliard}, Bruno and {Siegmund}, Oswald H.~W. and {Bianchi}, Luciana and {Byun}, Yong-Ik and {Donas}, Jose and {Forster}, Karl and {Heckman}, Timothy M. and {Lee}, Young-Wook and {Madore}, Barry F. and {Malina}, Roger F. and {Neff}, Susan G. and {Rich}, R. Michael and {Small}, Todd and {Surber}, Frank and {Szalay}, Alex S. and {Welsh}, Barry and {Wyder}, Ted K.},
        title = "{The Galaxy Evolution Explorer: A Space Ultraviolet Survey Mission}",
      journal = {\apjl},
         year = 2005,
        month = jan,
       volume = {619},
       number = {1},
        pages = {L1-L6},
          doi = {10.1086/426387},
archivePrefix = {arXiv},
       eprint = {astro-ph/0411302},
 primaryClass = {astro-ph},
       adsurl = {https://ui.adsabs.harvard.edu/abs/2005ApJ...619L...1M}
}

@ARTICLE{Wright2010,
       author = {{Wright}, Edward L. and {Eisenhardt}, Peter R.~M. and {Mainzer}, Amy K. and {Ressler}, Michael E. and {Cutri}, Roc M. and {Jarrett}, Thomas and {Kirkpatrick}, J. Davy and {Padgett}, Deborah and {McMillan}, Robert S. and {Skrutskie}, Michael and {Stanford}, S.~A. and {Cohen}, Martin and {Walker}, Russell G. and {Mather}, John C. and {Leisawitz}, David and {Gautier}, Thomas N., III and {McLean}, Ian and {Benford}, Dominic and {Lonsdale}, Carol J. and {Blain}, Andrew and {Mendez}, Bryan and {Irace}, William R. and {Duval}, Valerie and {Liu}, Fengchuan and {Royer}, Don and {Heinrichsen}, Ingolf and {Howard}, Joan and {Shannon}, Mark and {Kendall}, Martha and {Walsh}, Amy L. and {Larsen}, Mark and {Cardon}, Joel G. and {Schick}, Scott and {Schwalm}, Mark and {Abid}, Mohamed and {Fabinsky}, Beth and {Naes}, Larry and {Tsai}, Chao-Wei},
        title = "{The Wide-field Infrared Survey Explorer (WISE): Mission Description and Initial On-orbit Performance}",
      journal = {\aj},
         year = 2010,
        month = dec,
       volume = {140},
       number = {6},
        pages = {1868-1881},
          doi = {10.1088/0004-6256/140/6/1868},
archivePrefix = {arXiv},
       eprint = {1008.0031},
 primaryClass = {astro-ph.IM},
       adsurl = {https://ui.adsabs.harvard.edu/abs/2010AJ....140.1868W}
}

@ARTICLE{Salim2016,
       author = {{Salim}, Samir and {Lee}, Janice C. and {Janowiecki}, Steven and {da Cunha}, Elisabete and {Dickinson}, Mark and {Boquien}, M{\'e}d{\'e}ric and {Burgarella}, Denis and {Salzer}, John J. and {Charlot}, St{\'e}phane},
        title = "{GALEX-SDSS-WISE Legacy Catalog (GSWLC): Star Formation Rates, Stellar Masses, and Dust Attenuations of 700,000 Low-redshift Galaxies}",
      journal = {\apjs},
         year = 2016,
        month = nov,
       volume = {227},
       number = {1},
          eid = {2},
        pages = {2},
          doi = {10.3847/0067-0049/227/1/2},
archivePrefix = {arXiv},
       eprint = {1610.00712},
 primaryClass = {astro-ph.GA},
       adsurl = {https://ui.adsabs.harvard.edu/abs/2016ApJS..227....2S}
}

@ARTICLE{Simard2011,
       author = {{Simard}, Luc and {Mendel}, J. Trevor and {Patton}, David R. and {Ellison}, Sara L. and {McConnachie}, Alan W.},
        title = "{A Catalog of Bulge+disk Decompositions and Updated Photometry for 1.12 Million Galaxies in the Sloan Digital Sky Survey}",
      journal = {\apjs},
         year = 2011,
        month = sep,
       volume = {196},
       number = {1},
          eid = {11},
        pages = {11},
          doi = {10.1088/0067-0049/196/1/11},
archivePrefix = {arXiv},
       eprint = {1107.1518},
 primaryClass = {astro-ph.CO},
       adsurl = {https://ui.adsabs.harvard.edu/abs/2011ApJS..196...11S}
}

@ARTICLE{Chary2001,
       author = {{Chary}, R. and {Elbaz}, D.},
        title = "{Interpreting the Cosmic Infrared Background: Constraints on the Evolution of the Dust-enshrouded Star Formation Rate}",
      journal = {\apj},
         year = 2001,
        month = aug,
       volume = {556},
       number = {2},
        pages = {562-581},
          doi = {10.1086/321609},
archivePrefix = {arXiv},
       eprint = {astro-ph/0103067},
 primaryClass = {astro-ph},
       adsurl = {https://ui.adsabs.harvard.edu/abs/2001ApJ...556..562C}
}

@ARTICLE{Bell2005,
       author = {{Bell}, Eric F. and {Papovich}, Casey and {Wolf}, Christian and {Le Floc'h}, Emeric and {Caldwell}, John A.~R. and {Barden}, Marco and {Egami}, Eiichi and {McIntosh}, Daniel H. and {Meisenheimer}, Klaus and {P{\'e}rez-Gonz{\'a}lez}, Pablo G. and {Rieke}, G.~H. and {Rieke}, M.~J. and {Rigby}, Jane R. and {Rix}, Hans-Walter},
        title = "{Toward an Understanding of the Rapid Decline of the Cosmic Star Formation Rate}",
      journal = {\apj},
         year = 2005,
        month = may,
       volume = {625},
       number = {1},
        pages = {23-36},
          doi = {10.1086/429552},
archivePrefix = {arXiv},
       eprint = {astro-ph/0502246},
 primaryClass = {astro-ph},
       adsurl = {https://ui.adsabs.harvard.edu/abs/2005ApJ...625...23B}
}

@ARTICLE{Calzetti2010,
       author = {{Calzetti}, D. and {Wu}, S. -Y. and {Hong}, S. and {Kennicutt}, R.~C. and {Lee}, J.~C. and {Dale}, D.~A. and {Engelbracht}, C.~W. and {van Zee}, L. and {Draine}, B.~T. and {Hao}, C. -N. and {Gordon}, K.~D. and {Moustakas}, J. and {Murphy}, E.~J. and {Regan}, M. and {Begum}, A. and {Block}, M. and {Dalcanton}, J. and {Funes}, J. and {Gil de Paz}, A. and {Johnson}, B. and {Sakai}, S. and {Skillman}, E. and {Walter}, F. and {Weisz}, D. and {Williams}, B. and {Wu}, Y.},
        title = "{The Calibration of Monochromatic Far-Infrared Star Formation Rate Indicators}",
      journal = {\apj},
         year = 2010,
        month = may,
       volume = {714},
       number = {2},
        pages = {1256-1279},
          doi = {10.1088/0004-637X/714/2/1256},
archivePrefix = {arXiv},
       eprint = {1003.0961},
 primaryClass = {astro-ph.CO},
       adsurl = {https://ui.adsabs.harvard.edu/abs/2010ApJ...714.1256C}
}

@ARTICLE{Bendo2012,
       author = {{Bendo}, G.~J. and {Boselli}, A. and {Dariush}, A. and {Pohlen}, M. and {Roussel}, H. and {Sauvage}, M. and {Smith}, M.~W.~L. and {Wilson}, C.~D. and {Baes}, M. and {Cooray}, A. and {Clements}, D.~L. and {Cortese}, L. and {Foyle}, K. and {Galametz}, M. and {Gomez}, H.~L. and {Lebouteiller}, V. and {Lu}, N. and {Madden}, S.~C. and {Mentuch}, E. and {O'Halloran}, B. and {Page}, M.~J. and {Remy}, A. and {Schulz}, B. and {Spinoglio}, L.},
        title = "{Investigations of dust heating in M81, M83 and NGC 2403 with the Herschel Space Observatory}",
      journal = {\mnras},
         year = 2012,
        month = jan,
       volume = {419},
       number = {3},
        pages = {1833-1859},
          doi = {10.1111/j.1365-2966.2011.19735.x},
archivePrefix = {arXiv},
       eprint = {1109.0237},
 primaryClass = {astro-ph.CO},
       adsurl = {https://ui.adsabs.harvard.edu/abs/2012MNRAS.419.1833B}
}

@ARTICLE{Bianchi2007,
       author = {{Bianchi}, S.},
        title = "{The dust distribution in edge-on galaxies. Radiative transfer fits of V and K'-band images}",
      journal = {\aap},
         year = 2007,
        month = sep,
       volume = {471},
       number = {3},
        pages = {765-773},
          doi = {10.1051/0004-6361:20077649},
archivePrefix = {arXiv},
       eprint = {0705.1471},
 primaryClass = {astro-ph},
       adsurl = {https://ui.adsabs.harvard.edu/abs/2007A&A...471..765B}
}

@ARTICLE{Smith2016,
       author = {{Smith}, Matthew W.~L. and {Eales}, Stephen A. and {De Looze}, Ilse and {Baes}, Maarten and {Bendo}, George J. and {Bianchi}, Simone and {Boquien}, M{\'e}d{\'e}ric and {Boselli}, Alessandro and {Buat}, Veronique and {Ciesla}, Laure and {Clemens}, Marcel and {Clements}, David L. and {Cooray}, Asantha R. and {Cortese}, Luca and {Davies}, Jonathan I. and {Fritz}, Jacopo and {Gomez}, Haley L. and {Hughes}, Thomas M. and {Karczewski}, Oskar {\L}. and {Lu}, Nanyao and {Oliver}, Seb J. and {Remy-Ruyer}, Aur{\'e}lie and {Spinoglio}, Luigi and {Viaene}, Sebastien},
        title = "{Far-reaching dust distribution in galaxy discs}",
      journal = {\mnras},
         year = 2016,
        month = oct,
       volume = {462},
       number = {1},
        pages = {331-344},
          doi = {10.1093/mnras/stw1611},
archivePrefix = {arXiv},
       eprint = {1607.01020},
 primaryClass = {astro-ph.GA},
       adsurl = {https://ui.adsabs.harvard.edu/abs/2016MNRAS.462..331S}
}

@ARTICLE{Casasola2017,
       author = {{Casasola}, V. and {Cassar{\`a}}, L.~P. and {Bianchi}, S. and {Verstocken}, S. and {Xilouris}, E. and {Magrini}, L. and {Smith}, M.~W.~L. and {De Looze}, I. and {Galametz}, M. and {Madden}, S.~C. and {Baes}, M. and {Clark}, C. and {Davies}, J. and {De Vis}, P. and {Evans}, R. and {Fritz}, J. and {Galliano}, F. and {Jones}, A.~P. and {Mosenkov}, A.~V. and {Viaene}, S. and {Ysard}, N.},
        title = "{Radial distribution of dust, stars, gas, and star-formation rate in DustPedia face-on galaxies}",
      journal = {\aap},
         year = 2017,
        month = sep,
       volume = {605},
          eid = {A18},
        pages = {A18},
          doi = {10.1051/0004-6361/201731020},
archivePrefix = {arXiv},
       eprint = {1706.05351},
 primaryClass = {astro-ph.GA},
       adsurl = {https://ui.adsabs.harvard.edu/abs/2017A&A...605A..18C}
}

@ARTICLE{Gadotti2010,
       author = {{Gadotti}, Dimitri A. and {Baes}, Maarten and {Falony}, Sarah},
        title = "{Radiative transfer in disc galaxies - IV. The effects of dust attenuation on bulge and disc structural parameters}",
      journal = {\mnras},
         year = 2010,
        month = apr,
       volume = {403},
       number = {4},
        pages = {2053-2062},
          doi = {10.1111/j.1365-2966.2010.16243.x},
archivePrefix = {arXiv},
       eprint = {1001.2303},
 primaryClass = {astro-ph.CO},
       adsurl = {https://ui.adsabs.harvard.edu/abs/2010MNRAS.403.2053G}
}

@ARTICLE{Pastrav2013a,
       author = {{Pastrav}, B.~A. and {Popescu}, C.~C. and {Tuffs}, R.~J. and {Sansom}, A.~E.},
        title = "{The effects of dust on the derived photometric parameters of disks and bulges in spiral galaxies}",
      journal = {\aap},
         year = 2013,
        month = may,
       volume = {553},
          eid = {A80},
        pages = {A80},
          doi = {10.1051/0004-6361/201220962},
archivePrefix = {arXiv},
       eprint = {1301.5602},
 primaryClass = {astro-ph.CO},
       adsurl = {https://ui.adsabs.harvard.edu/abs/2013A&A...553A..80P}
}

@ARTICLE{Kennicutt1998,
       author = {{Kennicutt}, Robert C., Jr.},
        title = "{Star Formation in Galaxies Along the Hubble Sequence}",
      journal = {\araa},
         year = 1998,
        month = jan,
       volume = {36},
        pages = {189-232},
          doi = {10.1146/annurev.astro.36.1.189},
archivePrefix = {arXiv},
       eprint = {astro-ph/9807187},
 primaryClass = {astro-ph},
       adsurl = {https://ui.adsabs.harvard.edu/abs/1998ARA&A..36..189K}
}

@ARTICLE{Popescu2011,
       author = {{Popescu}, C.~C. and {Tuffs}, R.~J. and {Dopita}, M.~A. and {Fischera}, J. and {Kylafis}, N.~D. and {Madore}, B.~F.},
        title = "{Modelling the spectral energy distribution of galaxies. V. The dust and PAH emission SEDs of disk galaxies}",
      journal = {\aap},
         year = 2011,
        month = mar,
       volume = {527},
          eid = {A109},
        pages = {A109},
          doi = {10.1051/0004-6361/201015217},
archivePrefix = {arXiv},
       eprint = {1011.2942},
 primaryClass = {astro-ph.CO},
       adsurl = {https://ui.adsabs.harvard.edu/abs/2011A&A...527A.109P}
}

@ARTICLE{Heckman1998,
       author = {{Heckman}, Timothy M. and {Robert}, Carmelle and {Leitherer}, Claus and {Garnett}, Donald R. and {van der Rydt}, Fabienne},
        title = "{The Ultraviolet Spectroscopic Properties of Local Starbursts: Implications at High Redshift}",
      journal = {\apj},
         year = 1998,
        month = aug,
       volume = {503},
       number = {2},
        pages = {646-661},
          doi = {10.1086/306035},
archivePrefix = {arXiv},
       eprint = {astro-ph/9803185},
 primaryClass = {astro-ph},
       adsurl = {https://ui.adsabs.harvard.edu/abs/1998ApJ...503..646H}
}

@ARTICLE{Meurer1999,
       author = {{Meurer}, Gerhardt R. and {Heckman}, Timothy M. and {Calzetti}, Daniela},
        title = "{Dust Absorption and the Ultraviolet Luminosity Density at z \raisebox{-0.5ex}\textasciitilde 3 as Calibrated by Local Starburst Galaxies}",
      journal = {\apj},
         year = 1999,
        month = aug,
       volume = {521},
       number = {1},
        pages = {64-80},
          doi = {10.1086/307523},
archivePrefix = {arXiv},
       eprint = {astro-ph/9903054},
 primaryClass = {astro-ph},
       adsurl = {https://ui.adsabs.harvard.edu/abs/1999ApJ...521...64M}
}

@ARTICLE{Heinis2013,
       author = {{Heinis}, S. and {Buat}, V. and {B{\'e}thermin}, M. and {Aussel}, H. and {Bock}, J. and {Boselli}, A. and {Burgarella}, D. and {Conley}, A. and {Cooray}, A. and {Farrah}, D. and {Ibar}, E. and {Ilbert}, O. and {Ivison}, R.~J. and {Magdis}, G. and {Marsden}, G. and {Oliver}, S.~J. and {Page}, M.~J. and {Rodighiero}, G. and {Roehlly}, Y. and {Schulz}, B. and {Scott}, Douglas and {Smith}, A.~J. and {Viero}, M. and {Wang}, L. and {Zemcov}, M.},
        title = "{HERMES: unveiling obscured star formation - the far-infrared luminosity function of ultraviolet-selected galaxies at z {\ensuremath{\sim}} 1.5}",
      journal = {\mnras},
         year = 2013,
        month = feb,
       volume = {429},
       number = {2},
        pages = {1113-1132},
          doi = {10.1093/mnras/sts397},
archivePrefix = {arXiv},
       eprint = {1211.4336},
 primaryClass = {astro-ph.CO},
       adsurl = {https://ui.adsabs.harvard.edu/abs/2013MNRAS.429.1113H}
}

@ARTICLE{Bourne2017,
       author = {{Bourne}, N. and {Dunlop}, J.~S. and {Merlin}, E. and {Parsa}, S. and {Schreiber}, C. and {Castellano}, M. and {Conselice}, C.~J. and {Coppin}, K.~E.~K. and {Farrah}, D. and {Fontana}, A. and {Geach}, J.~E. and {Halpern}, M. and {Knudsen}, K.~K. and {Micha{\l}owski}, M.~J. and {Mortlock}, A. and {Santini}, P. and {Scott}, D. and {Shu}, X.~W. and {Simpson}, C. and {Simpson}, J.~M. and {Smith}, D.~J.~B. and {van der Werf}, P.~P.},
        title = "{Evolution of cosmic star formation in the SCUBA-2 Cosmology Legacy Survey}",
      journal = {\mnras},
         year = 2017,
        month = may,
       volume = {467},
       number = {2},
        pages = {1360-1385},
          doi = {10.1093/mnras/stx031},
archivePrefix = {arXiv},
       eprint = {1607.04283},
 primaryClass = {astro-ph.GA},
       adsurl = {https://ui.adsabs.harvard.edu/abs/2017MNRAS.467.1360B}
}

@ARTICLE{Wild2011,
       author = {{Wild}, Vivienne and {Charlot}, St{\'e}phane and {Brinchmann}, Jarle and {Heckman}, Timothy and {Vince}, Oliver and {Pacifici}, Camilla and {Chevallard}, Jacopo},
        title = "{Empirical determination of the shape of dust attenuation curves in star-forming galaxies}",
      journal = {\mnras},
         year = 2011,
        month = nov,
       volume = {417},
       number = {3},
        pages = {1760-1786},
          doi = {10.1111/j.1365-2966.2011.19367.x},
archivePrefix = {arXiv},
       eprint = {1106.1646},
 primaryClass = {astro-ph.CO},
       adsurl = {https://ui.adsabs.harvard.edu/abs/2011MNRAS.417.1760W}
}

@ARTICLE{Reddy2015,
       author = {{Reddy}, Naveen A. and {Kriek}, Mariska and {Shapley}, Alice E. and {Freeman}, William R. and {Siana}, Brian and {Coil}, Alison L. and {Mobasher}, Bahram and {Price}, Sedona H. and {Sanders}, Ryan L. and {Shivaei}, Irene},
        title = "{The MOSDEF Survey: Measurements of Balmer Decrements and the Dust Attenuation Curve at Redshifts z \raisebox{-0.5ex}\textasciitilde 1.4-2.6}",
      journal = {\apj},
         year = 2015,
        month = jun,
       volume = {806},
       number = {2},
          eid = {259},
        pages = {259},
          doi = {10.1088/0004-637X/806/2/259},
archivePrefix = {arXiv},
       eprint = {1504.02782},
 primaryClass = {astro-ph.GA},
       adsurl = {https://ui.adsabs.harvard.edu/abs/2015ApJ...806..259R}
}

@ARTICLE{Lorenz2023,
       author = {{Lorenz}, Brian and {Kriek}, Mariska and {Shapley}, Alice E. and {Reddy}, Naveen A. and {Sanders}, Ryan L. and {Barro}, Guillermo and {Coil}, Alison L. and {Mobasher}, Bahram and {Price}, Sedona H. and {Runco}, Jordan N. and {Shivaei}, Irene and {Siana}, Brian and {Weisz}, Daniel R.},
        title = "{An Updated Dust-to-Star Geometry: Dust Attenuation Does Not Depend on Inclination in 1.3 {\ensuremath{\leq}}z {\ensuremath{\leq}}2.6 Star-forming Galaxies from MOSDEF}",
      journal = {\apj},
         year = 2023,
        month = jul,
       volume = {951},
       number = {1},
          eid = {29},
        pages = {29},
          doi = {10.3847/1538-4357/accdd1},
archivePrefix = {arXiv},
       eprint = {2304.08521},
 primaryClass = {astro-ph.GA},
       adsurl = {https://ui.adsabs.harvard.edu/abs/2023ApJ...951...29L}
}

@ARTICLE{Kong2004,
       author = {{Kong}, X. and {Charlot}, S. and {Brinchmann}, J. and {Fall}, S.~M.},
        title = "{Star formation history and dust content of galaxies drawn from ultraviolet surveys}",
      journal = {\mnras},
         year = 2004,
        month = apr,
       volume = {349},
       number = {3},
        pages = {769-778},
          doi = {10.1111/j.1365-2966.2004.07556.x},
archivePrefix = {arXiv},
       eprint = {astro-ph/0312474},
 primaryClass = {astro-ph},
       adsurl = {https://ui.adsabs.harvard.edu/abs/2004MNRAS.349..769K}
}

@ARTICLE{Baldwin1981,
       author = {{Baldwin}, J.~A. and {Phillips}, M.~M. and {Terlevich}, R.},
        title = "{Classification parameters for the emission-line spectra of extragalactic objects.}",
      journal = {\pasp},
         year = 1981,
        month = feb,
       volume = {93},
        pages = {5-19},
          doi = {10.1086/130766},
       adsurl = {https://ui.adsabs.harvard.edu/abs/1981PASP...93....5B}
}

@ARTICLE{Kewley2005,
       author = {{Kewley}, Lisa J. and {Jansen}, Rolf A. and {Geller}, Margaret J.},
        title = "{Aperture Effects on Star Formation Rate, Metallicity, and Reddening}",
      journal = {\pasp},
         year = 2005,
        month = mar,
       volume = {117},
       number = {829},
        pages = {227-244},
          doi = {10.1086/428303},
archivePrefix = {arXiv},
       eprint = {astro-ph/0501229},
 primaryClass = {astro-ph},
       adsurl = {https://ui.adsabs.harvard.edu/abs/2005PASP..117..227K}
}

@ARTICLE{Chruslinska2024,
       author = {{Chru{\'s}li{\'n}ska}, M. and {Pakmor}, R. and {Matthee}, J. and {Matsuno}, T.},
        title = "{Trading oxygen for iron. I. The [O/Fe]-specific star formation rate relation of galaxies}",
      journal = {\aap},
         year = 2024,
        month = jun,
       volume = {686},
          eid = {A186},
        pages = {A186},
          doi = {10.1051/0004-6361/202347602},
archivePrefix = {arXiv},
       eprint = {2308.00023},
 primaryClass = {astro-ph.GA},
       adsurl = {https://ui.adsabs.harvard.edu/abs/2024A&A...686A.186C}
}

@ARTICLE{Chevallard2013,
       author = {{Chevallard}, J. and {Charlot}, S. and {Wandelt}, B. and {Wild}, V.},
        title = "{Insights into the content and spatial distribution of dust from the integrated spectral properties of galaxies}",
      journal = {\mnras},
         year = 2013,
        month = jul,
       volume = {432},
       number = {3},
        pages = {2061-2091},
          doi = {10.1093/mnras/stt523},
archivePrefix = {arXiv},
       eprint = {1303.6631},
 primaryClass = {astro-ph.CO},
       adsurl = {https://ui.adsabs.harvard.edu/abs/2013MNRAS.432.2061C}
}

@ARTICLE{Galliano2021,
       author = {{Galliano}, Fr{\'e}d{\'e}ric and {Nersesian}, Angelos and {Bianchi}, Simone and {De Looze}, Ilse and {Roychowdhury}, Sambit and {Baes}, Maarten and {Casasola}, Viviana and {Cassar{\'a}}, Letizia P. and {Dobbels}, Wouter and {Fritz}, Jacopo and {Galametz}, Maud and {Jones}, Anthony P. and {Madden}, Suzanne C. and {Mosenkov}, Aleksandr and {Xilouris}, Emmanuel M. and {Ysard}, Nathalie},
        title = "{A nearby galaxy perspective on dust evolution. Scaling relations and constraints on the dust build-up in galaxies with the DustPedia and DGS samples}",
      journal = {\aap},
         year = 2021,
        month = may,
       volume = {649},
          eid = {A18},
        pages = {A18},
          doi = {10.1051/0004-6361/202039701},
archivePrefix = {arXiv},
       eprint = {2101.00456},
 primaryClass = {astro-ph.GA},
       adsurl = {https://ui.adsabs.harvard.edu/abs/2021A&A...649A..18G}
}

@ARTICLE{Calura2025,
       author = {{Calura}, Francesco},
        title = "{Interstellar dust production, destruction and effects of dust depletion in galaxies}",
      journal = {arXiv e-prints},
         year = 2025,
        month = jun,
          eid = {arXiv:2506.13851},
        pages = {arXiv:2506.13851},
          doi = {10.48550/arXiv.2506.13851},
archivePrefix = {arXiv},
       eprint = {2506.13851},
 primaryClass = {astro-ph.GA},
       adsurl = {https://ui.adsabs.harvard.edu/abs/2025arXiv250613851C}
}

@ARTICLE{Lower2022,
       author = {{Lower}, Sidney and {Narayanan}, Desika and {Leja}, Joel and {Johnson}, Benjamin D. and {Conroy}, Charlie and {Dav{\'e}}, Romeel},
        title = "{How Well Can We Measure Galaxy Dust Attenuation Curves? The Impact of the Assumed Star-dust Geometry Model in Spectral Energy Distribution Fitting}",
      journal = {\apj},
         year = 2022,
        month = may,
       volume = {931},
       number = {1},
          eid = {14},
        pages = {14},
          doi = {10.3847/1538-4357/ac6959},
archivePrefix = {arXiv},
       eprint = {2203.00074},
 primaryClass = {astro-ph.GA},
       adsurl = {https://ui.adsabs.harvard.edu/abs/2022ApJ...931...14L}
}

@ARTICLE{Iyer2020,
       author = {{Iyer}, Kartheik G. and {Tacchella}, Sandro and {Genel}, Shy and {Hayward}, Christopher C. and {Hernquist}, Lars and {Brooks}, Alyson M. and {Caplar}, Neven and {Dav{\'e}}, Romeel and {Diemer}, Benedikt and {Forbes}, John C. and {Gawiser}, Eric and {Somerville}, Rachel S. and {Starkenburg}, Tjitske K.},
        title = "{The diversity and variability of star formation histories in models of galaxy evolution}",
      journal = {\mnras},
         year = 2020,
        month = oct,
       volume = {498},
       number = {1},
        pages = {430-463},
          doi = {10.1093/mnras/staa2150},
archivePrefix = {arXiv},
       eprint = {2007.07916},
 primaryClass = {astro-ph.GA},
       adsurl = {https://ui.adsabs.harvard.edu/abs/2020MNRAS.498..430I}
}

@ARTICLE{Kouroumpatzakis2023,
       author = {{Kouroumpatzakis}, K. and {Zezas}, A. and {Kyritsis}, E. and {Salim}, S. and {Svoboda}, J.},
        title = "{Star formation rate and stellar mass calibrations based on infrared photometry and their dependence on stellar population age and extinction}",
      journal = {\aap},
         year = 2023,
        month = may,
       volume = {673},
          eid = {A16},
        pages = {A16},
          doi = {10.1051/0004-6361/202346054},
archivePrefix = {arXiv},
       eprint = {2303.10013},
 primaryClass = {astro-ph.GA},
       adsurl = {https://ui.adsabs.harvard.edu/abs/2023A&A...673A..16K}
}

@ARTICLE{Popping2022,
       author = {{Popping}, Gerg{\"o} and {P{\'e}roux}, C{\'e}line},
        title = "{Observed cosmic evolution of galaxy dust properties with metallicity and tensions with models}",
      journal = {\mnras},
         year = 2022,
        month = jun,
       volume = {513},
       number = {1},
        pages = {1531-1543},
          doi = {10.1093/mnras/stac695},
archivePrefix = {arXiv},
       eprint = {2203.03686},
 primaryClass = {astro-ph.GA},
       adsurl = {https://ui.adsabs.harvard.edu/abs/2022MNRAS.513.1531P}
}

@ARTICLE{Shivaei2020,
       author = {{Shivaei}, Irene and {Darvish}, Behnam and {Sattari}, Zahra and {Chartab}, Nima and {Mobasher}, Bahram and {Scoville}, Nick and {Rieke}, George},
        title = "{Dependence of the IRX-{\ensuremath{\beta}} Dust Attenuation Relation on Metallicity and Environment}",
      journal = {\apjl},
         year = 2020,
        month = nov,
       volume = {903},
       number = {2},
          eid = {L28},
        pages = {L28},
          doi = {10.3847/2041-8213/abc1ef},
archivePrefix = {arXiv},
       eprint = {2010.10538},
 primaryClass = {astro-ph.GA},
       adsurl = {https://ui.adsabs.harvard.edu/abs/2020ApJ...903L..28S}
}

@ARTICLE{Bowler2024,
       author = {{Bowler}, R.~A.~A. and {Inami}, H. and {Sommovigo}, L. and {Smit}, R. and {Algera}, H.~S.~B. and {Aravena}, M. and {Barrufet}, L. and {Bouwens}, R. and {da Cunha}, E. and {Cullen}, F. and {Dayal}, P. and {De Looze}, I. and {Dunlop}, J.~S. and {Fudamoto}, Y. and {Mauerhofer}, V. and {McLure}, R.~J. and {Stefanon}, M. and {Schneider}, R. and {Ferrara}, A. and {Graziani}, L. and {Hodge}, J.~A. and {Nanayakkara}, T. and {Palla}, M. and {Schouws}, S. and {Stark}, D.~P. and {van der Werf}, P.~P.},
        title = "{The ALMA REBELS survey: obscured star formation in massive Lyman-break galaxies at z= 4-8 revealed by the IRX-{\ensuremath{\beta}} and M$_{{\ensuremath{\star}}}$ relations}",
      journal = {\mnras},
         year = 2024,
        month = jan,
       volume = {527},
       number = {3},
        pages = {5808-5828},
          doi = {10.1093/mnras/stad3578},
archivePrefix = {arXiv},
       eprint = {2309.17386},
 primaryClass = {astro-ph.GA},
       adsurl = {https://ui.adsabs.harvard.edu/abs/2024MNRAS.527.5808B}
}

@ARTICLE{Zhang2023,
       author = {{Zhang}, Junkai and {Wuyts}, Stijn and {Cutler}, Sam E. and {Mowla}, Lamiya A. and {Brammer}, Gabriel B. and {Momcheva}, Ivelina G. and {Whitaker}, Katherine E. and {van Dokkum}, Pieter and {F{\"o}rster Schreiber}, Natascha M. and {Nelson}, Erica J. and {Schady}, Patricia and {Villforth}, Carolin and {Wake}, David and {van der Wel}, Arjen},
        title = "{Dust attenuation, dust content, and geometry of star-forming galaxies}",
      journal = {\mnras},
         year = 2023,
        month = sep,
       volume = {524},
       number = {3},
        pages = {4128-4147},
          doi = {10.1093/mnras/stad2066},
archivePrefix = {arXiv},
       eprint = {2307.02568},
 primaryClass = {astro-ph.GA},
       adsurl = {https://ui.adsabs.harvard.edu/abs/2023MNRAS.524.4128Z}
}

@ARTICLE{Sachdeva2022,
       author = {{Sachdeva}, Sonali and {Nath}, Biman B.},
        title = "{Star-dust geometry main determinant of dust attenuation in galaxies}",
      journal = {\mnras},
         year = 2022,
        month = jun,
       volume = {513},
       number = {1},
        pages = {L63-L67},
          doi = {10.1093/mnrasl/slac037},
archivePrefix = {arXiv},
       eprint = {2204.03478},
 primaryClass = {astro-ph.GA},
       adsurl = {https://ui.adsabs.harvard.edu/abs/2022MNRAS.513L..63S}
}

@ARTICLE{Shapley2020,
       author = {{Shapley}, Alice E. and {Cullen}, Fergus and {Dunlop}, James S. and {McLure}, Ross J. and {Kriek}, Mariska and {Reddy}, Naveen A. and {Sanders}, Ryan L.},
        title = "{The First Robust Constraints on the Relationship between Dust-to-gas Ratio and Metallicity in Luminous Star-forming Galaxies at High Redshift}",
      journal = {\apjl},
         year = 2020,
        month = nov,
       volume = {903},
       number = {1},
          eid = {L16},
        pages = {L16},
          doi = {10.3847/2041-8213/abc006},
archivePrefix = {arXiv},
       eprint = {2009.10091},
 primaryClass = {astro-ph.GA},
       adsurl = {https://ui.adsabs.harvard.edu/abs/2020ApJ...903L..16S}
}

@ARTICLE{Popping2023,
       author = {{Popping}, Gerg{\"o} and {Shivaei}, Irene and {Sanders}, Ryan L. and {Jones}, Tucker and {Pope}, Alexandra and {Reddy}, Naveen A. and {Shapley}, Alice E. and {Coil}, Alison L. and {Kriek}, Mariska},
        title = "{The dust-to-gas mass ratio of luminous galaxies as a function of their metallicity at cosmic noon}",
      journal = {\aap},
         year = 2023,
        month = feb,
       volume = {670},
          eid = {A138},
        pages = {A138},
          doi = {10.1051/0004-6361/202243817},
archivePrefix = {arXiv},
       eprint = {2204.08483},
 primaryClass = {astro-ph.GA},
       adsurl = {https://ui.adsabs.harvard.edu/abs/2023A&A...670A.138P}
}

@ARTICLE{Konstantopoulou2024,
       author = {{Konstantopoulou}, Christina and {De Cia}, Annalisa and {Ledoux}, C{\'e}dric and {Krogager}, Jens-Kristian and {Mattsson}, Lars and {Watson}, Darach and {Heintz}, Kasper E. and {P{\'e}roux}, C{\'e}line and {Noterdaeme}, Pasquier and {Andersen}, Anja C. and {Fynbo}, Johan P.~U. and {Jermann}, Iris and {Ramburuth-Hurt}, Tanita},
        title = "{Dust depletion of metals from local to distant galaxies. II. Cosmic dust-to-metal ratio and dust composition}",
      journal = {\aap},
         year = 2024,
        month = jan,
       volume = {681},
          eid = {A64},
        pages = {A64},
          doi = {10.1051/0004-6361/202347171},
archivePrefix = {arXiv},
       eprint = {2310.07709},
 primaryClass = {astro-ph.GA},
       adsurl = {https://ui.adsabs.harvard.edu/abs/2024A&A...681A..64K}
}

@ARTICLE{Markov2025,
       author = {{Markov}, V. and {Gallerani}, S. and {Pallottini}, A. and {Brada{\v{c}}}, M. and {Carniani}, S. and {Tripodi}, R. and {Noirot}, G. and {Di Mascia}, F. and {Parlanti}, E. and {Martis}, N.},
        title = "{Unveiling the trends between dust attenuation and galaxy properties at z {\ensuremath{\sim}} 2‑12 with the James Webb Space Telescope}",
      journal = {\aap},
         year = 2025,
        month = oct,
       volume = {702},
          eid = {A33},
        pages = {A33},
          doi = {10.1051/0004-6361/202555182},
archivePrefix = {arXiv},
       eprint = {2504.12378},
 primaryClass = {astro-ph.GA},
       adsurl = {https://ui.adsabs.harvard.edu/abs/2025A&A...702A..33M}
}

@ARTICLE{Salim2020,
       author = {{Salim}, Samir and {Narayanan}, Desika},
        title = "{The Dust Attenuation Law in Galaxies}",
      journal = {\araa},
         year = 2020,
        month = aug,
       volume = {58},
        pages = {529-575},
          doi = {10.1146/annurev-astro-032620-021933},
archivePrefix = {arXiv},
       eprint = {2001.03181},
 primaryClass = {astro-ph.GA},
       adsurl = {https://ui.adsabs.harvard.edu/abs/2020ARA&A..58..529S}
}

@ARTICLE{Mitsuhashi2024a,
       author = {{Mitsuhashi}, Ikki and {Harikane}, Yuichi and {Bauer}, Franz E. and {Bakx}, Tom J.~L.~C. and {Ferrara}, Andrea and {Fujimoto}, Seiji and {Hashimoto}, Takuya and {Inoue}, Akio K. and {Iwasawa}, Kazushi and {Nishimura}, Yuri and {Imanishi}, Masatoshi and {Ono}, Yoshiaki and {Saito}, Toshiki and {Sugahara}, Yuma and {Umehata}, Hideki and {Vallini}, Livia and {Wang}, Tao and {Zavala}, Jorge A.},
        title = "{SERENADE. II. An ALMA Multiband Dust Continuum Analysis of 28 Galaxies at 5 < z < 8 and the Physical Origin of the Dust Temperature Evolution}",
      journal = {\apj},
         year = 2024,
        month = aug,
       volume = {971},
       number = {2},
          eid = {161},
        pages = {161},
          doi = {10.3847/1538-4357/ad5675},
archivePrefix = {arXiv},
       eprint = {2311.16857},
 primaryClass = {astro-ph.GA},
       adsurl = {https://ui.adsabs.harvard.edu/abs/2024ApJ...971..161M}
}

@ARTICLE{Greener2020,
       author = {{Greener}, Michael J. and {Arag{\'o}n-Salamanca}, Alfonso and {Merrifield}, Michael R. and {Peterken}, Thomas G. and {Fraser-McKelvie}, Amelia and {Masters}, Karen L. and {Krawczyk}, Coleman M. and {Boardman}, Nicholas F. and {Boquien}, M{\'e}d{\'e}ric and {Andrews}, Brett H. and {Brinkmann}, Jonathan and {Drory}, Niv},
        title = "{SDSS-IV MaNGA: spatially resolved dust attenuation in spiral galaxies}",
      journal = {\mnras},
         year = 2020,
        month = jun,
       volume = {495},
       number = {2},
        pages = {2305-2320},
          doi = {10.1093/mnras/staa1300},
archivePrefix = {arXiv},
       eprint = {2005.02772},
 primaryClass = {astro-ph.GA},
       adsurl = {https://ui.adsabs.harvard.edu/abs/2020MNRAS.495.2305G}
}

@ARTICLE{Maheson2025,
       author = {{Maheson}, Gabriel and {Tacchella}, Sandro and {Belli}, Sirio and {Park}, Minjung and {Danhaive}, A. Lola and {Bugiani}, Letizia and {Davies}, Rebecca and {Emami}, Razieh and {Khoram}, Amir H. and {Lam}, Laurence and {Leja}, Joel and {Mendel}, Trevor and {Nelson}, Erica June},
        title = "{Big, Dusty Galaxies in Blue Jay: Insights into the Relationship Between Morphology and Dust Attenuation at Cosmic Noon}",
      journal = {arXiv e-prints},
         year = 2025,
        month = apr,
          eid = {arXiv:2504.15346},
        pages = {arXiv:2504.15346},
          doi = {10.48550/arXiv.2504.15346},
archivePrefix = {arXiv},
       eprint = {2504.15346},
 primaryClass = {astro-ph.GA},
       adsurl = {https://ui.adsabs.harvard.edu/abs/2025arXiv250415346M}
}

@ARTICLE{Lu2022,
       author = {{Lu}, Jiafeng and {Shen}, Shiyin and {Yuan}, Fang-Ting and {Shao}, Zhengyi and {Hou}, Jinliang and {Zheng}, Xianzhong},
        title = "{The Chocolate Chip Cookie Model: Dust Geometry of Milky Way-like Disk Galaxies}",
      journal = {\apj},
         year = 2022,
        month = oct,
       volume = {938},
       number = {2},
          eid = {139},
        pages = {139},
          doi = {10.3847/1538-4357/ac92e9},
archivePrefix = {arXiv},
       eprint = {2209.08515},
 primaryClass = {astro-ph.GA},
       adsurl = {https://ui.adsabs.harvard.edu/abs/2022ApJ...938..139L}
}

@ARTICLE{Lu2023,
       author = {{Lu}, Jiafeng and {Shen}, Shiyin and {Yuan}, Fang-Ting and {Zeng}, Qi},
        title = "{The Chocolate Chip Cookie Model: Dust-to-metal Ratio of H II Regions}",
      journal = {\apjl},
         year = 2023,
        month = mar,
       volume = {946},
       number = {1},
          eid = {L7},
        pages = {L7},
          doi = {10.3847/2041-8213/acc4b7},
archivePrefix = {arXiv},
       eprint = {2303.12303},
 primaryClass = {astro-ph.GA},
       adsurl = {https://ui.adsabs.harvard.edu/abs/2023ApJ...946L...7L}
}

@ARTICLE{Dale2005,
       author = {{Dale}, D.~A. and {Bendo}, G.~J. and {Engelbracht}, C.~W. and {Gordon}, K.~D. and {Regan}, M.~W. and {Armus}, L. and {Cannon}, J.~M. and {Calzetti}, D. and {Draine}, B.~T. and {Helou}, G. and {Joseph}, R.~D. and {Kennicutt}, R.~C. and {Li}, A. and {Murphy}, E.~J. and {Roussel}, H. and {Walter}, F. and {Hanson}, H.~M. and {Hollenbach}, D.~J. and {Jarrett}, T.~H. and {Kewley}, L.~J. and {Lamanna}, C.~A. and {Leitherer}, C. and {Meyer}, M.~J. and {Rieke}, G.~H. and {Rieke}, M.~J. and {Sheth}, K. and {Smith}, J.~D.~T. and {Thornley}, M.~D.},
        title = "{Infrared Spectral Energy Distributions of Nearby Galaxies}",
      journal = {\apj},
         year = 2005,
        month = nov,
       volume = {633},
       number = {2},
        pages = {857-870},
          doi = {10.1086/491642},
archivePrefix = {arXiv},
       eprint = {astro-ph/0507645},
 primaryClass = {astro-ph},
       adsurl = {https://ui.adsabs.harvard.edu/abs/2005ApJ...633..857D}
}

@ARTICLE{Calzetti2000,
       author = {{Calzetti}, Daniela and {Armus}, Lee and {Bohlin}, Ralph C. and {Kinney}, Anne L. and {Koornneef}, Jan and {Storchi-Bergmann}, Thaisa},
        title = "{The Dust Content and Opacity of Actively Star-forming Galaxies}",
      journal = {\apj},
         year = 2000,
        month = apr,
       volume = {533},
       number = {2},
        pages = {682-695},
          doi = {10.1086/308692},
archivePrefix = {arXiv},
       eprint = {astro-ph/9911459},
 primaryClass = {astro-ph},
       adsurl = {https://ui.adsabs.harvard.edu/abs/2000ApJ...533..682C}
}

@ARTICLE{BC03,
       author = {{Bruzual}, G. and {Charlot}, S.},
        title = "{Stellar population synthesis at the resolution of 2003}",
      journal = {\mnras},
         year = 2003,
        month = oct,
       volume = {344},
       number = {4},
        pages = {1000-1028},
          doi = {10.1046/j.1365-8711.2003.06897.x},
archivePrefix = {arXiv},
       eprint = {astro-ph/0309134},
 primaryClass = {astro-ph},
       adsurl = {https://ui.adsabs.harvard.edu/abs/2003MNRAS.344.1000B}
}

@ARTICLE{Narayanan2018,
       author = {{Narayanan}, Desika and {Dav{\'e}}, Romeel and {Johnson}, Benjamin D. and {Thompson}, Robert and {Conroy}, Charlie and {Geach}, James},
        title = "{The IRX-{\ensuremath{\beta}} dust attenuation relation in cosmological galaxy formation simulations}",
      journal = {\mnras},
         year = 2018,
        month = feb,
       volume = {474},
       number = {2},
        pages = {1718-1736},
          doi = {10.1093/mnras/stx2860},
archivePrefix = {arXiv},
       eprint = {1705.05858},
 primaryClass = {astro-ph.GA},
       adsurl = {https://ui.adsabs.harvard.edu/abs/2018MNRAS.474.1718N}
}

@ARTICLE{Trayford2020,
       author = {{Trayford}, James W. and {Lagos}, Claudia del P. and {Robotham}, Aaron S.~G. and {Obreschkow}, Danail},
        title = "{Fade to grey: systematic variation of galaxy attenuation curves with galaxy properties in the EAGLE simulations}",
      journal = {\mnras},
         year = 2020,
        month = jan,
       volume = {491},
       number = {3},
        pages = {3937-3951},
          doi = {10.1093/mnras/stz3234},
archivePrefix = {arXiv},
       eprint = {1908.08956},
 primaryClass = {astro-ph.GA},
       adsurl = {https://ui.adsabs.harvard.edu/abs/2020MNRAS.491.3937T}
}

@ARTICLE{Charlot2000,
       author = {{Charlot}, St{\'e}phane and {Fall}, S. Michael},
        title = "{A Simple Model for the Absorption of Starlight by Dust in Galaxies}",
      journal = {\apj},
         year = 2000,
        month = aug,
       volume = {539},
       number = {2},
        pages = {718-731},
          doi = {10.1086/309250},
archivePrefix = {arXiv},
       eprint = {astro-ph/0003128},
 primaryClass = {astro-ph},
       adsurl = {https://ui.adsabs.harvard.edu/abs/2000ApJ...539..718C}
}

@ARTICLE{SalimNarayanan2020,
       author = {{Salim}, Samir and {Narayanan}, Desika},
        title = "{The Dust Attenuation Law in Galaxies}",
      journal = {\araa},
         year = 2020,
        month = aug,
       volume = {58},
        pages = {529-575},
          doi = {10.1146/annurev-astro-032620-021933},
archivePrefix = {arXiv},
       eprint = {2001.03181},
 primaryClass = {astro-ph.GA},
       adsurl = {https://ui.adsabs.harvard.edu/abs/2020ARA&A..58..529S}
}

@ARTICLE{Fudamoto2020,
       author = {{Fudamoto}, Y. and {Oesch}, P.~A. and {Faisst}, A. and {B{\'e}thermin}, M. and {Ginolfi}, M. and {Khusanova}, Y. and {Loiacono}, F. and {Le F{\`e}vre}, O. and {Capak}, P. and {Schaerer}, D. and {Silverman}, J.~D. and {Cassata}, P. and {Yan}, L. and {Amorin}, R. and {Bardelli}, S. and {Boquien}, M. and {Cimatti}, A. and {Dessauges-Zavadsky}, M. and {Fujimoto}, S. and {Gruppioni}, C. and {Hathi}, N.~P. and {Ibar}, E. and {Jones}, G.~C. and {Koekemoer}, A.~M. and {Lagache}, G. and {Lemaux}, B.~C. and {Maiolino}, R. and {Narayanan}, D. and {Pozzi}, F. and {Riechers}, D.~A. and {Rodighiero}, G. and {Talia}, M. and {Toft}, S. and {Vallini}, L. and {Vergani}, D. and {Zamorani}, G. and {Zucca}, E.},
        title = "{The ALPINE-ALMA [CII] survey. Dust attenuation properties and obscured star formation at z {\ensuremath{\sim}} 4.4-5.8}",
      journal = {\aap},
         year = 2020,
        month = nov,
       volume = {643},
          eid = {A4},
        pages = {A4},
          doi = {10.1051/0004-6361/202038163},
archivePrefix = {arXiv},
       eprint = {2004.10760},
 primaryClass = {astro-ph.GA},
       adsurl = {https://ui.adsabs.harvard.edu/abs/2020A&A...643A...4F}
}

@ARTICLE{Gomez-Guijarro2023,
       author = {{G{\'o}mez-Guijarro}, Carlos and {Magnelli}, Benjamin and {Elbaz}, David and {Wuyts}, Stijn and {Daddi}, Emanuele and {Le Bail}, Aur{\'e}lien and {Giavalisco}, Mauro and {Dickinson}, Mark and {P{\'e}rez-Gonz{\'a}lez}, Pablo G. and {Arrabal Haro}, Pablo and {Bagley}, Micaela B. and {Bisigello}, Laura and {Buat}, V{\'e}ronique and {Burgarella}, Denis and {Calabr{\`o}}, Antonello and {Casey}, Caitlin M. and {Cheng}, Yingjie and {Ciesla}, Laure and {Dekel}, Avishai and {Ferguson}, Henry C. and {Finkelstein}, Steven L. and {Franco}, Maximilien and {Grogin}, Norman A. and {Holwerda}, Benne W. and {Jin}, Shuowen and {Kartaltepe}, Jeyhan S. and {Koekemoer}, Anton M. and {Kokorev}, Vasily and {Long}, Arianna S. and {Lucas}, Ray A. and {Magdis}, Georgios E. and {Papovich}, Casey and {Pirzkal}, Nor and {Seill{\'e}}, Lise-Marie and {Tacchella}, Sandro and {Tarrasse}, Maxime and {Valentino}, Francesco and {de la Vega}, Alexander and {Wilkins}, Stephen M. and {Xiao}, Mengyuan and {Yung}, L.~Y. Aaron},
        title = "{JWST CEERS probes the role of stellar mass and morphology in obscuring galaxies}",
      journal = {\aap},
         year = 2023,
        month = sep,
       volume = {677},
          eid = {A34},
        pages = {A34},
          doi = {10.1051/0004-6361/202346673},
archivePrefix = {arXiv},
       eprint = {2304.08517},
 primaryClass = {astro-ph.GA},
       adsurl = {https://ui.adsabs.harvard.edu/abs/2023A&A...677A..34G}
}

@ARTICLE{Junais2024,
       author = {{Junais} and {Weilbacher}, P.~M. and {Epinat}, B. and {Boissier}, S. and {Galaz}, G. and {Johnston}, E.~J. and {Puzia}, T.~H. and {Amram}, P. and {Ma{\l}ek}, K.},
        title = "{MUSE observations of the giant low surface brightness galaxy Malin 1: Numerous HII regions, star formation rate, metallicity, and dust attenuation}",
      journal = {\aap},
         year = 2024,
        month = jan,
       volume = {681},
          eid = {A100},
        pages = {A100},
          doi = {10.1051/0004-6361/202347669},
archivePrefix = {arXiv},
       eprint = {2310.11872},
 primaryClass = {astro-ph.GA},
       adsurl = {https://ui.adsabs.harvard.edu/abs/2024A&A...681A.100J}
}

@ARTICLE{Tacchella2019,
       author = {{Tacchella}, Sandro and {Diemer}, Benedikt and {Hernquist}, Lars and {Genel}, Shy and {Marinacci}, Federico and {Nelson}, Dylan and {Pillepich}, Annalisa and {Rodriguez-Gomez}, Vicente and {Sales}, Laura V. and {Springel}, Volker and {Vogelsberger}, Mark},
        title = "{Morphology and star formation in IllustrisTNG: the build-up of spheroids and discs}",
      journal = {\mnras},
         year = 2019,
        month = aug,
       volume = {487},
       number = {4},
        pages = {5416-5440},
          doi = {10.1093/mnras/stz1657},
archivePrefix = {arXiv},
       eprint = {1904.12860},
 primaryClass = {astro-ph.GA},
       adsurl = {https://ui.adsabs.harvard.edu/abs/2019MNRAS.487.5416T}
}

@ARTICLE{Lorenz2024,
       author = {{Lorenz}, Brian and {Kriek}, Mariska and {Shapley}, Alice E. and {Sanders}, Ryan L. and {Coil}, Alison L. and {Leja}, Joel and {Mobasher}, Bahram and {Nelson}, Erica and {Price}, Sedona H. and {Reddy}, Naveen A. and {Runco}, Jordan N. and {Suess}, Katherine A. and {Shivaei}, Irene and {Siana}, Brian and {Weisz}, Daniel R.},
        title = "{Stacking and Analyzing MOSDEF Galaxies by Spectral Types: Implications for Dust Geometry and Galaxy Evolution}",
      journal = {\apj},
         year = 2024,
        month = nov,
       volume = {975},
       number = {2},
          eid = {187},
        pages = {187},
          doi = {10.3847/1538-4357/ad7de8},
archivePrefix = {arXiv},
       eprint = {2409.18179},
 primaryClass = {astro-ph.GA},
       adsurl = {https://ui.adsabs.harvard.edu/abs/2024ApJ...975..187L}
}

@ARTICLE{Mawatari2026,
       author = {{Mawatari}, Ken and {Costantin}, Luca and {Usui}, Mitsutaka and {Hashimoto}, Takuya and {{\'A}lvarez-M{\'a}rquez}, Javier and {Sugahara}, Yuma and {Colina}, Luis and {Inoue}, Akio K. and {Osone}, Wataru and {Arribas}, Santiago and {Marques-Chaves}, Rui and {Nakazato}, Yurina and {Hagimoto}, Masato and {Hashigaya}, Takeshi and {Ceverino}, Daniel and {Yoshida}, Naoki and {Bakx}, Tom J.~L.~C. and {Fudamoto}, Yoshinobu and {Crespo G{\'o}mez}, Alejandro and {Matsuo}, Hiroshi and {Pereira-Santaella}, Miguel and {Blanco-Prieto}, Carmen and {Ren}, Yi W. and {Tamura}, Yoichi},
        title = "{RIOJA. A Clumpy Galaxy Assembly at Redshift 6.81 Revealed by JWST}",
      journal = {\apj},
         year = 2026,
        month = feb,
       volume = {998},
       number = {1},
          eid = {119},
        pages = {119},
          doi = {10.3847/1538-4357/ae3151},
archivePrefix = {arXiv},
       eprint = {2507.02053},
 primaryClass = {astro-ph.GA},
       adsurl = {https://ui.adsabs.harvard.edu/abs/2026ApJ...998..119M}
}

@ARTICLE{Tsujita2026,
       author = {{Tsujita}, Akiyoshi and {Fujimoto}, Seiji and {Faisst}, Andreas and {Boquien}, M{\'e}d{\'e}ric and {Li}, Juno and {Ferrara}, Andrea and {Battisti}, Andrew J. and {Dam}, Poulomi and {Aravena}, Manuel and {B{\'e}thermin}, Matthieu and {Casey}, Caitlin M. and {Cooper}, Olivia R. and {Finkelstein}, Steven L. and {Ginolfi}, Michele and {G{\'o}mez-Espinoza}, Diego A. and {Hadi}, Ali and {Herrera-Camus}, Rodrigo and {Ibar}, Edo and {Inami}, Hanae and {Jones}, Gareth C. and {Koekemoer}, Anton M. and {Kohno}, Kotaro and {Lemaux}, Brian C. and {Liu}, Zhaoxuan and {de Looze}, Ilse and {Mitsuhashi}, Ikki and {Mobasher}, Bahram and {Molina}, Juan and {Nanni}, Ambra and {Pozzi}, Francesca and {Reddy}, Naveen A. and {Relano}, Monica and {Rodighiero}, Giulia and {Romano}, Michael and {Sanders}, David B. and {Sawant}, Prasad and {Solimano}, Manuel and {Sommovigo}, Laura and {Spilker}, Justin and {Tadaki}, Ken-Ichi and {Vallini}, Livia and {Villanueva}, Vicente and {Wang}, Wuji and {Zamorani}, Giovanni and {Alpine+Cristal Collaborations}},
        title = "{The ALPINE-CRISTAL-JWST Survey: Stellar and Nebular Dust Attenuation of Main-sequence Galaxies at z {\ensuremath{\sim}} 4─6}",
      journal = {\apj},
         year = 2026,
        month = feb,
       volume = {997},
       number = {2},
          eid = {319},
        pages = {319},
          doi = {10.3847/1538-4357/ae22d8},
archivePrefix = {arXiv},
       eprint = {2510.18248},
 primaryClass = {astro-ph.GA},
       adsurl = {https://ui.adsabs.harvard.edu/abs/2026ApJ...997..319T}
}

@ARTICLE{Reddy2026,
       author = {{Reddy}, Naveen A. and {Shapley}, Alice E. and {Sanders}, Ryan L. and {Topping}, Michael W. and {Ellis}, Richard S. and {Pettini}, Max and {Brammer}, Gabriel and {Cullen}, Fergus and {F{\"o}rster Schreiber}, Natascha M. and {Khostovan}, Ali A. and {McLeod}, Derek J. and {McLure}, Ross J. and {Narayanan}, Desika and {Oesch}, Pascal A. and {Pahl}, Anthony J. and {Steidel}, Charles C. and {Berg}, Danielle A.},
        title = "{The AURORA Survey: Multiple Balmer and Paschen Emission Lines for Individual Star-forming Galaxies at z = 1.5─4.4. I. A Diversity of Nebular Attenuation Curves and Evidence for Non-unity Dust Covering Fractions}",
      journal = {\apj},
         year = 2026,
        month = mar,
       volume = {999},
       number = {1},
          eid = {15},
        pages = {15},
          doi = {10.3847/1538-4357/ae38da},
archivePrefix = {arXiv},
       eprint = {2506.17396},
 primaryClass = {astro-ph.GA},
       adsurl = {https://ui.adsabs.harvard.edu/abs/2026ApJ...999...15R}
}

@ARTICLE{Tailor2025,
       author = {{Tailor}, Vidhi and {Casasola}, Viviana and {Pozzi}, Francesca and {Calura}, Francesco and {Bianchi}, Simone and {Relano}, Monica and {Fritz}, Jacopo and {Galliano}, Fr{\'e}d{\'e}ric and {Bonato}, Matteo and {Lara-L{\'o}pez}, Maritza A. and {Paspaliaris}, Evangelos Dimitrios and {Traina}, Alberto},
        title = "{The role of young and evolved stars in the heating of dust in local galaxies}",
      journal = {\aap},
         year = 2025,
        month = sep,
       volume = {701},
          eid = {A74},
        pages = {A74},
          doi = {10.1051/0004-6361/202555091},
archivePrefix = {arXiv},
       eprint = {2507.12275},
 primaryClass = {astro-ph.GA},
       adsurl = {https://ui.adsabs.harvard.edu/abs/2025A&A...701A..74T}
}

@ARTICLE{Maheson2024,
       author = {{Maheson}, Gabriel and {Maiolino}, Roberto and {Curti}, Mirko and {Sanders}, Ryan and {Tacchella}, Sandro and {Sandles}, Lester},
        title = "{Unravelling the dust attenuation scaling relations and their evolution}",
      journal = {\mnras},
         year = 2024,
        month = jan,
       volume = {527},
       number = {3},
        pages = {8213-8233},
          doi = {10.1093/mnras/stad3685},
archivePrefix = {arXiv},
       eprint = {2306.00069},
 primaryClass = {astro-ph.GA},
       adsurl = {https://ui.adsabs.harvard.edu/abs/2024MNRAS.527.8213M}
}

@ARTICLE{Salim2018,
       author = {{Salim}, Samir and {Boquien}, M{\'e}d{\'e}ric and {Lee}, Janice C.},
        title = "{Dust Attenuation Curves in the Local Universe: Demographics and New Laws for Star-forming Galaxies and High-redshift Analogs}",
      journal = {\apj},
         year = 2018,
        month = may,
       volume = {859},
       number = {1},
          eid = {11},
        pages = {11},
          doi = {10.3847/1538-4357/aabf3c},
archivePrefix = {arXiv},
       eprint = {1804.05850},
 primaryClass = {astro-ph.GA},
       adsurl = {https://ui.adsabs.harvard.edu/abs/2018ApJ...859...11S}
}

@ARTICLE{Nersesian2019,
       author = {{Nersesian}, A. and {Xilouris}, E.~M. and {Bianchi}, S. and {Galliano}, F. and {Jones}, A.~P. and {Baes}, M. and {Casasola}, V. and {Cassar{\`a}}, L.~P. and {Clark}, C.~J.~R. and {Davies}, J.~I. and {Decleir}, M. and {Dobbels}, W. and {De Looze}, I. and {De Vis}, P. and {Fritz}, J. and {Galametz}, M. and {Madden}, S.~C. and {Mosenkov}, A.~V. and {Tr{\v{c}}ka}, A. and {Verstocken}, S. and {Viaene}, S. and {Lianou}, S.},
        title = "{Old and young stellar populations in DustPedia galaxies and their role in dust heating}",
      journal = {\aap},
         year = 2019,
        month = apr,
       volume = {624},
          eid = {A80},
        pages = {A80},
          doi = {10.1051/0004-6361/201935118},
archivePrefix = {arXiv},
       eprint = {1903.05933},
 primaryClass = {astro-ph.GA},
       adsurl = {https://ui.adsabs.harvard.edu/abs/2019A&A...624A..80N}
}

@ARTICLE{Nersesian2020,
       author = {{Nersesian}, Angelos and {Verstocken}, Sam and {Viaene}, S{\'e}bastien and {Baes}, Maarten and {Xilouris}, Emmanuel M. and {Bianchi}, Simone and {Casasola}, Viviana and {Clark}, Christopher J.~R. and {Davies}, Jonathan I. and {De Looze}, Ilse and {De Vis}, Pieter and {Dobbels}, Wouter and {Fritz}, Jacopo and {Galametz}, Maud and {Galliano}, Fr{\'e}d{\'e}ric and {Jones}, Anthony P. and {Madden}, Suzanne C. and {Mosenkov}, Aleksandr V. and {Tr{\v{c}}ka}, Ana and {Ysard}, Nathalie},
        title = "{High-resolution, 3D radiative transfer modelling. III. The DustPedia barred galaxies}",
      journal = {\aap},
         year = 2020,
        month = may,
       volume = {637},
          eid = {A25},
        pages = {A25},
          doi = {10.1051/0004-6361/201936176},
archivePrefix = {arXiv},
       eprint = {2004.03616},
 primaryClass = {astro-ph.GA},
       adsurl = {https://ui.adsabs.harvard.edu/abs/2020A&A...637A..25N}
}

@ARTICLE{Law2021,
       author = {{Law}, Ka-Hei and {Gordon}, Karl D. and {Misselt}, Karl A.},
        title = "{DirtyGrid II: An Analysis of the Dust and Stellar Properties in Nearby Star-forming Galaxies}",
      journal = {\apj},
         year = 2021,
        month = oct,
       volume = {920},
       number = {2},
          eid = {96},
        pages = {96},
          doi = {10.3847/1538-4357/ac1427},
       adsurl = {https://ui.adsabs.harvard.edu/abs/2021ApJ...920...96L}
}

@ARTICLE{Calzetti2025,
       author = {{Calzetti}, Daniela and {Kennicutt}, Robert C. and {Adamo}, Angela and {Sandstrom}, Karin and {Dale}, Daniel A. and {Elmegreen}, Bruce and {Gallagher}, John S. and {Gregg}, Benjamin and {Bajaj}, Varun and {B{\"o}ker}, Torsten and {Bortolini}, Giacomo and {Boyer}, Martha and {Correnti}, Matteo and {De Looze}, Ilse and {Draine}, Bruce T. and {Duarte-Cabral}, Ana and {Faustino Vieira}, Helena and {Grasha}, Kathryn and {Hunt}, L.~K. and {Johnson}, Kelsey E. and {Klessen}, Ralf S. and {Krumholz}, Mark R. and {Lai}, Thomas S.-Y. and {Lapeer}, Drew and {Linden}, Sean T. and {Messa}, Matteo and {{\"O}stlin}, G{\"o}ran and {Pedrini}, Alex and {Rela{\~n}o}, M{\`o}nica and {Sabbi}, Elena and {Schinnerer}, Eva and {Skillman}, Evan and {Smith}, Linda J. and {Tosi}, Monica and {Walter}, Fabian and {Weinbeck}, Tony D.},
        title = "{Quantification of the Age Dependence of Mid-infrared Star Formation Rate Indicators}",
      journal = {\apj},
         year = 2025,
        month = oct,
       volume = {991},
       number = {2},
          eid = {198},
        pages = {198},
          doi = {10.3847/1538-4357/adfbe0},
archivePrefix = {arXiv},
       eprint = {2508.08451},
 primaryClass = {astro-ph.GA},
       adsurl = {https://ui.adsabs.harvard.edu/abs/2025ApJ...991..198C}
}

@ARTICLE{Paspaliaris2021,
       author = {{Paspaliaris}, E.-D. and {Xilouris}, E.~M. and {Nersesian}, A. and {Masoura}, V.~A. and {Plionis}, M. and {Georgantopoulos}, I. and {Bianchi}, S. and {Katsioli}, S. and {Mountrichas}, G.},
        title = "{The physical properties of local (U)LIRGs: A comparison with nearby early- and late-type galaxies}",
      journal = {\aap},
         year = 2021,
        month = may,
       volume = {649},
          eid = {A137},
        pages = {A137},
          doi = {10.1051/0004-6361/202038605},
archivePrefix = {arXiv},
       eprint = {2102.03913},
 primaryClass = {astro-ph.GA},
       adsurl = {https://ui.adsabs.harvard.edu/abs/2021A&A...649A.137P}
}

@ARTICLE{Fujimoto2025,
       author = {{Fujimoto}, S. and {Ouchi}, M. and {Kohno}, K. and {Valentino}, F. and {Gim{\'e}nez-Arteaga}, C. and {Brammer}, G.~B. and {Furtak}, L.~J. and {Kohandel}, M. and {Oguri}, M. and {Pallottini}, A. and {Richard}, J. and {Zitrin}, A. and {Bauer}, F.~E. and {Boylan-Kolchin}, M. and {Dessauges-Zavadsky}, M. and {Egami}, E. and {Finkelstein}, S.~L. and {Ma}, Z. and {Smail}, I. and {Watson}, D. and {Hutchison}, T.~A. and {Rigby}, J.~R. and {Welch}, B.~D. and {Ao}, Y. and {Bradley}, L.~D. and {Caminha}, G.~B. and {Caputi}, K.~I. and {Espada}, D. and {Endsley}, R. and {Fudamoto}, Y. and {Gonz{\'a}lez-L{\'o}pez}, J. and {Hatsukade}, B. and {Koekemoer}, A.~M. and {Kokorev}, V. and {Laporte}, N. and {Lee}, M. and {Magdis}, G.~E. and {Ono}, Y. and {Rizzo}, F. and {Shibuya}, T. and {Shimasaku}, K. and {Sun}, F. and {Toft}, S. and {Umehata}, H. and {Wang}, T. and {Yajima}, H.},
        title = "{Primordial rotating disk composed of at least 15 dense star-forming clumps at cosmic dawn}",
      journal = {Nature Astronomy},
         year = 2025,
        month = aug,
       volume = {9},
        pages = {1553-1567},
          doi = {10.1038/s41550-025-02592-w},
archivePrefix = {arXiv},
       eprint = {2402.18543},
 primaryClass = {astro-ph.GA},
       adsurl = {https://ui.adsabs.harvard.edu/abs/2025NatAs...9.1553F}
}

@ARTICLE{Lorenz2026,
       author = {{Lorenz}, Brian and {Suess}, Katherine A. and {Kriek}, Mariska and {Price}, Sedona H. and {Leja}, Joel and {Atek}, Hakim and {Barailee}, Abhiyan and {Bezanson}, Rachel and {Brammer}, Gabriel and {Cutler}, Sam E. and {Dayal}, Pratika and {de Graaf}, Anna and {Greene}, Jenny E. and {Furtak}, Lukas J. and {Labbe}, Ivo and {Marchesini}, Danilo and {Maseda}, Michael V. and {Miller}, Tim B. and {Mintz}, Abby and {Mitsuhashi}, Ikki and {Nanayakkara}, Themiya and {Nelson}, Erica and {Pan}, Richard and {Porraz Barrera}, Natalia and {Wang}, Bingjie and {Weaver}, John R. and {Williams}, Christina C. and {Whitaker}, Katherine E.},
        title = "{Evidence for Shallow Nebular Attenuation Curves and Patchy Dust Geometry at z\raisebox{-0.5ex}\textasciitilde2 with Pa-beta/H-alpha Measurements from JWST-MegaScience Medium Band Photometry}",
      journal = {arXiv e-prints},
         year = 2026,
        month = feb,
          eid = {arXiv:2602.11418},
        pages = {arXiv:2602.11418},
          doi = {10.48550/arXiv.2602.11418},
archivePrefix = {arXiv},
       eprint = {2602.11418},
 primaryClass = {astro-ph.GA},
       adsurl = {https://ui.adsabs.harvard.edu/abs/2026arXiv260211418L}
}

@ARTICLE{Mitsuhashi2026,
       author = {{Mitsuhashi}, Ikki and {Zavala}, Jorge A. and {Bakx}, Tom J.~L.~C. and {Inoue}, Akio K. and {Castellano}, Marco and {Calabr{\`o}}, Antonello and {Casey}, Caitlin M. and {Franco}, Maximilien and {Hatsukade}, Bunyo and {Hathi}, Nimish P. and {Ikeda}, Ryota and {Koekemoer}, Anton M. and {Kartaltepe}, Jeyhan and {Knudsen}, Kirsten K. and {Santini}, Paola and {Saito}, Toshiki and {Terlevich}, Elena and {Terlevich}, Roberto and {Yung}, L.~Y. Aaron},
        title = "{Low Dust Mass and High Star Formation Efficiency at z > 12 from Deep ALMA Observations}",
      journal = {\apj},
         year = 2026,
        month = apr,
       volume = {1000},
       number = {2},
          eid = {159},
        pages = {159},
          doi = {10.3847/1538-4357/ae4511},
archivePrefix = {arXiv},
       eprint = {2501.19384},
 primaryClass = {astro-ph.GA},
       adsurl = {https://ui.adsabs.harvard.edu/abs/2026ApJ..1000..159M}
}

@ARTICLE{Shapley2023,
       author = {{Shapley}, Alice E. and {Sanders}, Ryan L. and {Reddy}, Naveen A. and {Topping}, Michael W. and {Brammer}, Gabriel B.},
        title = "{JWST/NIRSpec Balmer-line Measurements of Star Formation and Dust Attenuation at z   3-6}",
      journal = {\apj},
         year = 2023,
        month = sep,
       volume = {954},
       number = {2},
          eid = {157},
        pages = {157},
          doi = {10.3847/1538-4357/acea5a},
archivePrefix = {arXiv},
       eprint = {2301.03241},
 primaryClass = {astro-ph.GA},
       adsurl = {https://ui.adsabs.harvard.edu/abs/2023ApJ...954..157S}
}

@ARTICLE{Karthikeyan2026,
       author = {{Karthikeyan}, Shreya and {Clarke}, Leonardo and {Shapley}, Alice E. and {Lam}, Natalie and {Sanders}, Ryan L. and {Reddy}, Naveen A. and {Topping}, Michael W. and {Brammer}, Gabriel B.},
        title = "{Balmer Decrements and Nebular-Stellar Reddening in JADES Galaxies at $2.7<z<7$}",
      journal = {arXiv e-prints},
         year = 2026,
        month = mar,
          eid = {arXiv:2603.11338},
        pages = {arXiv:2603.11338},
          doi = {10.48550/arXiv.2603.11338},
archivePrefix = {arXiv},
       eprint = {2603.11338},
 primaryClass = {astro-ph.GA},
       adsurl = {https://ui.adsabs.harvard.edu/abs/2026arXiv260311338K}
}

@ARTICLE{Shivaei2025,
       author = {{Shivaei}, Irene and {Naidu}, Rohan P. and {Rodriguez Montero}, Francisco and {Matsumoto}, Kosei and {Leja}, Joel and {Matthee}, Jorryt and {Johnson}, Benjamin D. and {Oesch}, Pascal A. and {Chevallard}, Jacopo and {Adamo}, Angela and {Bodansky}, Sarah and {Bunker}, Andrew J. and {Covelo Paz}, Alba and {Di Cesare}, Claudia and {Egami}, Eiichi and {Furtak}, Lukas J. and {Heintz}, Kasper E. and {Kramarenko}, Ivan and {Meyer}, Romain A. and {Reddy}, Naveen A. and {Rinaldi}, Pierluigi and {Tacchella}, Sandro and {Torralba}, Alberto and {Witstok}, Joris and {Wozniak}, Michael A. and {Xiao}, Mengyuan},
        title = "{Diversity and Evolution of Dust Attenuation Curves from Redshift z \raisebox{-0.5ex}\textasciitilde 1 to 9}",
      journal = {arXiv e-prints},
         year = 2025,
        month = sep,
          eid = {arXiv:2509.01795},
        pages = {arXiv:2509.01795},
          doi = {10.48550/arXiv.2509.01795},
archivePrefix = {arXiv},
       eprint = {2509.01795},
 primaryClass = {astro-ph.GA},
       adsurl = {https://ui.adsabs.harvard.edu/abs/2025arXiv250901795S}
}

@ARTICLE{Wijesekera2026,
       author = {{Wijesekera}, J.~V. and {Koprowski}, M.~P. and {Dunlop}, J.~S. and {Lisiecki}, K. and {McLeod}, D.~J. and {McLure}, R.~J. and {Micha{\l}owski}, M.~J. and {Solar}, M.},
        title = "{Evolution of dust attenuation in star-forming galaxies with UV slope, stellar mass, and redshift out to $z \sim 5$}",
      journal = {arXiv e-prints},
         year = 2026,
        month = feb,
          eid = {arXiv:2602.04765},
        pages = {arXiv:2602.04765},
          doi = {10.48550/arXiv.2602.04765},
archivePrefix = {arXiv},
       eprint = {2602.04765},
 primaryClass = {astro-ph.GA},
       adsurl = {https://ui.adsabs.harvard.edu/abs/2026arXiv260204765W}
}

@ARTICLE{Shivaei2024,
       author = {{Shivaei}, Irene and {Alberts}, Stacey and {Florian}, Michael and {Rieke}, George and {Wuyts}, Stijn and {Bodansky}, Sarah and {Bunker}, Andrew J. and {Cameron}, Alex J. and {Curti}, Mirko and {D'Eugenio}, Francesco and {Dudzevi{\v{c}}i{\={u}}t{\.{e}}}, Ugn{\.{e}} and {Ji}, Zhiyuan and {Johnson}, Benjamin D. and {Kramarenko}, Ivan and {Lyu}, Jianwei and {Matthee}, Jorryt and {Morrison}, Jane and {Naidu}, Rohan and {P{\'e}rez-Gonz{\'a}lez}, Pablo G. and {Reddy}, Naveen and {Robertson}, Brant and {Sun}, Yang and {Tacchella}, Sandro and {Whitaker}, Katherine and {Williams}, Christina C. and {Willmer}, Christopher N.~A. and {Witstok}, Joris and {Xiao}, Mengyuan and {Zhu}, Yongda},
        title = "{A new census of dust and polycyclic aromatic hydrocarbons at z = 0.7─2 with JWST MIRI}",
      journal = {\aap},
         year = 2024,
        month = oct,
       volume = {690},
          eid = {A89},
        pages = {A89},
          doi = {10.1051/0004-6361/202449579},
archivePrefix = {arXiv},
       eprint = {2402.07989},
 primaryClass = {astro-ph.GA},
       adsurl = {https://ui.adsabs.harvard.edu/abs/2024A&A...690A..89S}
}

@ARTICLE{Qin2022,
       author = {{Qin}, Jianbo and {Zheng}, Xian Zhong and {Fang}, Min and {Pan}, Zhizheng and {Wuyts}, Stijn and {Shi}, Yong and {Peng}, Yingjie and {Gonzalez}, Valentino and {Bian}, Fuyan and {Huang}, Jia-Sheng and {Gu}, Qiu-Sheng and {Liu}, Wenhao and {Tan}, Qinghua and {Shi}, Dong Dong and {Ren}, Jian and {Zhang}, Yuheng and {Qiao}, Man and {Wen}, Run and {Liu}, Shuang},
        title = "{Systematic biases in determining dust attenuation curves through galaxy SED fitting}",
      journal = {\mnras},
         year = 2022,
        month = mar,
       volume = {511},
       number = {1},
        pages = {765-783},
          doi = {10.1093/mnras/stac132},
archivePrefix = {arXiv},
       eprint = {2201.05467},
 primaryClass = {astro-ph.GA},
       adsurl = {https://ui.adsabs.harvard.edu/abs/2022MNRAS.511..765Q}
}

@ARTICLE{Jain2024,
       author = {{Jain}, Shweta and {Tacchella}, Sandro and {Mosleh}, Moein},
        title = "{Self-regulated growth of galaxy sizes along the star-forming main sequence}",
      journal = {The Open Journal of Astrophysics},
         year = 2024,
        month = dec,
       volume = {7},
          eid = {113},
        pages = {113},
          doi = {10.33232/001c.126775},
archivePrefix = {arXiv},
       eprint = {2412.00599},
 primaryClass = {astro-ph.GA},
       adsurl = {https://ui.adsabs.harvard.edu/abs/2024OJAp....7E.113J}
}

@ARTICLE{Barro2013,
       author = {{Barro}, Guillermo and {Faber}, S.~M. and {P{\'e}rez-Gonz{\'a}lez}, Pablo G. and {Koo}, David C. and {Williams}, Christina C. and {Kocevski}, Dale D. and {Trump}, Jonathan R. and {Mozena}, Mark and {McGrath}, Elizabeth and {van der Wel}, Arjen and {Wuyts}, Stijn and {Bell}, Eric F. and {Croton}, Darren J. and {Ceverino}, Daniel and {Dekel}, Avishai and {Ashby}, M.~L.~N. and {Cheung}, Edmond and {Ferguson}, Henry C. and {Fontana}, Adriano and {Fang}, Jerome and {Giavalisco}, Mauro and {Grogin}, Norman A. and {Guo}, Yicheng and {Hathi}, Nimish P. and {Hopkins}, Philip F. and {Huang}, Kuang-Han and {Koekemoer}, Anton M. and {Kartaltepe}, Jeyhan S. and {Lee}, Kyoung-Soo and {Newman}, Jeffrey A. and {Porter}, Lauren A. and {Primack}, Joel R. and {Ryan}, Russell E. and {Rosario}, David and {Somerville}, Rachel S. and {Salvato}, Mara and {Hsu}, Li-Ting},
        title = "{CANDELS: The Progenitors of Compact Quiescent Galaxies at z \raisebox{-0.5ex}\textasciitilde 2}",
      journal = {\apj},
         year = 2013,
        month = mar,
       volume = {765},
       number = {2},
          eid = {104},
        pages = {104},
          doi = {10.1088/0004-637X/765/2/104},
archivePrefix = {arXiv},
       eprint = {1206.5000},
 primaryClass = {astro-ph.CO},
       adsurl = {https://ui.adsabs.harvard.edu/abs/2013ApJ...765..104B}
}

@ARTICLE{Barrufet2023,
       author = {{Barrufet}, L. and {Oesch}, P.~A. and {Weibel}, A. and {Brammer}, G. and {Bezanson}, R. and {Bouwens}, R. and {Fudamoto}, Y. and {Gonzalez}, V. and {Gottumukkala}, R. and {Illingworth}, G. and {Heintz}, K.~E. and {Holden}, B. and {Labbe}, I. and {Magee}, D. and {Naidu}, R.~P. and {Nelson}, E. and {Stefanon}, M. and {Smit}, R. and {van Dokkum}, P. and {Weaver}, J.~R. and {Williams}, C.~C.},
        title = "{Unveiling the nature of infrared bright, optically dark galaxies with early JWST data}",
      journal = {\mnras},
         year = 2023,
        month = jun,
       volume = {522},
       number = {1},
        pages = {449-456},
          doi = {10.1093/mnras/stad947},
archivePrefix = {arXiv},
       eprint = {2207.14733},
 primaryClass = {astro-ph.GA},
       adsurl = {https://ui.adsabs.harvard.edu/abs/2023MNRAS.522..449B}
}

@ARTICLE{Sun2506.06418,
       author = {{Sun}, Fengwu and {Yang}, Jinyi and {Wang}, Feige and {Eisenstein}, Daniel J. and {Decarli}, Roberto and {Fan}, Xiaohui and {Rieke}, George H. and {Ba{\~n}ados}, Eduardo and {Bosman}, Sarah E.~I. and {Cai}, Zheng and {Champagne}, Jaclyn B. and {Colina}, Luis and {D'Eugenio}, Francesco and {Fudamoto}, Yoshinobu and {Li}, Mingyu and {Lin}, Xiaojing and {Liu}, Weizhe and {Lyu}, Jianwei and {Mazzucchelli}, Chiara and {Jin}, Xiangyu and {Jun}, Hyunsung D. and {Wu}, Yunjing and {Zhang}, Huanian},
        title = "{The Identification of Two JWST/NIRCam-Dark Starburst Galaxies at $z=6.6$ with ALMA}",
      journal = {arXiv e-prints},
         year = 2025,
        month = jun,
          eid = {arXiv:2506.06418},
        pages = {arXiv:2506.06418},
          doi = {10.48550/arXiv.2506.06418},
archivePrefix = {arXiv},
       eprint = {2506.06418},
 primaryClass = {astro-ph.GA},
       adsurl = {https://ui.adsabs.harvard.edu/abs/2025arXiv250606418S}
}

@ARTICLE{Reddy2018,
       author = {{Reddy}, Naveen A. and {Oesch}, Pascal A. and {Bouwens}, Rychard J. and {Montes}, Mireia and {Illingworth}, Garth D. and {Steidel}, Charles C. and {van Dokkum}, Pieter G. and {Atek}, Hakim and {Carollo}, Marcella C. and {Cibinel}, Anna and {Holden}, Brad and {Labb{\'e}}, Ivo and {Magee}, Dan and {Morselli}, Laura and {Nelson}, Erica J. and {Wilkins}, Steve},
        title = "{The HDUV Survey: A Revised Assessment of the Relationship between UV Slope and Dust Attenuation for High-redshift Galaxies}",
      journal = {\apj},
         year = 2018,
        month = jan,
       volume = {853},
       number = {1},
          eid = {56},
        pages = {56},
          doi = {10.3847/1538-4357/aaa3e7},
archivePrefix = {arXiv},
       eprint = {1705.09302},
 primaryClass = {astro-ph.GA},
       adsurl = {https://ui.adsabs.harvard.edu/abs/2018ApJ...853...56R}
}

@ARTICLE{Xilouris1999,
       author = {{Xilouris}, E.~M. and {Byun}, Y.~I. and {Kylafis}, N.~D. and {Paleologou}, E.~V. and {Papamastorakis}, J.},
        title = "{Are spiral galaxies optically thin or thick?}",
      journal = {\aap},
         year = 1999,
        month = apr,
       volume = {344},
        pages = {868-878},
          doi = {10.48550/arXiv.astro-ph/9901158},
archivePrefix = {arXiv},
       eprint = {astro-ph/9901158},
 primaryClass = {astro-ph},
       adsurl = {https://ui.adsabs.harvard.edu/abs/1999A&A...344..868X}
}

@ARTICLE{Yoachim2006,
       author = {{Yoachim}, Peter and {Dalcanton}, Julianne J.},
        title = "{Structural Parameters of Thin and Thick Disks in Edge-on Disk Galaxies}",
      journal = {\aj},
         year = 2006,
        month = jan,
       volume = {131},
       number = {1},
        pages = {226-249},
          doi = {10.1086/497970},
archivePrefix = {arXiv},
       eprint = {astro-ph/0508460},
 primaryClass = {astro-ph},
       adsurl = {https://ui.adsabs.harvard.edu/abs/2006AJ....131..226Y}
}

@ARTICLE{DeGeyter2014,
       author = {{De Geyter}, Gert and {Baes}, Maarten and {Camps}, Peter and {Fritz}, Jacopo and {De Looze}, Ilse and {Hughes}, Thomas M. and {Viaene}, S{\'e}bastien and {Gentile}, Gianfranco},
        title = "{The distribution of interstellar dust in CALIFA edge-on galaxies via oligochromatic radiative transfer fitting}",
      journal = {\mnras},
         year = 2014,
        month = jun,
       volume = {441},
       number = {1},
        pages = {869-885},
          doi = {10.1093/mnras/stu612},
archivePrefix = {arXiv},
       eprint = {1403.7527},
 primaryClass = {astro-ph.GA},
       adsurl = {https://ui.adsabs.harvard.edu/abs/2014MNRAS.441..869D}
}

@ARTICLE{Schulz2020,
       author = {{Schulz}, Sebastian and {Popping}, Gerg{\"o} and {Pillepich}, Annalisa and {Nelson}, Dylan and {Vogelsberger}, Mark and {Marinacci}, Federico and {Hernquist}, Lars},
        title = "{A redshift-dependent IRX-{\ensuremath{\beta}} dust attenuation relation for TNG50 galaxies}",
      journal = {\mnras},
         year = 2020,
        month = oct,
       volume = {497},
       number = {4},
        pages = {4773-4794},
          doi = {10.1093/mnras/staa1900},
archivePrefix = {arXiv},
       eprint = {2001.04992},
 primaryClass = {astro-ph.GA},
       adsurl = {https://ui.adsabs.harvard.edu/abs/2020MNRAS.497.4773S}
}

@ARTICLE{Wuyts2011,
       author = {{Wuyts}, Stijn and {F{\"o}rster Schreiber}, Natascha M. and {van der Wel}, Arjen and {Magnelli}, Benjamin and {Guo}, Yicheng and {Genzel}, Reinhard and {Lutz}, Dieter and {Aussel}, Herv{\'e} and {Barro}, Guillermo and {Berta}, Stefano and {Cava}, Antonio and {Graci{\'a}-Carpio}, Javier and {Hathi}, Nimish P. and {Huang}, Kuang-Han and {Kocevski}, Dale D. and {Koekemoer}, Anton M. and {Lee}, Kyoung-Soo and {Le Floc'h}, Emeric and {McGrath}, Elizabeth J. and {Nordon}, Raanan and {Popesso}, Paola and {Pozzi}, Francesca and {Riguccini}, Laurie and {Rodighiero}, Giulia and {Saintonge}, Amelie and {Tacconi}, Linda},
        title = "{Galaxy Structure and Mode of Star Formation in the SFR-Mass Plane from z \raisebox{-0.5ex}\textasciitilde 2.5 to z \raisebox{-0.5ex}\textasciitilde 0.1}",
      journal = {\apj},
         year = 2011,
        month = dec,
       volume = {742},
       number = {2},
          eid = {96},
        pages = {96},
          doi = {10.1088/0004-637X/742/2/96},
archivePrefix = {arXiv},
       eprint = {1107.0317},
 primaryClass = {astro-ph.CO},
       adsurl = {https://ui.adsabs.harvard.edu/abs/2011ApJ...742...96W}
}

@ARTICLE{Wangec2018,
       author = {{Wang}, Enci and {Kong}, Xu and {Pan}, Zhizheng},
        title = "{Connecting Compact Star-forming and Extended Star-forming Galaxies at Low Redshift: Implications for Galaxy Compaction and Quenching}",
      journal = {\apj},
         year = 2018,
        month = sep,
       volume = {865},
       number = {1},
          eid = {49},
        pages = {49},
          doi = {10.3847/1538-4357/aadb9e},
archivePrefix = {arXiv},
       eprint = {1808.05929},
 primaryClass = {astro-ph.GA},
       adsurl = {https://ui.adsabs.harvard.edu/abs/2018ApJ...865...49W}
}

@ARTICLE{Brennan2017,
       author = {{Brennan}, Ryan and {Pandya}, Viraj and {Somerville}, Rachel S. and {Barro}, Guillermo and {Bluck}, Asa F.~L. and {Taylor}, Edward N. and {Wuyts}, Stijn and {Bell}, Eric F. and {Dekel}, Avishai and {Faber}, Sandra and {Ferguson}, Henry C. and {Koekemoer}, Anton M. and {Kurczynski}, Peter and {McIntosh}, Daniel H. and {Newman}, Jeffrey A. and {Primack}, Joel},
        title = "{The relationship between star formation activity and galaxy structural properties in CANDELS and a semi-analytic model}",
      journal = {\mnras},
         year = 2017,
        month = feb,
       volume = {465},
       number = {1},
        pages = {619-640},
          doi = {10.1093/mnras/stw2690},
archivePrefix = {arXiv},
       eprint = {1607.06075},
 primaryClass = {astro-ph.GA},
       adsurl = {https://ui.adsabs.harvard.edu/abs/2017MNRAS.465..619B}
}

@ARTICLE{Qiao2024,
       author = {{Qiao}, Man and {Zheng}, Xian Zhong and {Katsianis}, Antonios and {Qin}, Jianbo and {Pan}, Zhizheng and {Liu}, Wenhao and {Tan}, Qing-Hua and {An}, Fang Xia and {Shi}, Dong Dong and {Lyu}, Zongfei and {Zhang}, Yuheng and {Wen}, Run and {Liu}, Shuang and {Yang}, Chao},
        title = "{The dust attenuation scaling relation of star-forming galaxies in the EAGLE simulations}",
      journal = {\mnras},
         year = 2024,
        month = feb,
       volume = {528},
       number = {1},
        pages = {997-1015},
          doi = {10.1093/mnras/stae047},
archivePrefix = {arXiv},
       eprint = {2401.02875},
 primaryClass = {astro-ph.GA},
       adsurl = {https://ui.adsabs.harvard.edu/abs/2024MNRAS.528..997Q}
}

@ARTICLE{Graham2005,
       author = {{Graham}, Alister W. and {Driver}, Simon P.},
        title = "{A Concise Reference to (Projected) S{\'e}rsic R$^{1/n}$ Quantities, Including Concentration, Profile Slopes, Petrosian Indices, and Kron Magnitudes}",
      journal = {\pasa},
         year = 2005,
        month = jan,
       volume = {22},
       number = {2},
        pages = {118-127},
          doi = {10.1071/AS05001},
archivePrefix = {arXiv},
       eprint = {astro-ph/0503176},
 primaryClass = {astro-ph},
       adsurl = {https://ui.adsabs.harvard.edu/abs/2005PASA...22..118G}
}

@ARTICLE{Shen2003,
       author = {{Shen}, Shiyin and {Mo}, H.~J. and {White}, Simon D.~M. and {Blanton}, Michael R. and {Kauffmann}, Guinevere and {Voges}, Wolfgang and {Brinkmann}, J. and {Csabai}, Istvan},
        title = "{The size distribution of galaxies in the Sloan Digital Sky Survey}",
      journal = {\mnras},
         year = 2003,
        month = aug,
       volume = {343},
       number = {3},
        pages = {978-994},
          doi = {10.1046/j.1365-8711.2003.06740.x},
archivePrefix = {arXiv},
       eprint = {astro-ph/0301527},
 primaryClass = {astro-ph},
       adsurl = {https://ui.adsabs.harvard.edu/abs/2003MNRAS.343..978S}
}

@ARTICLE{Katsianis2021,
       author = {{Katsianis}, Antonios and {Yang}, Xiaohu and {Zheng}, Xianzhong},
        title = "{The Observed Cosmic Star Formation Rate Density Has an Evolution that Resembles a {\ensuremath{\Gamma}}(a, bt) Distribution and Can Be Described Successfully by Only Two Parameters}",
      journal = {\apj},
         year = 2021,
        month = oct,
       volume = {919},
       number = {2},
          eid = {88},
        pages = {88},
          doi = {10.3847/1538-4357/ac11f2},
archivePrefix = {arXiv},
       eprint = {2107.02733},
 primaryClass = {astro-ph.GA},
       adsurl = {https://ui.adsabs.harvard.edu/abs/2021ApJ...919...88K}
}

@ARTICLE{Zhang2026,
       author = {{Zhang}, Junkai and {Ramnichal}, Steven and {Wuyts}, Stijn and {Li}, Cheng},
        title = "{SE3D: Testing the recovery of stellar population, dust and structural properties on mock-observed toy model and simulated galaxies}",
      journal = {\mnras},
         year = 2026,
        month = may,
          doi = {10.1093/mnras/stag884},
archivePrefix = {arXiv},
       eprint = {2511.19614},
 primaryClass = {astro-ph.GA},
       adsurl = {https://ui.adsabs.harvard.edu/abs/2026MNRAS.tmp..853Z}
}

@ARTICLE{Ramnichal2026,
       author = {{Ramnichal}, Steven and {Zhang}, Junkai and {Wuyts}, Stijn and {Li}, Cheng},
        title = "{SE3D: Building a radiative transfer emulator to fit panchromatic resolved galaxy observations with 3D models of dust and stars}",
      journal = {\mnras},
         year = 2026,
        month = may,
       volume = {548},
       number = {1},
          eid = {stag533},
        pages = {stag533},
          doi = {10.1093/mnras/stag533},
archivePrefix = {arXiv},
       eprint = {2511.19623},
 primaryClass = {astro-ph.GA},
       adsurl = {https://ui.adsabs.harvard.edu/abs/2026MNRAS.548ag533R}
}

@ARTICLE{Casasola2020,
       author = {{Casasola}, V. and {Bianchi}, S. and {De Vis}, P. and {Magrini}, L. and {Corbelli}, E. and {Clark}, C.~J.~R. and {Fritz}, J. and {Nersesian}, A. and {Viaene}, S. and {Baes}, M. and {Cassar{\`a}}, L.~P. and {Davies}, J. and {De Looze}, I. and {Dobbels}, W. and {Galametz}, M. and {Galliano}, F. and {Jones}, A.~P. and {Madden}, S.~C. and {Mosenkov}, A.~V. and {Tr{\v{c}}ka}, A. and {Xilouris}, E.},
        title = "{The ISM scaling relations in DustPedia late-type galaxies: A benchmark study for the Local Universe}",
      journal = {\aap},
         year = 2020,
        month = jan,
       volume = {633},
          eid = {A100},
        pages = {A100},
          doi = {10.1051/0004-6361/201936665},
archivePrefix = {arXiv},
       eprint = {1911.09187},
 primaryClass = {astro-ph.GA},
       adsurl = {https://ui.adsabs.harvard.edu/abs/2020A&A...633A.100C}
}

@ARTICLE{DeVis2019,
       author = {{De Vis}, P. and {Jones}, A. and {Viaene}, S. and {Casasola}, V. and {Clark}, C.~J.~R. and {Baes}, M. and {Bianchi}, S. and {Cassara}, L.~P. and {Davies}, J.~I. and {De Looze}, I. and {Galametz}, M. and {Galliano}, F. and {Lianou}, S. and {Madden}, S. and {Manilla-Robles}, A. and {Mosenkov}, A.~V. and {Nersesian}, A. and {Roychowdhury}, S. and {Xilouris}, E.~M. and {Ysard}, N.},
        title = "{A systematic metallicity study of DustPedia galaxies reveals evolution in the dust-to-metal ratios}",
      journal = {\aap},
         year = 2019,
        month = mar,
       volume = {623},
          eid = {A5},
        pages = {A5},
          doi = {10.1051/0004-6361/201834444},
archivePrefix = {arXiv},
       eprint = {1901.09040},
 primaryClass = {astro-ph.GA},
       adsurl = {https://ui.adsabs.harvard.edu/abs/2019A&A...623A...5D}
}

@ARTICLE{Lee2025,
       author = {{Lee}, Jong Chul and {Lee}, Joon Hyeop and {Jeong}, Hyunjin and {Pak}, Mina and {Oh}, Sree},
        title = "{Spatially Resolved Star Formation Rate and Dust Attenuation of Nearby Galaxies from CALIFA, GALEX, and WISE Data}",
      journal = {\apj},
         year = 2025,
        month = jun,
       volume = {986},
       number = {2},
          eid = {143},
        pages = {143},
          doi = {10.3847/1538-4357/add1d6},
archivePrefix = {arXiv},
       eprint = {2504.19509},
 primaryClass = {astro-ph.GA},
       adsurl = {https://ui.adsabs.harvard.edu/abs/2025ApJ...986..143L}
}

@ARTICLE{Battisti2022,
       author = {{Battisti}, A.~J. and {Bagley}, M.~B. and {Baronchelli}, I. and {Dai}, Y.~S. and {Henry}, A.~L. and {Malkan}, M.~A. and {Alavi}, A. and {Calzetti}, D. and {Colbert}, J. and {McCarthy}, P.~J. and {Mehta}, V. and {Rafelski}, M. and {Scarlata}, C. and {Shivaei}, I. and {Wisnioski}, E.},
        title = "{The average dust attenuation curve at z   1.3 based on HST grism surveys}",
      journal = {\mnras},
         year = 2022,
        month = jul,
       volume = {513},
       number = {3},
        pages = {4431-4450},
          doi = {10.1093/mnras/stac1052},
archivePrefix = {arXiv},
       eprint = {2204.05553},
 primaryClass = {astro-ph.GA},
       adsurl = {https://ui.adsabs.harvard.edu/abs/2022MNRAS.513.4431B}
}

@ARTICLE{Allain1996,
       author = {{Allain}, T. and {Leach}, S. and {Sedlmayr}, E.},
        title = "{Photodestruction of PAHs in the interstellar medium. I. Photodissociation rates for the loss of an acetylenic group.}",
      journal = {\aap},
         year = 1996,
        month = jan,
       volume = {305},
        pages = {602},
       adsurl = {https://ui.adsabs.harvard.edu/abs/1996A&A...305..602A}
}

@ARTICLE{Jones2013,
       author = {{Jones}, A.~P. and {Fanciullo}, L. and {K{\"o}hler}, M. and {Verstraete}, L. and {Guillet}, V. and {Bocchio}, M. and {Ysard}, N.},
        title = "{The evolution of amorphous hydrocarbons in the ISM: dust modelling from a new vantage point}",
      journal = {\aap},
         year = 2013,
        month = oct,
       volume = {558},
          eid = {A62},
        pages = {A62},
          doi = {10.1051/0004-6361/201321686},
archivePrefix = {arXiv},
       eprint = {1411.6293},
 primaryClass = {astro-ph.GA},
       adsurl = {https://ui.adsabs.harvard.edu/abs/2013A&A...558A..62J}
}

@ARTICLE{Zhou2023,
       author = {{Zhou}, Shuang and {Li}, Cheng and {Li}, Niu and {Mo}, Houjun and {Yan}, Renbin and {Eracleous}, Michael and {Molina}, Mallory and {Gronwall}, Caryl and {Ajgaonkar}, Nikhil and {Cheng}, Zhuo and {Guo}, Ruonan},
        title = "{Mapping Dust Attenuation and the 2175 {\r{A}} Bump at Kiloparsec Scales in Nearby Galaxies}",
      journal = {\apj},
         year = 2023,
        month = nov,
       volume = {957},
       number = {2},
          eid = {75},
        pages = {75},
          doi = {10.3847/1538-4357/acfb80},
archivePrefix = {arXiv},
       eprint = {2212.01918},
 primaryClass = {astro-ph.GA},
       adsurl = {https://ui.adsabs.harvard.edu/abs/2023ApJ...957...75Z}
}

@ARTICLE{Salim2019,
       author = {{Salim}, Samir and {Boquien}, M{\'e}d{\'e}ric},
        title = "{Diversity of Galaxy Dust Attenuation Curves Drives the Scatter in the IRX-{\ensuremath{\beta}} Relation}",
      journal = {\apj},
         year = 2019,
        month = feb,
       volume = {872},
       number = {1},
          eid = {23},
        pages = {23},
          doi = {10.3847/1538-4357/aaf88a},
archivePrefix = {arXiv},
       eprint = {1812.05606},
 primaryClass = {astro-ph.GA},
       adsurl = {https://ui.adsabs.harvard.edu/abs/2019ApJ...872...23S}
}

\appendix
\onecolumn
\section{DENSITY-WEIGHTED FITTING METHODOLOGY}\label{AppendixA}
Here, we evaluate the efficacy of a density-weighted fitting scheme in mitigating the bias induced by the non-uniform sampling of galaxies in parameter space.  The fitting function f IRX relation includes multiple parameters (IRX, $12+\log({\mathrm{O/H}})$, $L_{\mathrm{IR}}$, $R_{\mathrm{e}}$, and $b/a$).  The maximum likelihood estimation (MLE) method used to derive the IRX relation can be dominated by data points in the high-density regime in the sample.  The residuals between observed and predicted IRX are directly used optimise the fitting. Consequently, the method preferentially minimises fitting residuals within the high-density regime of the sample. Approximately 91.2\,per\,cent of our sample galaxies have IRX between 0.3 and 1.5 in logarithm. At $\log {\mathrm{IRX}}<0.3$ and  $\log {\mathrm{IRX}}>1.5$, only 2.2\,per\,cent and 6.6\,per\,cent of the sample are distributed, respectively. This non-uniform distribution could heavily bias unweighted fits towards the dominant sample interval. Thus, we evaluate this bias by implementing a density-weighted MLE scheme compared with the unweighted MLE scheme adopted in \citetalias{Qin2019a}. A negative logarithm likelihood function is introduced by 
\begin{equation}
	\mathcal{L} = -\frac{1}{2}\sum_{i=1}^N\left[ w_i \cdot \frac{\left(\log\text{IRX}_{{\mathrm{obs}}, i} - \log\text{IRX}_{{\mathrm{mod}}, i}\right)^2}{\sigma_i^2} +\log\sigma_i^2 \right],
\end{equation}
where $\sigma_i^2$ is the total error variance for each data point, $1/\sigma_i^2$ is the intrinsic weight corresponding to the observational uncertainty, and $w_i$ is the additional density weight coefficient (set to $w_i=1$ in \citetalias{Qin2019a}). The weight $w_i$ is determined by the number density distribution of $\log {\mathrm{IRX}}$. To suppress biases by sample selection, we test three weight calculation schemes: inverse linear, inverse square root, and inverse logarithmic. All three schemes make $w_i$ inversely correlated with the $\log {\mathrm{IRX}}$ number density, which weakens the dominance of the high-density regime and enhances the fitting contribution of the sparse regimes at the low- and high-IRX ends. To prevent excessive amplification of weights in sparse regimes from introducing outlier interference, we applied a weight truncation threshold to all schemes. 

\begin{figure*}
	\centering
	\includegraphics[width=0.95\textwidth]{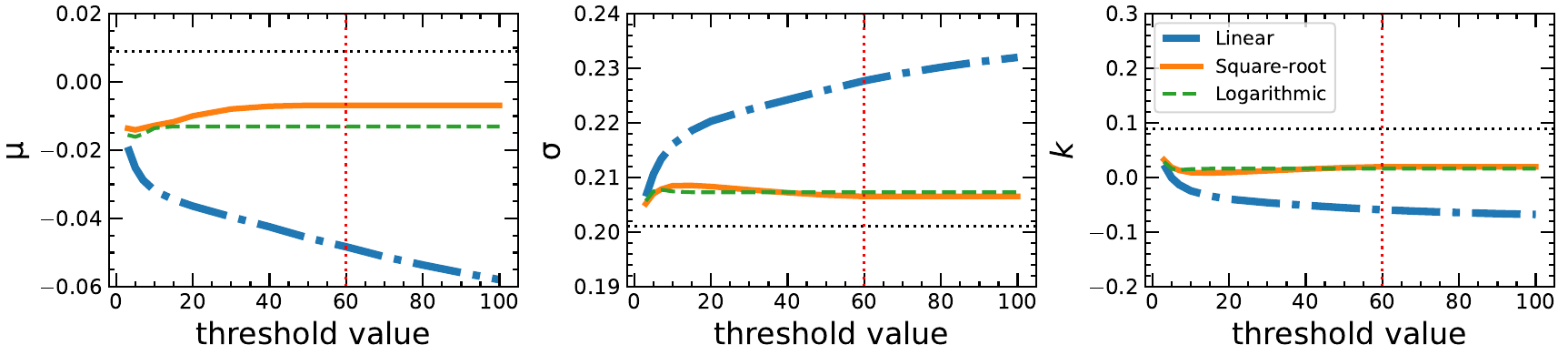}
	\caption{Three statistical metrics of fitting residuals, the residual mean $\mu$ (left), the residual standard deviation $\sigma$ (middle), and the residual tilt $k$ (right), as a function of weight truncation threshold, for three weight calculation methods:  inverse linear (blue dash-dotted), inverse square-root (orange solid), and inverse logarithmic (green dashed). The vertical dotted lines mark the adopted weight truncation threshold. The black dashed lines mark the corresponding values from the unweighted fit.}
	\label{figA1}
\end{figure*}

Figure~\ref{figA1} presents the test results of the three weighting schemes across different truncation thresholds. We use three statistical metrics to assess the best fit: the residual mean, the residual standard deviation, and a tilt (or slope) parameter $k$. These metrics evaluate the systematic bias, the overall scatter, and the residual trend of the fit, respectively. For an ideal fit, residuals would evenly distribute along the 1:1 equality line in the observed versus predicted $\log {\mathrm{IRX}}$ plane. However, we notice that the unweighted fitting from \citetalias{Qin2019a} resulted in a biased distribution: the fitting residuals are systematically positive at the high-IRX end and negative  at the low-IRX end. To rigorously quantify the systematic tilt of the data relative to this line, we apply a $45\degr$ geometric rotation to the coordinate system. We define $X'$ as the orthogonal projection coordinate of a data point along the 1:1 line, and $Y'$ as its orthogonal deviation perpendicular to the 1:1 line: 
\begin{equation}
	X' = \frac{\log {\mathrm{IRX}}_{\mathrm{predicted}} + \log {\mathrm{IRX}}_{\mathrm{observed}}}{\sqrt{2}},
\end{equation}
\begin{equation}
	Y' = \frac{\log {\mathrm{IRX}}_{\mathrm{observed}} - \log {\mathrm{IRX}}_{\mathrm{predicted}}}{\sqrt{2}} = \frac{\Delta \log {\mathrm{IRX}}}{\sqrt{2}}.
\end{equation}
We then perform a linear regression in this rotated parameter space: 
\begin{equation}
Y' = k \cdot X' + b.
\end{equation}
Here, the tilt parameter $k$ quantitatively measures the systematic trend of the geometric residuals. A completely unbiased fit uniformly distributed along the 1:1 line would yield $k = 0$. Our test results reveal that the inverse square root scheme performs best among all three schemes. While ensuring the residual mean remains close to zero and the standard deviation stays low, it significantly reduces the residual slope $k$ towards zero. Furthermore, all three statistical metrics would remain stable once the truncation threshold exceeds 60 times the median weight. We therefore adopt the inverse square root density weighting scheme, utilizing a truncation threshold of 60 times the median weight, as the optimal density-weighted MLE method.

\begin{figure}
	\centering
	\includegraphics[width=0.45\textwidth]{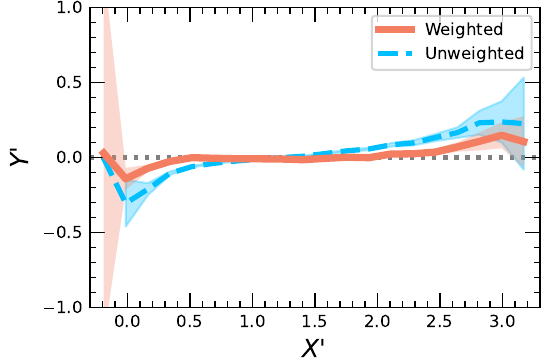}
	\caption{Distribution of orthogonal fitting residuals ($Y'$) as a function of the projection coordinate along the 1:1 line ($X'$). Here, $X' = (\log {\mathrm{IRX}}_{\mathrm{predicted}} + \log {\mathrm{IRX}}_{\mathrm{observed}})/\sqrt{2}$ represents the coordinate along the 1:1 line, and $Y' = \Delta \log {\mathrm{IRX}}/\sqrt{2}$ denotes the orthogonal deviation from it. The data are binned along the $X'$-axis to highlight the geometric tilt of the residuals. The orange solid line shows the mean orthogonal residual for the density-weighted fit using the inverse square-root scheme (with a truncation threshold of 60 times the median weight), whereas the grey solid line represents the unweighted fitting results. Shaded regions indicate the 95\,per\,cent confidence intervals.}
	\label{figA2}
\end{figure}

Figure~\ref{figA2} shows the residual trend for the optimal weighting scheme. Compared to the unweighted fitting scheme, the tilt parameter $k$ decreases from 0.09 to 0.02. This reduction significantly flattens the systematic trend in the residuals. Simultaneously, the residual mean shifts from 0.009 to $-$0.007. The residual standard deviation experiences a marginal increase from 0.201 to 0.207. These results indicate that the density-weighted MLE method effectively corrects the  residual trend across $\log {\mathrm{IRX}}$ distribution, with only a negligible impact on the overall bias and scatter. Therefore, when our sample populates a non-uniform distribution in parameter space, it is reasonable to sacrifice a minor degree of accuracy for the dominant population. This approach can help improve the applicability of the fitting results across the entire parameter space.

\section{SIZE DISTRIBUTIONS ACROSS THE $M_\ast$--SFR PLANE}\label{AppendixB}

To verify that the systematic size differences observed among the UV-luminosity subsamples are not artefacts of the stellar mass--size relation or the star-forming main sequence, we compare the three subsamples at fixed $M_\ast$ and SFR. We divide the $M_\ast$--SFR plane into a $4 \times 3$ grid comprising four stellar mass bins ($\log(M_\ast/{\rm M_\odot}) < 9.5$, $[9.5, 10.0)$,  $[10.0, 10.5)$, and $\ge 10.5$) and three regimes of star formation activity defined by the MS offset ($\Delta{\rm MS} \equiv \log{\rm SFR} - \log{\rm SFR}_{\rm MS}$): below ($\Delta{\rm MS} < -0.10$), on ($-0.10 \le \Delta{\rm MS} < 0.09$), and above the MS ($\Delta{\rm MS} \ge 0.09$).

\begin{figure*}
	\centering
	\includegraphics[width=0.95\textwidth]{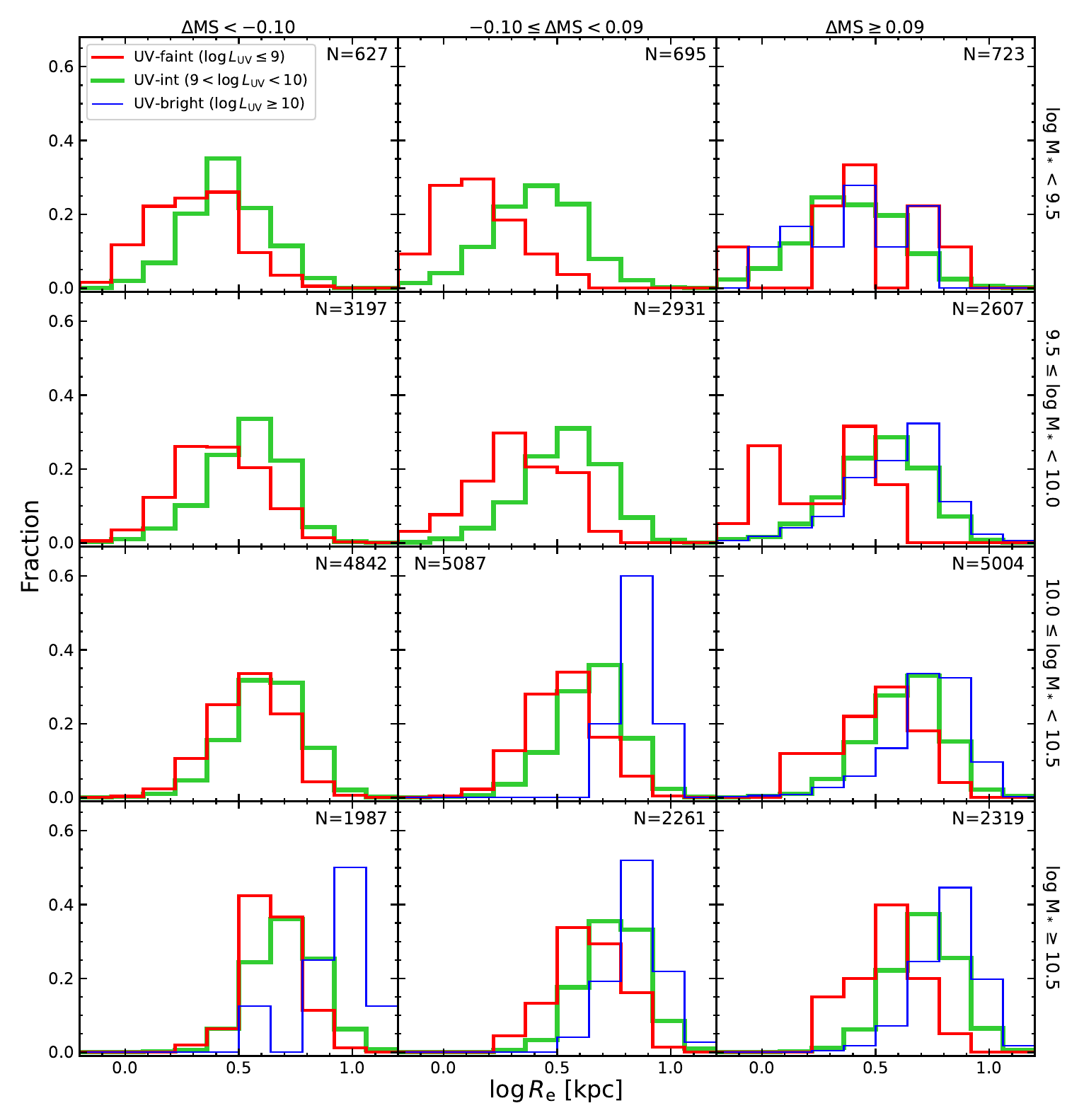}
	\caption{Comparison of the $R_{\rm e}$ distributions among the UV-faint (medium red line), UV-intermediate (thick green line), and UV-bright (thin blue line) subsamples, compared within a $4 \times 3$ grid across the $M_\ast$--SFR plane. Columns from left to right correspond to galaxies below the star-forming main sequence ($\Delta{\rm MS} < -0.10$), on the MS ($-0.10 \le \Delta{\rm MS} < 0.09$), and above the MS ($\Delta{\rm MS} \ge 0.09$). Rows from top to bottom indicate increasing stellar mass bins: $\log(M_\ast/{\rm M_\odot}) < 9.5$, $[9.5, 10.0)$, $[10.0, 10.5)$, and $\ge 10.5$. The total number of sample galaxies $N$ is shown in each panel. Bins lacking blue histograms contain fewer than 20 UV-bright galaxies, such that 7 of the 12 bins permit a three-way comparison across all UV brightness regimes.}
	\label{figB1}
\end{figure*}

Figure~\ref{figB1} presents the effective radius distributions of the UV-faint, UV-intermediate, and UV-bright SFGs across all 12 parameter cells. Across the entire grid, size differences persist at fixed $M_\ast$ and SFR, with UV-faint galaxies remaining systematically more compact, and UV-bright galaxies more extended than the UV-intermediate population, showing that these structural differences are not driven by the stellar mass--size relation or the star-forming main sequence. 

Taking the dominant UV-intermediate population as the baseline, the relative size offsets of the UV-faint and UV-bright populations vary systematically across the plane, primarily with stellar mass: in the low-mass regime ($\log(M_\ast/{\rm M_\odot}) < 10.0$), the size contrast is dominated by the pronounced compactness of UV-faint galaxies, whose median $\log R_{\rm e}$ falls $\sim$0.14--0.28\,dex below intermediate SFGs (median physical sizes of $R_{\rm e} \approx 1.4$--$2.5$\,kpc, compared to $2.5$--$3.6$\,kpc for the intermediate population). Towards higher stellar masses ($\log(M_*/{\rm M_\odot}) \ge 10.0$), UV-faint galaxies remain moderately smaller than intermediate SFGs (by $\sim$0.07--0.16\,dex), while UV-bright galaxies appear as extended systems that exceed the intermediate baseline by $\sim$0.1--0.2\,dex in median $\log R_{\rm e}$. We note that UV-bright galaxies are omitted in 5 of the 12 cells (primarily at lower masses and on or below the MS) due to an insufficient number of objects ($N < 20$). These systematic size offsets indicate that the star--dust geometry tracks the evolutionary state of SFGs across the $M_*$--SFR plane, and this geometry is closely linked to the effective dust optical depth and hence to the systematic deviations from the IRX scaling relation (see Section~\ref{sec4.5}).

\bsp	
\label{lastpage}
\end{document}